\documentclass[oneside,11pt]{article}
\usepackage[headheight=1in,margin=1.1in]{geometry}
\usepackage{lipsum}
\usepackage{amsfonts}
\usepackage{graphicx}
\usepackage{epstopdf}
\usepackage{natbib}
\usepackage{amssymb}
\usepackage{fancyhdr}
\usepackage{algorithmic}
\usepackage[colorlinks=false,allbordercolors={1 1 1}]{hyperref}
\usepackage[symbol]{footmisc}

\usepackage{natbib}
\usepackage[acronym]{glossaries}
\usepackage{bm}

\newacronym{ntd}{NNTuck}{nonnegative Tucker decomposition}

\newacronym{ntds}{NNTucks}{nonnegative Tucker decompositions}

\newacronym{sbm}{SBM}{stochastic block model}
\newacronym{dcmmsbm}{dc-mm-SBM}{degree-corrected, mixed-membership SBM}
\newacronym{sbms}{SBMs}{stochastic block models}
\newacronym{lrt}{LRT}{likelihood ratio test}
\newacronym{slrt}{split-LRT}{split likelihood ratio test}
\newacronym{lrts}{LRTs}{likelihood ratio tests}
\newacronym{caplrts}{LRTs}{Likelihood ratio tests}
\newacronym{css}{CSS}{cognitive social structure}
\newacronym{mt}{MT}{MULTITENSOR}

\newacronym{nmf}{NMF}{nonnegative matrix factorization}

\newacronym{em}{EM}{expectation maximization}

\newacronym{pmf}{PMF}{Poisson matrix factorization}
\newacronym{kld}{KL-divergence}{Kullback-Leibler divergence}
\newacronym{xval}{cross-validation}{cross-validation}

\newacronym{capntd}{NNTuck}{Nonnegative Tucker Decomposition}

\newacronym{capsbm}{SBM}{Stochastic block model}
\newacronym{capsbms}{SBMs}{Stochastic block models}

\newacronym{capnmf}{NMF}{Nonnegative Matrix Factorization}

\newacronym{capem}{EM}{Expectation Maximization}

\newacronym{cappmf}{PMF}{Poisson Matrix Factorization}

\newtheorem{example}{Example} 
\newtheorem{theorem}{Theorem}
\usepackage{multirow}
\usepackage{fancyvrb}
 
\newtheorem{proposition}{Proposition} 
\newtheorem{remark}[theorem]{Remark}

\newtheorem{definition}{Definition}
\newtheorem{proof}{Proof}

\usepackage{ctable}

\newcommand{\ten}[1]{\boldsymbol{\mathcal{#1}}}
\newcommand{\mat}[1]{\bm{#1}}
\renewcommand{\vec}[1]{\mathbf{#1}}
\newcommand{\Lam}{\boldsymbol{\Lambda}}
\newcommand{\R}{\mathbb{R}}
\newcommand{\G}{\boldsymbol{\mathcal{G}}}
\newcommand{\A}{\boldsymbol{\mathcal{A}}}
\newcommand{\hatA}{\boldsymbol{\mathcal{\hat{A}}}}

\newcommand{\U}{\bm{U}}
\newcommand{\V}{\bm{V}}
\newcommand{\Y}{\bm{Y}}
\newcommand{\Ysub}[1]{\bm{Y}_{#1}}
\newcommand{\Adims}{N \times N \times L}
\newcommand{\Gdims}{K \times K \times C}
\newcommand{\Udims}{N \times K}
\newcommand{\Ydims}{L \times C}
\newcommand{\Aspace}{\mathbb{Z}_{0+}}
\newcommand{\Gspace}{\mathbb{R}_{+}}
\newcommand{\by}{\times}
\newcommand{\uf}[1]{_{(#1)}}
\newcommand{\Ahat}{\boldsymbol{\mathcal{\hat{A}}}}
\newcommand{\Yhat}{\hat{\Y}}

\usepackage{listings}
\newcommand{\acs}[1]{\acrshort{#1}}
\newcommand{\acf}[1]{\acrfull{#1}}
\newcommand{\acl}[1]{\acrlong{#1}}
\usepackage{algorithm}[0.1]
\usepackage[capitalize, nameinlink]{cleveref}[0.19]

\newenvironment{keywords}
{\bgroup\leftskip 20pt\rightskip 20pt \small\noindent{\bf Keywords:} }%
{\par\egroup\vskip 0.25ex}
\glsdisablehyper
\begin{document}
 \title{A tensor factorization model of multilayer network interdependence}
 
\author{Izabel Aguiar\\ Stanford University \\ \texttt{izzya@stanford.edu} 
      \and
      Dane Taylor\\University of Wyoming\\  \texttt{dane.taylor@uwyo.edu}
      \and
      Johan Ugander\\ Stanford University\\ \texttt{jugander@stanford.edu}}

\maketitle

\begin{abstract}%
Multilayer networks describe the rich ways in which nodes are related by accounting for different relationships in separate layers. These multiple relationships are naturally represented by an adjacency tensor. In this work we study the use of the \acf{ntd} of such tensors under a KL loss as an expressive factor model that naturally generalizes existing stochastic block models of multilayer networks. Quantifying interdependencies between layers can identify redundancies in the structure of a network, indicate relationships between disparate layers, and potentially inform survey instruments for collecting social network data. We propose definitions of layer independence, dependence, and redundancy based on likelihood ratio tests between nested \acl{ntds}. Using both synthetic and real-world data, we evaluate the use and interpretation of the \acs{ntd} as a model of multilayer networks. Algorithmically, we show that using \acf{em} to maximize the log-likelihood under the \acs{ntd} is step-by-step equivalent to tensorial multiplicative updates for the \acs{ntd} under a KL loss, extending a previously known equivalence from nonnegative matrices to nonnegative tensors.
\end{abstract}

\section{Introduction}
\label{sec:intro}
Multilayer networks capture the many ways that a set of units can be connected:
through different types of relationships in a social network \citep{banerjee2013, power2017social, Breiger1975, sampson_crisis_1969}; at different time steps \citep{Carlen2019, finn2019use, taylor2021tunable}; through different types of interactions between genes or proteins \citep{de2015structural, larremore2013network}; or by different modes of transit in a transportation network \citep{de2014navigability, gallotti2015multilayer}. For more examples see \citet{kivela2014multilayer,boccaletti2014structure}. As more and more data take on a multilayer network form, tools for network analysis are being steadily adapted to multilayer contexts.
\begin{figure}
    \centering
    \includegraphics[width = 0.35\textwidth]{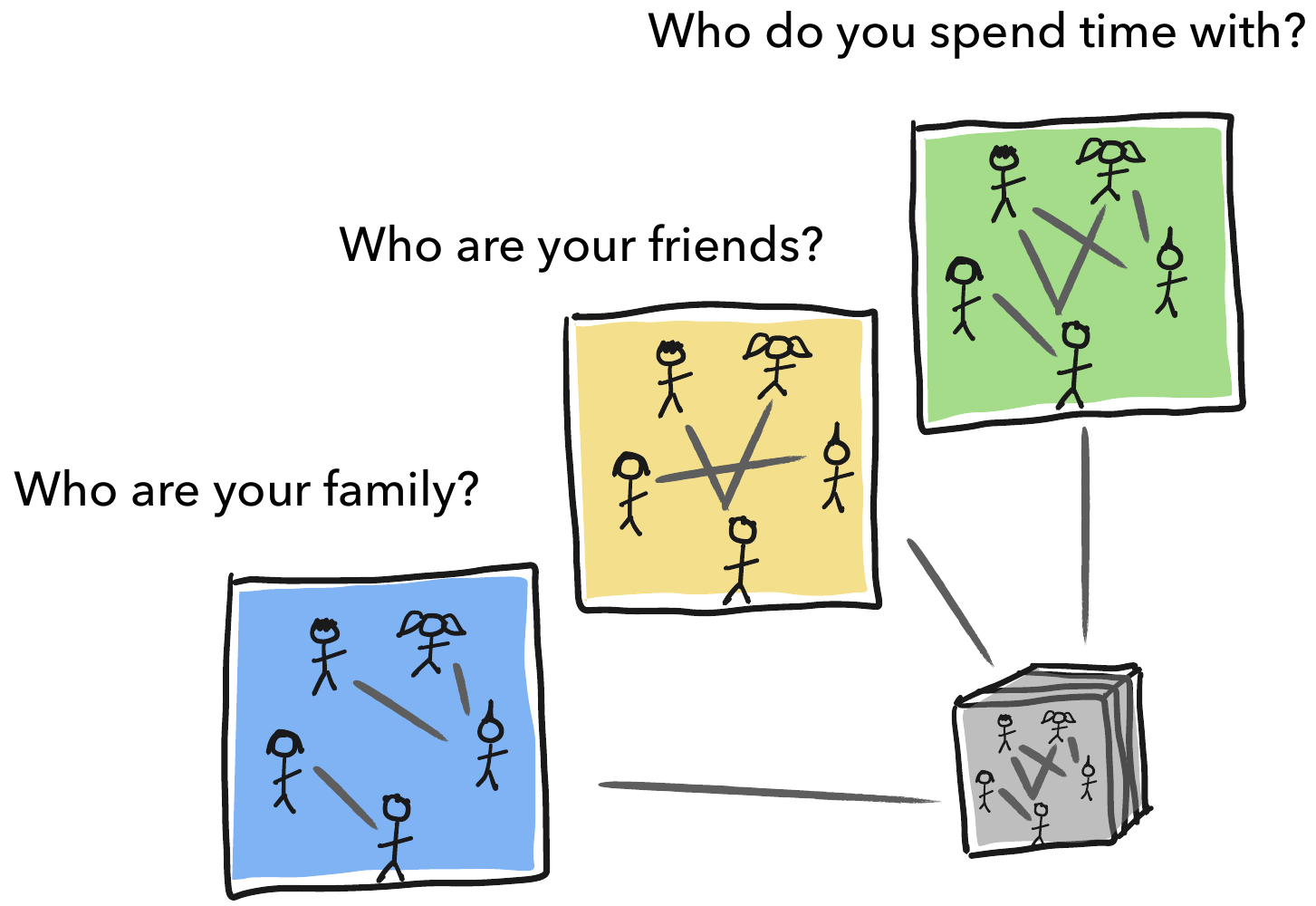}
    \includegraphics[width = 0.6\textwidth]{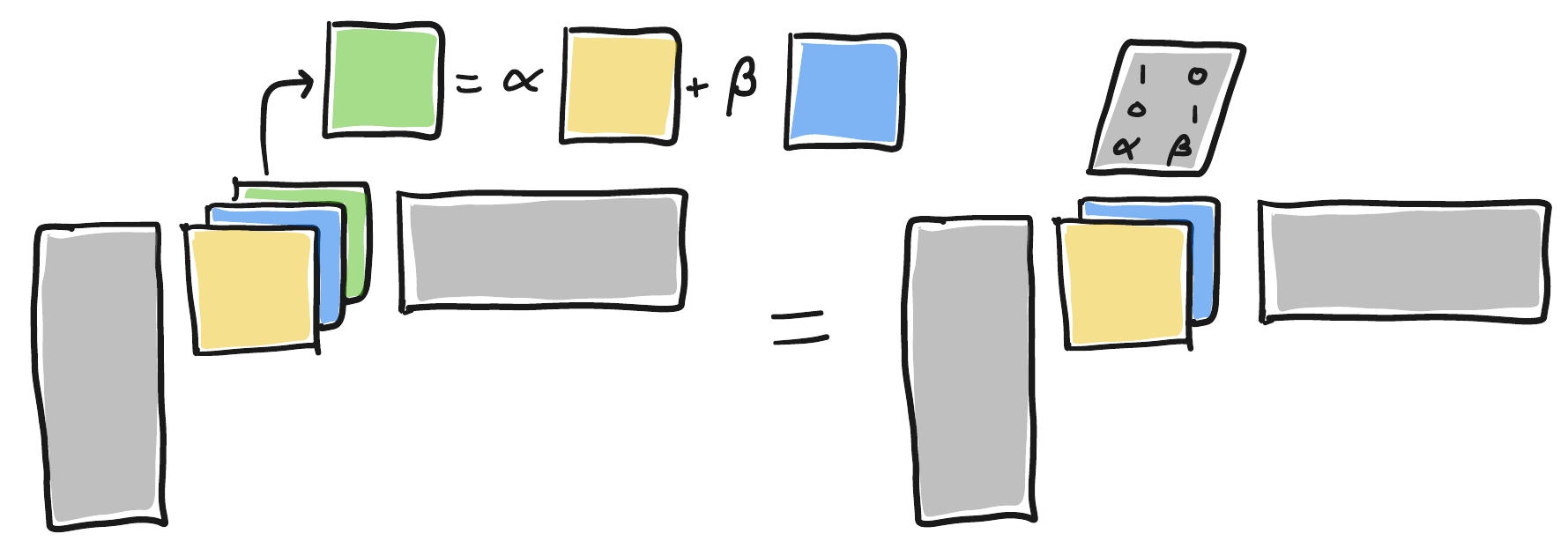}
    \caption{Multilayer networks account for the reality and variety of ways in which nodes interact in a system. In this example, a social network is complexly defined by three different types of social interaction and is represented by a tensor with three frontal slices. In this example, the process generating the ``spend time with'' layer is a linear combination of those processes generating the ``family'' and ``friends'' layers. On the right, we see a visual representation of the \acl{ntd} of this network and how the third factor matrix accounts for these linearly dependent layers. 
} 
    \label{fig:mln}
\end{figure}

Recent work has productively cast the study of multilayer community structure in the language of multilinear algebra \citep{wu2019tensor}, furnishing tensor-based definitions of multilayer \acf{sbms} \citep{schein2016bayesian, deBacco, gauvin2014detecting, Carlen2019, tarres2019tensorial}. 
We extend and generalize these efforts, connecting the tensorial \acf{ntd} with \acs{kld} to the statistical inference of multilayer \acs{sbms}. 
We show that minimizing the \acs{kld} of the \acs{ntd} is exactly equivalent to maximizing the log-likelihood of observing a multilayer network assumed to have been generated from a Poisson model with parameters defined by the \acs{ntd}. 
In this sense the \acs{ntd} identifies a natural generalization of existing multilayer \acs{sbms}, and as such, can be used for community detection and link prediction.

We investigate the use of the \acl{ntd} to identify and statistically define layer interdependence. 
The vocabulary around assessing interdependence amongst the layers of a multilayer network is scattered across the literature \citep{battiston2014structural, de2015structural, stanley2016}. In this work, we use the term \textit{interdependence} colloquially, to refer to the concept of dependence between layers in an abstract way without specificity about \textit{how} the layers are dependent. Conversely, we use and define the terms of layer dependence, independence, and redundance in specific ways, as defined either by a model or by a statistical test. These terms all specify \textit{what type} of interdependence is present in a multilayer network.

The ability to quantify and identify interdependencies between layers has the potential to inform survey instruments for collecting social network data, identify redundancies in the structure of a network, and indicate relationships between disparate layers. Reducing a multilayer network through identifying layer interdependence is both theoretically and practically appealing; although there are likely many situations where it is specifically useful, we highlight three. First, and as noted in \citet{de2015structural}, basic structural properties of multilayer networks\textemdash like centrality, clustering coefficient, and distance\textemdash ``scale superlinearly or even exponentially with the number of layers.'' Second, in addition to this compelling computational motivation to identify layer redundancies, there is practical motivation as well. As discussed in \cite{deBacco}, understanding layer redundancies in terms of social connections can help in data \textit{collection}, as well as analysis. For example, if layers of a social network are identified as redundant, it could justify the choice to not collect those layers for future data collection in similar settings. Third, identifying redundant layers, and aggregating those layers, can help enhance structural features of networks \citep{nayar2015, taylor2016enhanced, taylor2017super}. 

We build upon these motivations from previous work \citep{schein2016bayesian, de2015structural, stanley2016,  de2016spectral, deBacco, kao2018layer} and develop the \acs{ntd} as a natural way to identify a latent space in the dimension of the layers. Analogous to how the factor matrices in the single layer \acs{sbm} identify {\it node communities}, the additional third factor matrix in the \acs{ntd} identifies {\it layer communities} (see \Cref{fig:mln} for a visualization). As such, the third factor matrix of the \acs{ntd} allows for the adjacency tensor to be low rank in the layer dimension. 

Analyzing the third factor matrix is a significant focus of our work, and we propose three methods for interpreting it to quantify layer interdependence based on the structure of that factor matrix. Furthermore, we propose definitions of layer interdependence based on \acf{lrts} between different models of the data differing in the structure of that third factor matrix. To address concerns with using the traditional \acs{lrt} in latent factor models, we also implement the \acs{slrt} from \citet{wasserman2020}, which requires no regularity conditions. We use these models and tests to classify a variety of empirical networks as layer independent, dependent, or redundant, and find layer independence in a biological multilayer network, layer dependence in a cognitive social structure, and layer redundance in a collection of multilayer social support networks.

The structure of this work is as follows. In \Cref{background} we discuss and define the notation of \acf{sbms}, multilayer networks, and previous and related approaches to multilayer \acs{sbm}s. In \Cref{NNTuck} we define the \acf{ntd} and its notation, discuss the connection of its definition under KL-divergence to \acl{sbms}, motivate using the multiplicative updates algorithm from \cite{kimChoi}, and offer an interpretation of the low-dimensional third factor matrix as describing the dependence between the layers of a multilayer network. In \Cref{deflation} we discuss the use of the \acs{ntd} to empirically validate layer interdependence and define \acl{lrt}-based definitions.

In \Cref{applications} we use \acs{xval} to select the \acs{ntd}'s hyper-parameters $K$ and $C$. We discuss \acs{xval} based on two link prediction tasks:
\textit{independent link prediction}, in which elements of the adjacency tensor are missing independently and according to an identical uniform distribution, and \textit{tubular link prediction}, in which entire tubes of the adjacency tensor (see \cref{fibers} for a visualization of \textit{tube fibers} in a third-order tensor) are missing (i.i.d.). In \Cref{empirical} we use the \acs{ntd} to analyze layer dependence in practice for: two synthetic networks; the \acl{css} dataset from \citet{krackhardt1987}; a biological multilayer network from \citet{larremore2013network}; a social support multilayer network from \citet{banerjee2013};
and 113 other multilayer social support networks from \citet{banerjee2013, banerjee2019}.
We conclude in \Cref{conclusion} by discussing our work and indicating future directions of research. 
\section{Background}
\label{background}
In this section we discuss related work and define notation and vocabulary.  For easy reference, the primary notation is organized in \Cref{tab:notation}. We present \acf{sbms} in \Cref{sec:sbm} and \acf{nmf} in \Cref{sec:nmf}. In \Cref{subsec:tensorVocab} we introduce tensor vocabulary and notation used throughout the work and review the Tucker decomposition. In \Cref{sec:mln} we discuss multilayer networks and in \Cref{sec:related} we present a brief survey of related work, summarized in \Cref{tab:mlSBM-summary}. 

\begin{table}[t]
    \centering
    \begin{tabular}{p{2.2cm}|| p{10cm}}
        $\mat{A}$ & The $(N \by N)$ \textbf{adjacency matrix} of a single-layer network with $N$ nodes. \\
        $\A $ & The $(N \by N \by L)$ \textbf{adjacency tensor} of a multilayer network with $N$ nodes and $L$ layers. The \textbf{Tucker Decomposition} of $\A$ is given by $\A = \G \times_1 \U \times_2 \V \times_3 \Y$.\\
        $\mat{U}$ & The $(N \by K)$ \textbf{\textit{outgoing} community membership matrix} in an \acs{sbm} for directed networks, where $K$ is the number of communities generating the network. For an undirected networks $\mat{U} = \mat{V}$ (see below) and we simply call $\mat{U}$ the \textbf{community membership matrix}.\\
        $\mat{V}$ & The $(N \by K)$ \textbf{\textit{incoming} community membership matrix} in an \acs{sbm} for a directed network. For an undirected network, $\mat{U} = \mat{V}$.\\
        $\Y$ & The \textbf{third factor matrix} in the Tucker Decomposition, also referred to as the \textbf{layer interdependence matrix}. \\
        $\mat{G}$ & The $(K \by K)$\textbf{ affinity matrix} describing the rate at which nodes in different communities form edges with one another in an \acs{sbm}.\\
        $\G$ & The \textbf{core tensor} in the Tucker Decomposition.  \\
    \end{tabular}
    \caption{The notation and definitions for vocabulary that will be used throughout this paper.}
    \label{tab:notation}
\end{table}

\subsection{Stochastic Block Model (SBM)}\label{sec:sbm}

\acf{capsbms} identify latent groups of nodes and the density of connections between nodes in these groups as a descriptive and/or generative tool for analyzing networks. Introduced by \citet{White1976} and expanded by \citet{Holland1983},
\acs{sbms} decompose a network into factors that aim to uncover group structure, identify to which groups each node belongs, and describe how nodes in these groups form connections with one another. Beyond these context-specific questions,
\acs{sbms} identify low-dimensional structure in a network by grouping nodes into latent communities. Extensions of the original \acs{sbm} have allowed for the model to account for heterogeneous degree distributions \citep{KarrerNewman}, nodes belonging to multiple \textit{overlapping} communities (sometimes referred to as \textit{mixed-membership)} \citep{BallKarrerNewman2011}, and Bayesian approaches \citep{Airoldi2008}. 
Here, we focus on generalizing the \textit{\acf{dcmmsbm}} \citep{BallKarrerNewman2011} to multilayer networks. Since we use much of the same framework to build our model for multilayer networks, we begin by describing the \acs{dcmmsbm} in depth.

For a network with adjacency matrix $\mat{A}\in \Aspace^{N \by N}$, the \acs{dcmmsbm} assumes that each node $i$ has outgoing and incoming nonnegative membership row vectors of dimension $K$, $\vec{u}_i$ and $\vec{v}_i$ respectively, describing node $i$'s membership to $K$ different groups when forming outgoing and incoming edges ($\vec{u}_i = \vec{v}_i$ when the network is undirected). A nonnegative \textit{affinity matrix} $\mat{G}$ describes the rate at which nodes in different groups form edges with one another. Given $\U,\V\in \Gspace^{\Udims}$ and $\mat{G} \in \Gspace^{K \by K}$ the \acs{dcmmsbm} assumes each edge is an independent realization of the Poisson distribution, 
\begin{equation*}
    a_{ij} \sim \text{Poisson}(\vec{u}_i \vec{G} \vec{v}_j^\top), \text{ for all } i, j = 1, \dots, N.
\end{equation*}
For a more detailed discussion on the common modeling choice to  use the Poisson distribution (as opposed to, e.g., a Bernoulli distribution), see \citet{zhao2012consistency}. Written this way, we see that $\vec{u}_i \vec{G} \vec{v}_j^T$ must be positive in order to specify a valid  Poisson rate. By requiring all the elements $\vec{u}_i, \vec{G},  \vec{v}_i$ to be independently nonnegative, the parameters are all interpretable as membership weights and affinities. In matrix form, we then have, 
\begin{equation}
    \vec{A} \sim \text{Poisson}(\U \vec{G} \V^\top).
    \label{bkn-sbm}
\end{equation}
We estimate $\U, \V$, and $\mat{G}$ by maximizing the log-likelihood of observing $\vec{A}$ under this model. Note that both weighted and unweighted networks can be described using this model and the likelihood maximized all the same.

The formulation given by \eqref{bkn-sbm} incorporates both the degree-corrected \acs{sbm} (dc-SBM) \citep{KarrerNewman} and the mixed-membership \acs{sbm} (mm-SBM) \citep{BallKarrerNewman2011}. 
In the dc-SBM each node may only belong to one of $K$ groups but may have heterogeneous degree distribution. In the mm-SBM each node may belong, in part, to each of $K$ groups but their memberships must sum to one. To account for both, the \acs{dcmmsbm} assumes that each node has a scalar parameter $\theta_i > 0$ describing its gregariousness. Equivalently, each node's membership vector absorbs this degree parameter such that $\vec{u}_i = \theta_i \vec{s}_i$ and $\vec{v}_i = \theta_i \vec{t}_i$ for normalized membership vectors $\sum_{k} \vec{s}_{k} = 1$ and $\sum_{k} \vec{t}_{k} = 1$ for $\vec{s}_k, \vec{t}_k \geq 0$. We will henceforth describe this type of community membership as \textit{soft membership}. Such an approach allows nodes to have membership across multiple groups while also allowing for a heterogeneous degree distribution across nodes. 

There is a direct connection between the \acs{dcmmsbm} and \acf{pmf} \citep{gopalan2013scalable}. \acs{pmf} assumes that the entries of $\mat{A}$ are realizations of a Poisson distribution with rate parameters given by the product of $\mat{W} \in \Gspace^{N \by K}$ and $\mat{H} \in \Gspace^{K \by N}$. That is,
$
    a_{ij} \sim \text{Poisson}\left(\sum_k w_{ik}h_{kj}\right).
$
Dropping the constant term, the log-likelihood of observing $\mat{A}$ under this distribution is given by,
\begin{equation}
    \mathcal{L}(\mat{A} | \mat{W},\mat{H}) = \sum_{ij} a_{ij}\,\,\text{log} \sum_{k} w_{ik}h_{k j} - \sum_{k} w_{ik}h_{k j}.
    \label{eq:PMF_likeli}
\end{equation}
The factors $\U, \mat{G},$ and $\V$ can be grouped together such that we can consider the \acs{dcmmsbm} as equivalent to the \acl{pmf} (e.g., if we define $\mat{W}= \U \mat{G}$ and $\mat{H} = \V^\top$, then the nonnegative factors of the \acs{dcmmsbm} also describe the nonnegative factors of a \acl{pmf}).
\subsection{Nonnegative Matrix Factorization (NMF)}\label{sec:nmf}
A related approach for finding latent structure in a matrix, \acf{nmf} \citep{nmf1994, nmf1999}, aims to factor nonnegative matrix $\mat{A}$ into two nonnegative factor matrices, $\mat{W}$ and $\mat{H}$, for $\mat{W}\in \Gspace^{\Udims}$ and $\mat{H}\in \Gspace^{K \times N}$. 

Not every matrix can be exactly factorized in this way, and although for these cases we are actually finding the nonnegative matrix \textit{approximation}, we will henceforth refer to both problems as \acs{nmf}. When estimating the \acs{nmf} of a matrix $\mat{A}$ there are many loss functions with respect to which the factorization may be optimized. For reasons that will be clearly motivated in the following sections, we focus on minimizing the \textit{\acs{kld}} between the matrix and its factorization, defined as
\begin{equation}
    \begin{aligned}
    &  \text{KL}(\mat{A} \|  \mat{WH}) = \sum_{ij}\left( a_{ij} \,\,\text{log} \frac{a_{ij}}{(\mat{WH})_{ij}} \,- \,a_{ij} + (\mat{WH})_{ij} \right).
    \end{aligned}
    \label{eq:kl-div}
\end{equation}
An algorithm based on multiplicative updates for minimizing KL-divergence was developed by \citet{LeeSeung} and is widely used to find local optima of the non-convex optimization problem given by \cref{eq:kl-div}. This algorithm guarantees that, given nonnegative initializations, factor matrices $\mat{W}$ and $\mat{H}$ remain nonnegative throughout the optimization. Furthermore, the algorithm guarantees monotonic convergence to a local minimum. 

It is known that \textit{maximizing} the log-likelihood in \acs{pmf} \cref{eq:PMF_likeli} is equivalent to \textit{minimizing} the KL-divergence in \acs{nmf}:
\begin{equation}
    \begin{aligned}
& \underset{\mat{W},\mat{H}}{\text{minimize}} \,\,\,\text{KL}(\mat{A}||\mat{WH}) \\&\Leftrightarrow \underset{\mat{W},\mat{H}}{\text{minimize}} \,\, \sum_{ij}\left( a_{ij} \,\,\text{log} {a_{ij}} - a_{ij} \,\, \text{log} {(\mat{WH})_{ij}} \,- \,a_{ij} + (\mat{WH})_{ij} \right)\\
&\Leftrightarrow \underset{\mat{W},\mat{H}}{\text{minimize}} - \sum_{ij}\left( a_{ij} \text{log} {(\mat{WH})_{ij}} \, - (\mat{WH})_{ij} \right)\\
&\Leftrightarrow \underset{\mat{W},\mat{H}}{\text{maximize}} \,\,\mathcal{L}(\mat{A}|\mat{W},\mat{H}).
\end{aligned}
\label{eq:kleqLog}
\end{equation}
Furthermore, as was first noted in \citet{fevotte}, using \acf{em} to find a local maximum of the log-likelihood for \acs{pmf} is step-by-step equivalent to using the multiplicative updates given in \citet{LeeSeung} to minimize KL-divergence. This equivalence does not hold when comparing \acs{em} updates under a Gaussian generative model to the multiplicative updates under a Frobenius loss. This observation, in combination with the equivalence in \cref{eq:kleqLog}, gives \acs{nmf} with KL-divergence a strong statistical foundation. Specifically, there are two important connections to be made: (i) to \textit{factorize} matrix $\mat{A}$ into a product of two nonnegative matrices by maximizing log-likelihood in \acs{pmf} and minimizing KL-divergence in \acs{nmf} are equivalent optimization problems, and (ii) the \textit{algorithms} (Lee and Seung's multiplicative updates and \acs{em}) by which to find the model associated with a local minimum of the shared loss function is the same for \acs{pmf} and \acs{nmf}. 

\subsection{Tensor Notation and Tucker Decomposition}\label{subsec:tensorVocab}

To facilitate a clear analysis and discussion of the tensor-based model in \Cref{NNTuck}, we now define the tensor-specific vocabulary and notation that we will use throughout this paper.
For a more thorough overview of tensor vocabulary, methods, decompositions, and definitions, see \citet{Kolda2009} for an excellent review.
We focus notation and terms to \textit{third-order} ``frontally square'' tensors $\ten{X}$ of dimension $N \times N \times L$.

\paragraph{Frontal slices} 
The frontal slices of $\ten{X}$ are the $L$ matrices of size $N \by N$ that, when stacked together, form the $N \times N \times L$ tensor. A depiction of frontal slices can be found at the bottom left of \cref{fig:unfolding}. We denote the $\ell$th frontal slice of $\ten{X}$ as $\vec{X}_\ell$. The frontal slice of an adjacency tensor $\ten{A}$ corresponds to the adjacency matrix $\mat{A}_\ell$ of a particular layer $\ell$ of the multilayer network, and thus we will make frequent references to it.

\paragraph{Tensor fibers}
Analogous to rows and columns in matrices, third-order tensors have what are called \textit{row}, \textit{column}, and \textit{tube} fibers, denoted $\ten{X}_{:jk}$, $\ten{X}_{i:k}$, and $\ten{X}_{ij:}$, respectively. See \Cref{fibers} for a visualization of each.

\paragraph{Unfoldings} A third-order tensor has three unfoldings: the 1-unfolding, 2-unfolding, and 3-unfolding. These are higher-dimensional equivalents to \textit{vectorizing} a matrix. The $n$-unfolding of a third-order tensor stacks its column, row, or tube fibers to form a matrix, and is denoted by $\vec{X}_{(n)}$. See \Cref{fig:unfolding} and Section~2 of \citet{Kolda2009} for helpful visualizations.

\paragraph{The tensor $n$-mode product ($\times_n$)} A third-order tensor can be multiplied by a matrix through a 1-, 2-, or 3-mode product. Dimensionally, for an $N \times N \times L$ tensor $\ten{X}$ one can take the 1-mode product with a $P \times N$ matrix, the 2-mode product with a $Q \times N$ matrix, and the 3-mode product with a $R \times L$ matrix. The resulting dimensions of these mode products would be $P \by N \by L$, $N \by Q \by L$, and $N \by N \by R$, respectively. Elementwise, the 1-mode product gives $(\ten{X} \times_1 \vec{B})_{ijk} = \sum_h x_{hjk} b_{ih}$.

\paragraph{Tucker decomposition} Although the most prominent of the many tensor decompositions are the CP decomposition \citep{carrollchang, harshman1970} and the Tucker decomposition \citep{tucker1966}, other notable decompositions include RESCAL \citep{Nickel2011}, DEDICOM \citep{harshman1978}, and PARATUCK2 \citep{harshman1996}, where both the CP and RESCAL decompositions are special cases of the Tucker decomposition. The Tucker decomposition decomposes an $n$th-order tensor $\ten{X}$ into an $n$th-order \textit{core tensor} $\G$ and $n$ \textit{factor matrices}. The Tucker decomposition of a third order $N \times N \times L$ tensor $\ten{X}$ is
\begin{equation}
    \ten{X} = \G \times_1 \U \times_2 \V \times_3 \Y.
    \label{eq:ntd}
\end{equation}
Here $\G$ is $P \by Q \by R$, $\U$ is $N \by P$, $\V$ is $N \by Q$, and $\Y$ is $L \by R$. A \textit{\acl{ntd}} is one where all elements of the factor matrices $\U, \V,$ and $\Y$ and core tensor $\G$ are nonnegative.
When fitting a \acl{ntd} to a data tensor $\ten{X}$, two common notions of approximation are the Frobenius loss
and the KL-divergence, where the latter is given by
\begin{equation}
    \text{KL}(\ten{X} \mid \G \times_1 \U \times_2 \V \times_3 \Y): = \sum_{ijk} x_{ijk} \log \frac{x_{ijk}}{\hat{x}_{ijk}} - x_{ijk} + \hat{x}_{ijk},
    \label{eq:ten_KL_div}
\end{equation}
for $\widehat{\ten{X}} = \G \times_1 \U \times_2 \V \times_3 \Y$. 
\subsection{Multilayer Networks}
\label{sec:mln}

Multilayer networks consist of a set of different network `layers' which each encode different types of edges (sometimes called \emph{intralayer} edges). These types of edges can represent, for example, different types of relationships in social networks \citep{banerjee2013, power2017social, banerjee2019}, different shared genetic subsequences in biological networks \citep{larremore2013network}, or different time steps in temporal networks \citep{gallotti2015multilayer}. In general, the node set can differ across layers, however we focus on a subclass called \emph{multiplex networks} in which the nodes are identical in each layer \citep{mucha2010community}. 

Distinct from \textit{heterogeneous networks} \citep[e.g.,][]{dong2020heterogeneous}, wherein there are different categories of nodes \textit{and} edges, multilayer networks have only one type of node and only distinguish between different types of edges. Similarly, \textit{multigraphs} allow for multiple edges to exist between nodes and \textit{labels} corresponding to nodes and/or edges identify the different types of relationships. For a more comprehensive survey of related terminologies, see \cite{kivela2014multilayer}.

Although some work about multilayer networks also models connections between layers using \emph{interlayer} edges, we do not assume or model the coupling of layers. In this case, there exists a one-to-one alignment of layers, allowing them to be encoded in a 3-dimensional \textit{adjacency tensor} $\A \in \mathbb{R}^{N \by N \by L}$, where $N$ and $L$ are the numbers of nodes and layers, respectively, and where each frontal slice  $\mat{A}_\ell$ is the adjacency matrix of a particular layer $\ell$ of the multilayer network. Here, $a_{ij\ell}>0$ if an only if there is an edge from $i$ to $j$ in layer $\ell$, and is otherwise zero. 

Multilayer networks can be either directed or undirected. In this work we assume that all layers within a given network are either directed or undirected. Extending our approach to networks that have a mixture of directed and undirected layers would be straightforward. We also assume that all edges are unweighted, $a_{ij\ell}\in\{0,1\}$, but this assumption can easily be relaxed. For a more comprehensive review of multilayer networks, see \citet{kivela2014multilayer,boccaletti2014structure}

\subsection{Related Work}\label{sec:related}
We now discuss related work as categorized by previous approaches for (i) tensor methods for multilayer networks, (ii) \acl{sbms} for multilayer networks, and (iii) addressing layer interdependence in multilayer networks.
\paragraph{Tensor methods for multilayer networks}
Multilayer networks have been studied since as far back as 1939 \citep{roethlisberger1939and}, and they have been mathematically represented by tensors since at least 1987, when Krackhardt introduced the concept of {\it cognitive social structures} \citep{krackhardt1987}. In fact, one of the foundational tensor decomposition papers by \citet{carrollchang}, although not a multilayer network, was a study of multilayer relational data. Since then, tensor methods have become more prominent in analysing multilayer networks \citep{bader2007temporal, Kolda2009, Nickel2011}, 
and \citet{de2013mathematical} formalized this tensorial framework by generalizing many network analysis tools to the multilayer setting.

The CP tensor decomposition \citep{carrollchang, harshman1970} and the Tucker decomposition \citep{tucker1966}, have been implemented for their use in analyzing multilayer networks. The CP decomposition, for example, is implemented to interpret a fourth-order tensor of multilayer network data in \citet{Schein2015}, for community detection and analysis of activity patterns in a temporal network in  \citet{gauvin2014detecting}, and to assess centrality of nodes in multilayer networks \citet{wang2018new}. The Tucker decomposition is used for community detection in a temporal multilayer network representing brain dynamics in \citet{al2018tensor} and to cluster keywords and communities in a multilayer email network in \citet{sun2009multivis}. 

\paragraph{Stochastic block models for multilayer networks}
There have been a wide range of approaches to generalize the \acs{sbm} to multilayer networks. In \citet{valles2016} a multilayer \acs{sbm} is developed by fitting a new \acs{sbm} to each layer, assuming that neither node-membership nor group-to-group connectivity is fixed across layers. \citet{stanley2016} develop a related model that assumes layers are sampled from a small set of \acs{sbms} and the set of layers generated from the same \acs{sbm} are referred to as belonging to the same \textit{strata}.  
In \citet{Carlen2019} and \citet{deBacco}, a node's membership vectors are held fixed across layers, but a new affinity matrix is fit for each layer. A similar model is proposed in  \citet{paul2016consistent} but with node membership vectors constrained to take on binary values and with a Bernoulli distribution assumption instead of Poisson. Conversely, in \citet{tarres2019tensorial} a Tucker decomposition accounting for layer community structure is fit with the aim to predict \textit{types} of links in a multilayer network. To do so, a new core tensor is fit for each type of link. Although layer community structure is addressed in \citet{tarres2019tensorial}, the number of node-communities is always fixed to equal the number of layer-communities: a missed opportunity to examine layer interdependence by examining the optimal number of layer-communities.

In \citet{wang2019multiway}, the authors propose using a Tucker decomposition as a multilayer \acs{sbm}, but limit their factor matrices to only take on binary values. Thus, the extent to which layer dependence is addressed is limited to the binary clustering of layers and is more similar to the strata work of \citet{stanley2016}. Furthermore, the core tensor is not constrained to be nonnegative, and the proposed algorithm focuses on minimizing the Frobenius norm of the difference between the tensor and the approximation given by the Tucker decomposition. 

Previous work explicitly using the full \acl{ntd} as a multilayer extension of the \acs{sbm} to study layer dependence is limited to that of \citet{schein2016bayesian}, wherein the authors propose the use of a Bayesian Poisson Tucker Decomposition (BPTD) as a generalization of the \acs{dcmmsbm} to study a multilayer network. They highlight the BPTD using a fourth-order tensor to study international relations between countries over time, and show how the BPTD can group together countries, actions, and time periods into communities. Our work extends this modeling framework by introducing a technical approach to studying layer dependence. We build upon \cite{schein2016bayesian}'s work to significantly expand the motivation, estimation, and interpretation of the \acl{ntd} as a sensible extension of the \acs{sbm} to multilayer networks. Distinct from their work, we aim to motivate the \acl{ntd} as a model for understanding layer dependence \textit{in general}, and we do so through extensive examples (see \Cref{model}). Furthermore, departing from their MCMC algorithm, we justify the use of an algorithm for estimating a \acl{ntd} by minimizing the \acs{kld} with multiplicative updates \citep{kimChoi} by connecting it to the pointwise maximum likelihood estimate of the log-likelihood using \acl{em}\textemdash the estimation method proposed in \cite{deBacco}.

Because we will often reference the work and the multilayer \acs{sbm} model \acf{mt} built in \citet{deBacco}, we define and discuss the work in more detail here. Consider a multilayer network with $N$ nodes and $L$ layers represented by adjacency tensor $\A \in \Aspace^{\Adims}$. Assume each node $i$ in the network has outgoing and incoming nonnegative membership vectors $\vec{u}_i$ and $\vec{v}_i$, respectively, representing their soft assignment to $K$ groups. The densities with which nodes in each community interact in layer $\ell$ is given by nonnegative affinity matrix $\mat{G}_\ell$. The \acs{mt} model then assumes the generative process whereby
\begin{equation}
   \A \sim \text{Poisson}(\Lam), \,\text{where} \, \Lam_\ell = \U \mat{G}_\ell \V^\top \, \text{for } \ell \in [1, L].
   \label{eq:MT}
\end{equation}
Written this way we see that \acs{mt} fits an \acs{sbm} to \textit{each} layer of the network, holding fixed the outgoing and incoming group memberships across layers. Parameters $\U, \V, \text{ and } \mat{G}_\ell$ are estimated by maximizing the log-likelihood of observing $\A$ via an \acs{em} algorithm.

\paragraph{Layer interdependence}
Understanding how the layers of a multilayer network interact with, represent, or are different from one another has been a relevant question ever since multilayer networks started being studied. As such, there have been a multitude of proposed methods to study and assess layer interdependence. \citet{krackhardt1987} suggested differentiating layer similarity by comparing individual layers to a \textit{consensus structure}. \citet{battiston2014structural} introduce the measure of \textit{edge overlap} which they propose to use for determining similarity between layers. This measure is built upon in \citet{kao2018layer} which uses a similarity measure based on edge overlap to identify layer communities. The authors construct a single layer network where each node is a layer and each edge is weighted by the similarity measure, and then find layer communities by doing community detection on this new network. 
\citet{de2015structural} and \citet{de2016spectral} develop information-theoretic tools to identify layer dependency and cluster similar layers. In \citet{stanley2016}, the authors study layer interdependence by categorizing layers into groups such that all layers were drawn from the same \acs{sbm}. In the \acf{mt} work of \citet{deBacco}, layer interdependence is studied through building multiple \acs{mt} models using different subsets of the layers, and the models' performance (measured by test-AUC) on a link prediction task is used to determine if there is layer interdependence in the model. In this setting, layer interdependence can be viewed as a specific application of transfer learning which assesses how a model built in one setting performs in an alternative one \citep{torrey2010transfer, altenburger2021which}. 
 Finally, while not explicitly developed for use in the multilayer setting, tools to compare similar structures across graphs such as those developed in \citet{wills2020metrics} and \citet{racz2021correlated} could be used to compare layers in a multilayer network. 
 
These various approaches to studying layer interdependence have been applied to various disciplinary contexts, and have resulted in varying discipline-specific conclusions, as well. In particular: \citet{de2015structural} identifies and interprets layer dependence in varying contexts\textemdash from the worldwide food import/export multilayer network to the London metropolitan public transportation multilayer network; \cite{battiston2014structural} interprets the dependencies between trust and communication, business, and operating partnerships within an Indonesian terrorist network; \cite{kao2018layer} interprets the dependencies between different research areas in the American Physical Society's collaboration network, the regional dependencies in an airline network, and the distinction between positive relationships, negative relationships, and temporally distinguished esteem in a social network; \cite{stanley2016} and \cite{de2016spectral} both discuss the interpretation of layer dependence in the context of the human microbiome, and the ability to use layer interdependence methods to identify similarly functioning regions within the body. Overall, previous work on studying layer interdependence in multilayer networks identifies many disciplines and contexts within which such methods have the potential to corroborate or identify socially, scientifically, or theoretically interesting findings. We aim to add to this body of work.

In contrast to these previous approaches to identify layer interdependence, we propose that the \acl{ntd} does so by identifying which layers can be described by shared generative \acl{sbms}. As we will discuss in \Cref{subsec:deflation}, the \acs{ntd} is a generalization of the strata \acs{sbm} model from \cite{stanley2016} that is similar to moving from fixed- to mixed-membership assignments in the single layer \acs{sbm}. 
 
\subsection{Contributions} \label{contributions}
Situated in this related work, the contributions of our work are as follows: (i) we use and expand the motivation of the \acl{ntd} with KL-divergence as a natural extension of the \acs{dcmmsbm} to multilayer networks by allowing for distinct latent structure in the nodes \textit{and} layers; (ii) we propose the \acs{ntd} as a generalization of many prior models and approaches for extending the \acs{sbm} to multilayer networks (see \Cref{tab:mlSBM-summary} to see how the \acs{ntd} generalizes related work); (iii) we propose the inspection of the third factor matrix in the \acs{ntd} for quantifying and assessing layer interdependence and discuss three specific methods for doing so; (iv) we show the equivalence in model, loss function, and algorithm between Poisson Tucker decomposition and \acl{ntd}; (v) we propose definitions of layer interdependence based on the \acl{lrt}; (vi) we define two link prediction tasks for multilayer networks to use \acs{xval} as a tool for model selection; and (vii) we use the \acs{ntd} to study layer interdependence in a variety of empirical multilayer networks: one biological, one \acl{css}, and 113 social support networks.
\section{Nonnegative Tucker Decomposition (NNTuck)}
\label{NNTuck}
We begin this section by outlining our approach to a multilayer \acs{sbm} that corresponds to a \acl{ntd} with KL-divergence. 
We will henceforth refer to the multilayer \acs{sbm} developed here as just the \acf{ntd}, although it's important to note that the \acs{sbm} interpretation only corresponds to the \acl{ntd} estimated with KL-divergence loss as in \Cref{eq:ten_KL_div}, \textit{not} Frobenius loss. We present the model of the \acs{ntd} as a multilayer \acs{sbm} in \Cref{model}, and discuss \textit{deflation} and layer dependence in the \acs{ntd} in \Cref{subsec:deflation}. We present an algorithm for estimating the \acs{ntd} of a multilayer network and discuss the algorithm's limitations in \Cref{alg}.

\subsection{The Model}
\label{model}
Consider a multilayer network with $N$ nodes and $L$ layers represented by adjacency tensor $\A \in \Aspace^{\Adims}$. We assume that each node $i$ has nonnegative membership vectors $\vec{u}_i \in \R_+^K$ and $\vec{v}_i \in \R_+^K$ representing its soft assignment to $K\leq N$ groups. Moreover, we assume that each layer $\ell$ has nonnegative vector $\vec{y}_\ell \in \R_+^C$ describing the layer's soft membership to each of $C\leq L$ layer communities. Just as matrices $\U$ and $\V$ in the \acs{sbm} describe latent community structure in the \textit{nodes} of single-layer networks, the factor matrix $\Y$ in the \acs{ntd} describes latent structure in the \textit{layers} of a multilayer network. Finally, we assume tensor $\G \in \Gspace^{\Gdims}$ defines $C$ different affinity matrices. 
Let $\vec{u}_i, \vec{v}_i, \vec{y}_\ell$ be the rows of nonnegative matrices $\U,\V,$ and $\Y$, respectively. The \acs{ntd} multilayer \acs{sbm} assumes
\begin{equation}
    \A \sim \text{Poisson}(\G \times_1 \U \times_2 \V \times_3 \Y).
    \label{Tucker}
\end{equation}
For an undirected network, we set $\U := \V$ and constrain the frontal slices of $\G$ to be symmetric. Maximizing the log-likelihood of observing $\A$ under the model given by \eqref{Tucker} is equivalent to minimizing the KL-divergence between $\A$ and $\Ahat = \G \times_1 \U \times_2 \V \times_3 \Y$. This is a tensorial generalization of the connection between \acs{pmf} and \acs{nmf} with KL-divergence referenced in \eqref{eq:kleqLog} and motivates the use of the KL-divergence for determining the \acs{ntd}. 

We now define vocabulary for three types of \acs{ntds}, each based on different assumptions of the structure and dimension of $\Y \in \R^{L \by C}$.

\begin{definition}[Layer independent \acs{ntd}]\label{def:ind}
A \textbf{layer independent \acs{ntd}} is a \acl{ntd} where $C = L$ and $\Y$ has the constraint $\Y = \mat{I}$.
\end{definition}
\begin{definition}[Layer dependent \acs{ntd}]\label{def:dep}
A \textbf{layer dependent \acs{ntd}} is a \acl{ntd} where $\Y$ has the constraint $C<L$.
\end{definition}
\begin{definition}[Layer redundant \acs{ntd}] \label{def:red}
A \textbf{layer redundant \acs{ntd}} is a \acl{ntd} where $C=1$ and we constrain $\Y$ to be the ones vector, $\Y = [1, \dots, 1]^{\top}$.
\end{definition}
\begin{table}
\begin{tabular}{p{5.5cm}| p{9cm}}
\textbf{Citation} & \vspace{0.05cm}\textbf{Approach}\\
\hline \\
\citet{schein2016bayesian} & The same model as the \acl{ntd}. The factor matrices and core tensor are estimated using an MCMC inference algorithm. \\ \vspace{0.05cm}
\citet{valles2016} & \vspace{0.05cm} A separate \acs{sbm} is estimated for each layer. \\ 
\vspace{0.05cm} \citet{stanley2016}  & \vspace{0.05cm} The model assumes a Bernoulli distribution and $K$ is not fixed across layers. Layers within the same \textit{strata} $s$ are drawn from the same \acs{sbm} with $\U^s := \V^s$ constrained to only take on binary values. \\
\vspace{0.05cm}\citet{paul2016consistent} & \vspace{0.05cm} $\U:=\V$ constrained to only take on binary values, $\Y$ is constrained so that $\Y:=\mat{I}$, and the model assumes a Bernoulli distribution.\\
\vspace{0.05cm}
\vspace{0.05cm}\citet{deBacco} and \citet{Carlen2019} & \vspace{0.05cm} $\Y$ is constrained so that $\Y:=\mat{I}$.\\
\vspace{0.05cm} \citet{tarres2019tensorial} & \vspace{0.05cm} $\U, \V$, and $\Y$ have constraint $C=K$ and a new core tensor $\G$ is estimated for each type of link in the network.  \\
\vspace{0.05cm} \citet{wang2019multiway}  & \vspace{0.05cm} $\U, \V$, and $\Y$ are constrained to only take on binary values.
\end{tabular}
\caption{Previous approaches to multilayer \acs{sbms}. Excepting the first citation, we write these approaches in relation to the \acf{ntd} which assumes $\A \sim \text{Poisson}(\G \times_1 \U \times_2 \V \times_3 \Y)$ for $\G \in \Gspace^{\Gdims}$, $\U, \V \in \Gspace^{\Udims}$, and $\Y \in \Gspace^{\Ydims}$. For descriptions of our novel contributions situated in this work see \Cref{contributions} and for more details on the \acs{ntd} see \Cref{NNTuck}.}
\label{tab:mlSBM-summary}
\end{table}

\subsection{Deflation and Layer Interdependence} \label{subsec:deflation}
\begin{figure}
    \centering
    \includegraphics[width=0.7\textwidth]{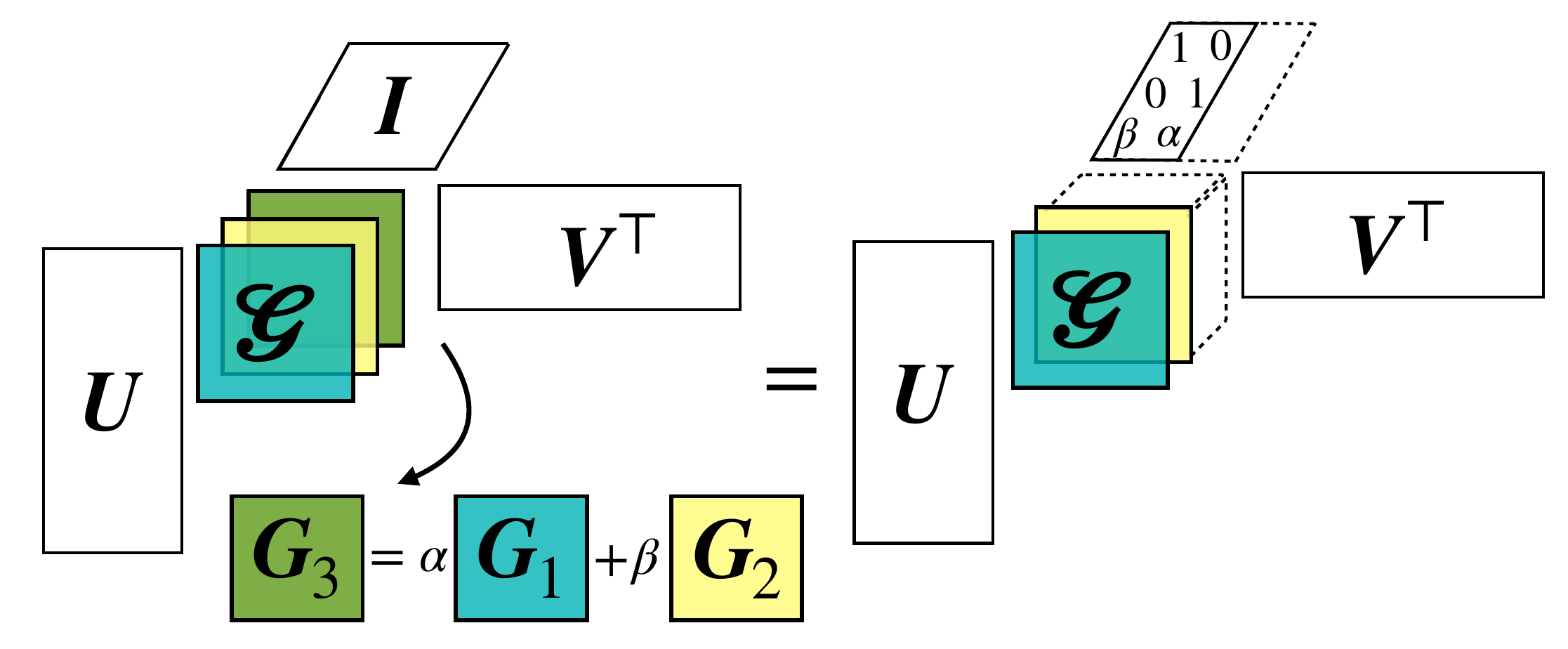}
    \caption{If one or more of the frontal slices of the core tensor are linear combinations of another, there is a \textit{deflation} of the core tensor. In this example, we show how a layer independent \acs{ntd} (left) can be equivalently written as a layer dependent \acs{ntd} (right). This figure shows how layer dependence is stored in the factor matrix $\Y$. 
    }
    \label{fig:lincomb}
\end{figure}

\begin{figure}
    \centering
    \includegraphics[width=0.7\textwidth]{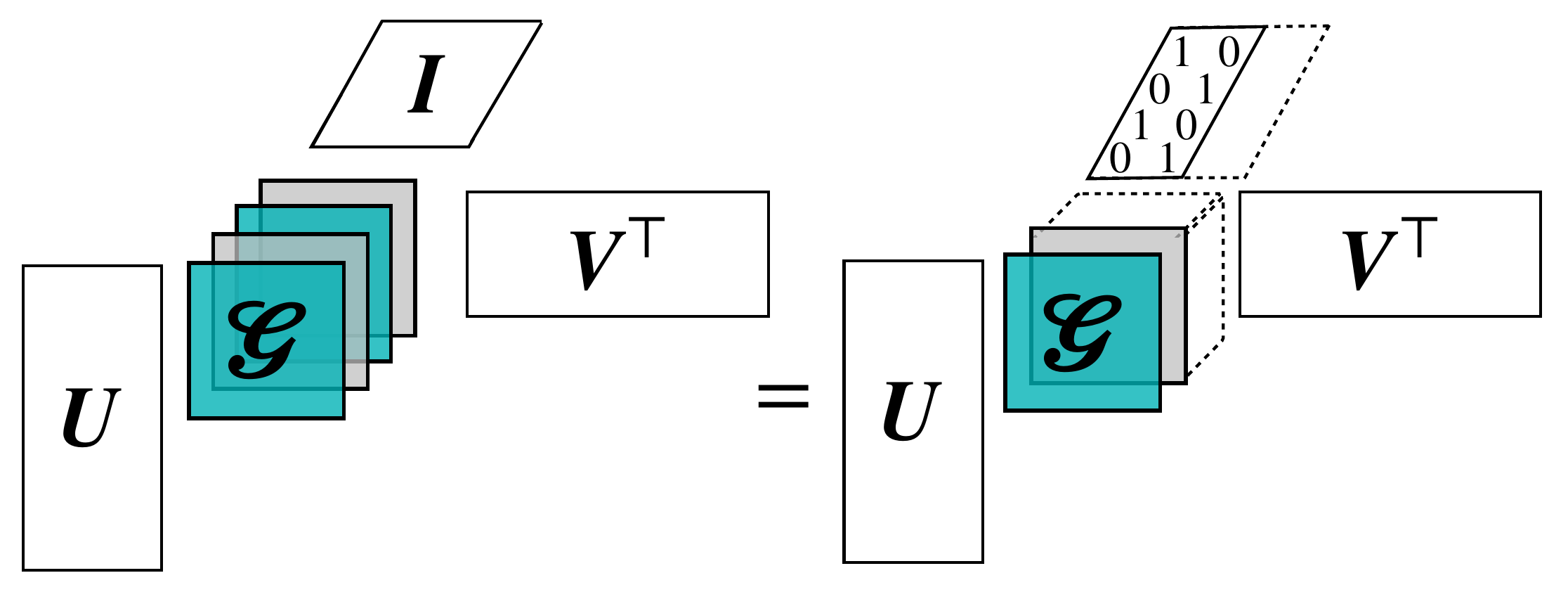}
    \caption{This figure shows how the \acs{ntd} generalizes the strata multilayer \acf{sbm} of \citet{stanley2016}. If, for example, all layers in a multilayer network are drawn from one of two \acs{sbm}s (with the same $\U$ and $\V$ across layers), the factor matrix $\Y$ has only zeros and ones.}
    \label{fig:strata}
\end{figure}

\begin{figure}
    \centering
    \includegraphics[width=0.7\textwidth]{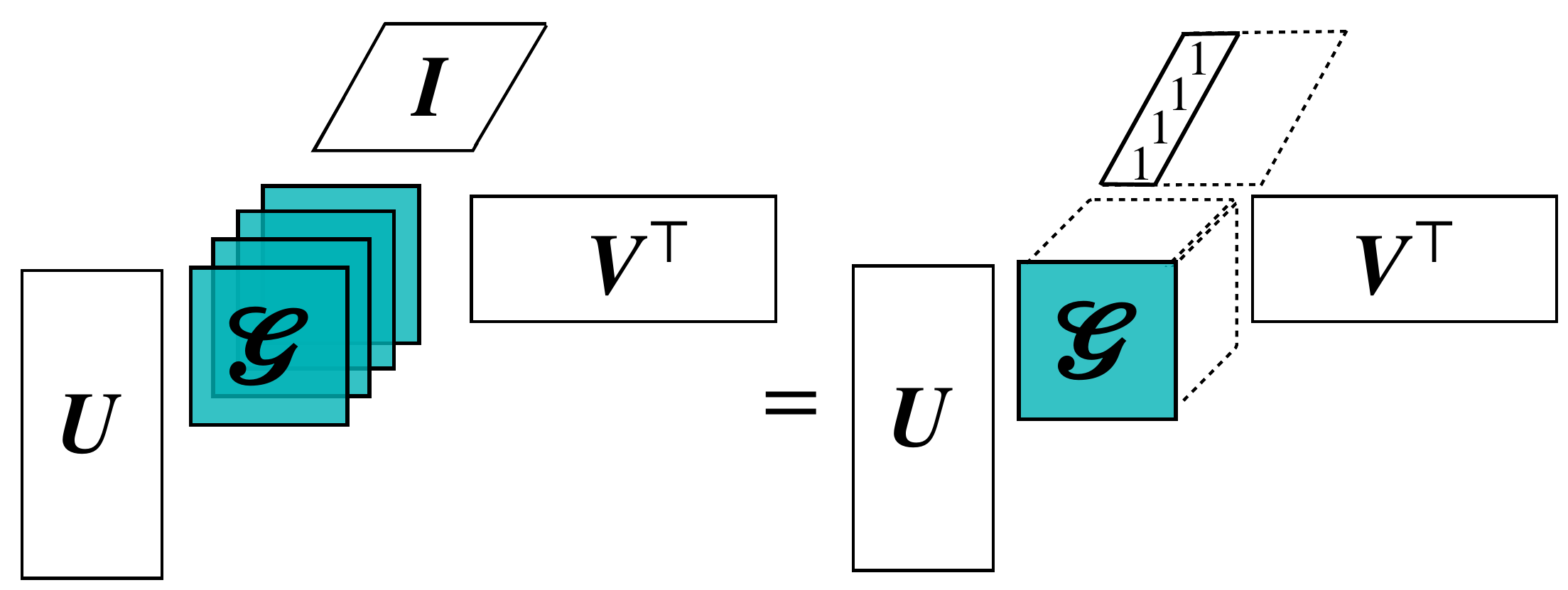}
    \caption{A visualization of a layer redundant \acs{ntd}. This figure shows how \acs{ntd} can model a multilayer network wherein each layer was drawn from the same \acs{sbm}. In such a case, the factor matrix $\Y$ is a vector of all ones and the core tensor $\G$ is of dimension $K \by K \by 1$.}
    \label{fig:repeat}
\end{figure}

In this section we discuss layer dependence in the \acs{ntd} through three examples and discuss how \textit{deflation} of the core tensor allows for latent structure to be identified in the layers of a multilayer network. 
\begin{definition}[Deflation]\label{def:deflation}
 We say there is a \textit{deflation} of the core tensor $\G \in \Gspace^{K \by K \by L}$ of a layer independent \acs{ntd} if there exists a tensor $\G^{'} \in \Gspace^{\Gdims}$ and a factor matrix $\Y \in \Gspace^{L \by C}$ for $C<L$ such that,
\begin{equation}
    \G \times_1 \U \times_2 \V = \G^{'} \times_1 \U \times_2 \V \times_3 \Y.
\end{equation}
\end{definition}
When the core tensor can be deflated the factor matrix $\Y$ in the \acs{ntd} captures the interdependence between layers. We examine deflation and the $\Y$ factor matrix through the three example model instances below, respectively depicted in Figures \ref{fig:lincomb}, \ref{fig:strata}, and \ref{fig:repeat}.
\begin{example}[Linearly dependent core tensor]
\label{ex:lindep} 
For a three-layer adjacency tensor $\A \in \{0,1\}^{N \by N \by 3}$ consider the layer independent \acs{ntd} given by $\A = \ten{G} \times_1 \U \times_2 \V \times_3 \vec{I}$ where the frontal slices of core tensor $\ten{G}\in \Gspace^{K \by K \by 3}$ are as follows:
\begin{equation*}
    \mat{G}_{1} = \begin{bmatrix}
    0.2 & 0.1\\
    0.1 & 0.2
    \end{bmatrix}, 
    \mat{G}_{2} = \begin{bmatrix}
    0.3 & 0.01 \\
    0.01 & 0
    \end{bmatrix}
    \mat{G}_{3} = \begin{bmatrix}
    0.35 & 0.105\\
    0.105 & 0.2
    \end{bmatrix}.
\end{equation*}
\end{example}
As may be evident, $\mat{G}_{3}$ is a linear combination of $\mat{G}_{1}$ and $\mat{G}_{2}$. Specifically, $\mat{G}_{3} = \mat{G}_{1} + \frac{1}{2} \mat{G}_{2}$. In the same sense that a rank-deficient matrix has one or more columns which are a linear combination of others, we can consider the inclusion of $\mat{G}_3$ in the core tensor redundant. If we have a limited data source from which we are estimating our model, ``wasting'' information to fit this redundant frontal slice could lead to a less efficiently estimated model. 
Instead, consider tensor $\G'$ whose frontal slices are $\mat{G'}_1 = \mat{G}_1$ and $\mat{G'}_2 = \mat{G}_2$, and define 
\begin{equation*}
    \Y := \begin{bmatrix}
    1 & 0 \\
    0 & 1 \\
    1 & 0.5
    \end{bmatrix},
\end{equation*}
where the third row contains the respective weights of $\mat{G}_1$ and $\mat{G}_2$ that sum to $\mat{G}_3$. Then 
\begin{equation*}
    \ten{G} \times_1 \U \times_2 \V \times_3 \vec{I} = \G' \times_1 \U \times_2 \V \times_3 \Y.
\end{equation*}
Note that whereas these two models are mathematically equivalent, the \textit{deflated} model allows for latent structure in the layers to be more efficiently identified. 
\begin{example}[strata \acs{sbm}] 
\label{ex:strata}
For $\A \in \Aspace^{N \by N \by 4}$ consider the nonnegative Tucker-2 model given by $\A = \ten{G} \times_1 \U \times_2 \V \times_3 \vec{I}$. Assume that the core tensor $\ten{G}\in \Gspace^{K \by K \by 4}$ has frontal slices $\mat{G}_3 := \mat{G}_1$ and  $\mat{G}_4 := \mat{G}_2$, for the same $\mat{G}_1$ and $\mat{G}_2$ in the previous example. For $\G' = \Gspace^{K \by K \by 2}$ with frontal slices $\mat{G'}_1 = \mat{G}_1$ and $\mat{G'}_2 = \mat{G}_2$ and factor matrix 
\begin{equation*}
    \Y = \begin{bmatrix}
    1 & 0 \\
    0 & 1 \\
    1 & 0 \\
    0 & 1
    \end{bmatrix},
\end{equation*}
then $\ten{G} \times_1 \U \times_2 \V \times_3 \vec{I} = \G' \times_1 \U \times_2 \V \times_3 \Y.$
\end{example}
The interpretation of this example is that layers 1 and 3 in the multilayer network were drawn from the same \acs{sbm}, one distinct from that which generated layers 2 and 4. Because node-membership matrices $\U$ and $\V$ are held to be fixed across layers, this means that communities interact with the same rates in layers 1 and 3, although these rates are different from those which determine interaction in layers 2 and 4. This example clusters layers generated from the same \acs{sbm}. 
This example is generatively equivalent to the strata \acs{sbm} \citep{stanley2016} if, in the example we fix $\U :=\V$, and if, in the strata \acs{sbm} $K$ and node-membership are fixed across layers.

\begin{example}[repeated \acs{sbm}s] 
\label{ex:ones}
For $\A \in \{0,1\}^{N \by N \by 4}$ consider the layer independent \acs{ntd} given by $\A = \ten{G} \times_1 \U \times_2 \V \times_3 \vec{I}$. In this example, consider that all of the frontal slices of $\ten{G}$ are equal: $\mat{G}_\ell = \mat{G}_1$ for $\ell = 1,2,3,4$. Define $\G' \in \Gspace^{K \by K \by 1} = \mat{G}_1$ and factor matrix $\Y = \mat{1} = [1, 1, 1, 1]^\top$.  Then $\ten{G} \times_1 \U \times_2 \V \times_3 \vec{I} = \mat{G}_1 \times_1 \U \times_2 \V \times_3 \vec{Y}$. 
\end{example}
This deflated model is a layer redundant \acs{ntd} and can be interpreted by considering that all layers of the network are different realizations of the \textit{exact same} \acs{sbm}. That is, the underlying process which is assumed to have generated the structure observed in layer 1 is the same as that which generated the structure observed in all other layers. In this sense, a multilayer network with this multilayer \acs{sbm} does not need to be represented as a multilayer network. However, see \citet{taylor2016enhanced} for a discussion of the detectibility limit when a multilayer network's layers are generated from a repeated \acs{sbm}. 

The following final example serves as a warning of the limitations of assessing layer interdependence using the \acs{ntd} model. 
\begin{example}[Linear basis as a cause for deflation]
Consider a network for which $K=2$, $L > 4$, and a layer independent \acs{ntd} with $2\by 2$ affinity matrices $\mat{G}_\ell$ for $\ell = 1, \dots, L$ for each layer of the network.
\end{example}
Note that there is a natural linear algebraic (as opposed to sociological or contextual) reason why this \acs{ntd} can be deflated, or more generally why an estimated model may perform well when $C < L$ without necessarily exhibiting contextually relevant layer dependence. 
Specifically, \textit{any} $2\by2$ matrix can be written as a linear combination of the following \textit{bases}:
\begin{equation*}
    \mat{B}_1 = \begin{bmatrix}
    1 & 0 \\ 0 & 0
    \end{bmatrix},
    \mat{B}_2 = \begin{bmatrix}
    0 & 1 \\ 0 & 0
    \end{bmatrix},
    \mat{B}_3 = \begin{bmatrix}
    0 & 0 \\ 1 & 0
    \end{bmatrix},
    \mat{B}_4 = \begin{bmatrix}
    0 & 0 \\ 0 & 1
    \end{bmatrix}.
\end{equation*}
For example, 
\begin{equation*}
    \mat{G} = \begin{bmatrix}
    a & b \\
    c & d
    \end{bmatrix} = a \mat{B}_1 + b \mat{B}_2 + c \mat{B}_3 + d \mat{B}_4.
\end{equation*}
For the \acs{ntd} where all matrices $\mat{G}_\ell$ have nonnegative entries, coefficients $a, b,c, d$ above will also be nonnegative. Thus for this $K = 2$ case, any core tensor $\ten{G}\in \Gspace^{2 \by 2 \by L}$ can be deflated to core tensor $\ten{B} \in \Gspace^{2 \by 2 \by 4}$ whose $\ell^{\text{th}}$ frontal slice is $\mat{B}_\ell$. In this sense, we can only hope to interpret deflation as a characteristic of the network (as opposed to a characteristic of the linear algebra) when $C < K^2$. For an undirected network, where we constrain the frontal slices of the core tensor to be symmetric, this constraint is $C < \frac{K (K+1)}{2}$. 

For the empirical examples we consider in the next section (where there are between $L=7$ and $L=21$ layers), we will need to consider this issue only in the case where $K=2,3,4,$ depending on the network, because $L < K^2$ for all larger $K$, and thus $C<K^2$ as well. As broader context, for the over 45 datasets provided in De Domenico's multilayer network database \citep{dedomenicoWeb}, only two datasets have more than 16 layers. This, however, is not always the case, and especially not so when considering temporal multilayer networks wherein a network is captured at many time steps. In such situations, one must bear in mind the constraint of $C< K^2$ for the purposes of interpreting the \acs{ntd}.

\begin{remark}[Relationship to MULTITENSOR (MT)]
If we collect the affinity matrices $\mat{G}_\ell$ from the \acs{mt} model to be the frontal slices of tensor $\G \in \Gspace^{K \by K \by L}$, then \eqref{eq:MT} is equivalent to
\begin{equation}
   \A \sim \text{Poisson}(\G \times_1 \U  \times_2 \V).
   \label{Tucker2}
\end{equation}
Note that $ \G \times_1 \U  \times_2 \V = \G \times_1 \U  \times_2 \V \times_3 \vec{I}$. This model is what we define as a \textit{layer independent \acs{ntd}} in \cref{def:ind} above and is sometimes called a Tucker-2 decomposition \citep[see][]{Kolda2009}. Since all factors are nonnegative, the \acs{mt} model seeks to find the nonnegative Tucker-2 decomposition by maximizing the log-likelihood through \acs{em}. Given that the interpretation of the $\Y$ factor matrix is that it describes layer communities in the network, constraining $\Y = \mat{I}$ as is done in \acs{mt} assumes that each layer of the multilayer network was drawn from a distinct \acs{sbm}, albeit with common membership matrices $\U$ and $\V$. That is, \acs{mt} assumes that there is no latent structure in the layers of the network.
\end{remark}

\subsection{Algorithmic Approach}\label{alg}
\citet{kimChoi} extend the multiplicative updates for \acs{nmf} from \citet{LeeSeung} to the \acl{ntd} for minimizing both KL-divergence and Frobenius loss. We reproduce the updates for minimizing KL-divergence in \Cref{MU-NTD}. The updates in \citet{kimChoi} are written for a general, $n$-th order tensor, so we rewrite them here for a 3rd order tensor in a setting wherein some data is masked as specified by a masking tensor $\ten{M} \in \{0,1\}^{\Adims}$. Note that if the data is not masked, the all ones masking tensor $\ten{M} = 1^{\Adims}$ recovers the original multiplicative updates. As we'll see in \cref{applications}, the masking tensor allows for \acs{xval} wherein the \acs{ntd} is only estimated from a portion of the network. Note that these updates are done sequentially, not in parallel. See the URL in \cref{code_repo} for a python implementation.

\begin{algorithm}
\caption{Multiplicative Updates for minimizing KL-Divergence in the \acs{ntd} \citep{kimChoi}}\label{MU-NTD} 
\textbf{Input:} $\A, K, C,$ \verb|Symmetric|, \verb|Masked|, $\ten{M}$, \verb|Independent|, \verb|Redundant|\\
\textbf{Initialize} $\U, \V \in \Gspace^{N \times K}, \Y \in \Gspace^{L \times C},$ and $\G \in \Gspace^{\Gdims}$ to have random, nonnegative entries. Initialize $\Ahat = \G \times_1 \U \times_2 \V \times_3 \Y.$\\
if \verb|Symmetric|: $\V \leftarrow \U$, $\mat{G}_\ell \leftarrow \mat{G}_\ell^\top \mat{G}_\ell$ for $\ell = 1,\dots, C$, and skip each $\V$ update step below.\\
if \verb|Independent|: $\Y \leftarrow \mat{I}$ and skip each $\Y$ update step below.\\
if \verb|Redundant|: $\Y \leftarrow$ \verb|ones(C)| and skip each $\Y$ update step below.\\
if not \verb|Masked|: $\ten{M} = 1^{\Adims}$ \\ \\
while $\frac{KL(\A || \Ahat_{t})- KL(\A || \Ahat_{t-1})}{KL(\A || \Ahat_{t})}< $ \verb|rel_tol|:
\begin{align*}
    \U &\leftarrow \U \circ \frac{[\mat{M}_{(1)} \circ \mat{A}\uf{1}/ \mat{\hat{A}}\uf{1}] [\G\times_2 \V\times_3\Y]\uf{1}^\top}{\mat{M}\uf{1}[\G\times_2 \V\times_3 \Y]\uf{1}^\top} \\
    \V &\leftarrow \V \circ \frac{[\mat{M}\uf{2} \circ \mat{A}\uf{2}/ \mat{\hat{A}}\uf{2}] [\G\times_1 \U\times_3\Y]\uf{2}^\top}{\mat{M}\uf{2}[\G\times_1 \U\times_3 \Y]\uf{2}^\top} \\
    \Y &\leftarrow \Y \circ \frac{[\mat{M}\uf{3} \circ \mat{A}\uf{3}/ \mat{\hat{A}}\uf{3}] [\G\times_1 \U\times_2\V]\uf{3}^\top}{\mat{M}\uf{3}[\G\times_1 \U\times_2 \V]\uf{3}^\top} \\
    \G &\leftarrow \G \circ \frac{[\ten{M} \circ \A/ \Ahat] \times_1  \U^\top\times_2\V^\top \times_3 \Y^\top}{\ten{M}\times_1 \U^\top \times_2 \V^\top \times_3 \Y^\top} \\
    \Ahat &\leftarrow \G \times_1 \U \by_2 \V \by_3 \Y
\end{align*}
\textbf{Return} $\U, \V, \Y, \G$.
\label{nntuck:alg}
\end{algorithm}

Because these updates are derived from the multiplicative updates for \acs{nmf} from \citet{LeeSeung}, they come with guaranteed monotonic convergence to a local minima. In practice, we declare that the algorithm has found a local minima if the KL-divergence has not decreased by more than a relative tolerance of $10^{-5}$ in ten steps. For the case of an undirected network, we initialize the core tensor to have symmetric frontal slices ($\mat{G}_\ell = \mat{G}_\ell^\top$), and initialize and fix $\U = \V$ throughout the updates. We then follow the multiplicative updates above by only making updates to $\U, \Y$, and $\G$. Doing so maintains the guaranteed monotonic convergence to a local minima while preserving the symmetric structure in $\G$, and ensures the constraint for the undirected case that $\U = \V$. A proof of the following Proposition appears in \Cref{SM:MTeqNTD}.

\begin{proposition} \label{nntuck_em_equiv_prop}
Determining factor matrices $\U, \V$, and $\Y$ and the core tensor $\G$ in the \acs{ntd} by maximizing the log-likelihood using \acf{em} is equivalent to using the multiplicative updates given in \citet{kimChoi} to minimize KL-divergence.

\end{proposition}
The significance of this proposition is noticing that not only is minimizing KL-divergence equivalent to maximizing log-likelihood, but also that the \textit{algorithm} by which to find a local minimum of the KL-divergence is the exact same as that to find a local maximum of the log-likelihood. Moreover, using \acs{em} to maximize the log-likelihood of observing $\A$ under the \citet{deBacco} \acf{mt} model is equivalent to minimizing the KL-divergence between $\A$ and a layer independent \acs{ntd}. That is, the \acs{em} steps given in \citet{deBacco} are equivalent to the multiplicative updates in \citet{kimChoi}, where at initialization $\Y = \mat{I}$ is fixed and at each step $\Y$ is not updated. The algorithmic equivalence between \acs{em} for a Poisson model and multiplicative updates has been noted for \acs{nmf} in \citet{fevotte} and for the CP decomposition in \citet{chi2012}.

\paragraph{Algorithmic limitations}
It is important to emphasize that the the KL-divergence given by \eqref{eq:ten_KL_div} (and thus the log-likelihood of observing $\A$ under \eqref{Tucker}) is non-convex. Therefore, although the multiplicative updates discussed above guarantee monotonic convergence, it is only to local optima. In practice, we use a multistart approach: run the algorithm multiple times with different initial conditions and select the \acs{ntd} with the maximal log-likelihood over these runs. See \cref{code_repo} for details in choosing the number of random initializations. Going forward, we use hat notation to denote the \acs{ntd} factors and core tensor estimated by \Cref{nntuck:alg} using a multistart approach. An evaluation of alternative optimization methods for nonnegative matrix and tensor factorizations, including mirror descent \citep{hien2020}, projected gradient descent \citep{cichocki2007}, and stochastic gradient descent \citep{kasai}, is beyond the scope of this work. Furthermore, \citet{chi2012} propose a related algorithm for nonnegative Poisson CP decomposition using multiplicative updates and discuss conditions under which the algorithm converges to KKT points. A similar analysis of the \acl{ntd} would be intriguing but is again outside the scope of this work.
\section{Statistical Tests for Validating Layer Interdependence}
\label{deflation}
In this section we introduce formal definitions of \textit{layer interdependence} by defining corresponding \acf{lrts}. We conclude with a presentation of three methods by which to interpret $\hat{\Y}$ of an \acs{ntd} estimated for an empirical multilayer network.

\subsection{Layer Interdependence and Likelihood Ratio Tests}
\label{subsec:independence}
\acl{caplrts} can be used to assess the performance of two models, where one model is nested within the other. 
In the context of evaluating different \acs{ntd} models we compare the layer independent \acs{ntd} to the nested models of layer dependent \acs{ntds} or a layer redundant \acs{ntd}. The null hypothesis of the \acs{lrt} is that the two models fit the data equally well, and the alternative hypothesis is that the richer model fits the data significantly better. If the resulting $p$-value rejects the null hypothesis, then the full model should be used. Otherwise, the nested model should be used. We use this framework to define three tests for multilayer networks.
 
\begin{definition}[Layer independence]
For a multilayer network let model I be the layer independent \acs{ntd} and let model II be the layer dependent \acs{ntd}. A multilayer network has \textbf{layer independence} at level $\alpha$ if the \acl{lrt} with $(L-C)K^2 - LC$ degrees of freedom is significant at level $\alpha$.
\end{definition}
\begin{definition}[Layer dependence]
A multilayer network has \textbf{layer dependence} at level $\alpha$ if the \acs{lrt} described above is \textit{not} significant at level $\alpha$ for a pre-specified $C$.
\end{definition}
\begin{definition}[Layer redundance]
A multilayer network has \textbf{layer redundance} at level $\alpha$, if the \acs{lrt} comparing the layer redundant \acs{ntd} to the $C=2$ layer dependent \acs{ntd} with $K^2 + 2L$ degrees of freedom is not significant at level $\alpha$.
\end{definition}

To use these \acs{lrts}, one must determine how many parameters are in the full model and how many are in the nested model. For example, to find the difference in the number of parameters between the layer dependent and independent \acs{ntd}, consider that the layer independent \acs{ntd} has $N\times K$ parameters in $\U$ and $N\times K$ parameters in $\V$. There are $K\times K \times L$ parameters in $\G$ and no free parameters in $\Y$ because it is fixed. The layer dependent \acs{ntd} has $N\times K$ parameters in each of $\U$ and $\V$, $L \times C$ parameters in $\Y$, and $K \times K \times C$ parameters in $\G$. Thus the difference in parameters between both models is $2NK + K^2L - 2NK - LC - K^2C = (L-C)K^2 - LC$. Likewise, when comparing two layer dependent \acs{ntds} with dimensions $C_f$ and $C_n$ for $C_n < C_f$ and fixed $K$, there is a difference of $(L+K^2)(C_f-C_n)$ parameters between the two models. When comparing the layer redundant \acs{ntd} nested under the $C=2$ layer dependent \acs{ntd}, the difference in number of parameters is $K^2 + 2L$. 

It is important to note that the theory underlying the \acl{lrt}, Wilks' theorem \citep{wilks1938}, necessarily depends on (i) the maximum likelihood being reached and (ii) the model being identifiable. These are conditions we cannot guarantee in our problem context. Moreover, we propose this method for determining layer interdependence constrained to the classes of models for which the difference in the degrees of freedom $d = (L-C)K^2 - LC> 0$. Because of the non-identifiability, the way by which these models are nested is nuanced, and thus this inequality is not always true for certain values of $L, K,$ and $C$. Furthermore, Wilks' theorem gives asymptotic analysis of how the difference between two likelihoods approaches a $\chi ^2$ distribution with degrees of freedom given by the difference in parameters. To explore how the number of samples\textemdash the number of nodes and layers, in the context of a multilayer network\textemdash impacts the power of these \acs{lrts}, we conduct a numerical experiment in \Cref{SM:LRT_test}.

To address both of these issues, we also utilize the \textit{\acs{slrt}}, developed by \citet{wasserman2020}, which requires no regularity conditions. The \acs{slrt}, however, still requires that the estimation of the nested model corresponds to the global maximum of the log-likelihood. \Cref{nntuck:alg} only guarantees convergence to a local maxima, so for comparing models using both the standard and the split \acs{lrt}, we use the \acs{ntd} corresponding to the highest log-likelihood over multiple initializations of \Cref{nntuck:alg}. See \Cref{fig:multistart} in \Cref{code_repo} to see how the maximal log-likelihood achieved by \Cref{nntuck:alg} varies across multiple initializations, and see \Cref{SM:UI} for more details on \acs{slrt}. For the datasets we discuss below, the layer independence, dependence, and redundancy tests only differ for one dataset when comparing the regular \acs{lrt} to the \acs{slrt}. This difference is consistent with the fact that the \acs{slrt} is lower powered than the regular \acs{lrt}. The low power of the \acs{slrt} may be intensified by the presence of \textit{nuisance parameters}, and proposed solutions have been discussed in \citet{tse2022note}, \citet{strieder2022choice}, and \citet{spectordiscussion}. Alternative \acs{lrt}s for latent variable models have also been proposed \citep[see][]{chen2020note}, and address other common issues that arise when using the \acs{lrt} to compare latent variable models.

\subsection{Layer Interdependence in an Estimated NNTuck} \label{subsec:yinterp}

If the layer dependence test determines that an empirical multilayer network has dependent layers, it is useful to investigate \textit{how} they are related. In the examples in \Cref{subsec:deflation} above, the frontal slices of the deflated core tensor correspond exactly to the affinity matrix of one or more of the layers. As an example, consider the frontal slices of the deflated \acs{ntd} in \Cref{fig:lincomb}. These frontal slices are the affinity matrices for the first and second layer of the multilayer network (beyond the color coding in the example, we can also see this in the first two rows of the $\Y$ factor matrix, which are $[ 1, 0]$ and $[ 0,1]$, respectively). For an \acs{ntd} estimated for an empirical multilayer network there is no constraint such that this must be true. As such, one must use certain heuristics to appropriately interpret $\hat{\Y}$ estimated from empirical data. 

The first approach is to row-normalize $\hat{\Y}$ such that $\hat{\vec{y}}^{(1)}_\ell = \hat{\vec{y}}_\ell / \| \hat{\vec{y}}_\ell \|_1$ and inspect the rows of $\hat{\Y}^{(1)}$ relative to one another. The second approach is to row-normalize $\hat{\Y}$  such that $\hat{\vec{y}}^{(2)}_\ell = \hat{\vec{y}}_\ell / \| \hat{\vec{y}}_\ell \|_2$ and inspect the entries of similarity matrix given by $\hat{\Y}^{(2)}\hat{\Y}^{(2)\top}$. The third approach uses \textit{reference layer bases}. In this approach, $C$ reference layers are chosen, $\hat{\G}$ is rewritten in the linear bases of those reference layers' affinity matrices, and corresponding $\hat{\Y}^*$ is defined in relation to the new core tensor. This last approach has the added benefit of interpreting each layer's dependence \textit{with respect to} the $C$ reference layers. For specifics on this process and guidance on choosing the reference layers, see \Cref{SM:Yinterp}. 
\section{Model Selection Through Cross-Validation}
\label{applications}
In this section we discuss the use of \acs{xval} in this work and define two link prediction tasks for tensors. 
We emphasize that we are more interested in how the \acs{xval} highlights interesting model choices than by the actual predictive performance of \acs{ntd} in these link prediction tasks.
Statistical factor models of networks are generally not competitive with machine learning classifiers that use even simple topological features \citep[see, e.g.,][]{clauset2008, liben2007, ghasemian2020}. As such, the absolute performance here should not be considered a metric of primary interest, but as a means of comparative inspection. That said, recall that in the following link prediction tasks the layer independent \acs{ntd} is equivalent to \acl{mt} from \citet{deBacco}; to compare the performance of \acs{ntd} to other link prediction methods, see \citet{deBacco}.

The construction of the \acs{xval} approach is as follows. For each link prediction task we construct five different masking tensors and estimate a model based on only observed entries of the data tensor. We select the \acs{ntd} with the highest test set log-likelihood from 50 different runs of the multiplicative updates algorithm with random initializations. Then, test-AUC is averaged across the five different maskings. This process is repeated for varying dimensions ($K$, $C$) in the \acs{ntd}.

We define the link prediction tasks via the structure of their \textit{masking tensors} $\ten{M} \in \{0,1\}^{N \by N \by L}$ where $M_{ij\ell}=0$ indicates that the presence or absence of an edge between nodes $i$ and $j$ in layer $\ell$ is missing, and $M_{ij\ell}=1$ otherwise. In undirected networks we enforce $M_{ij\ell} = M_{ji\ell}$ for both link-prediction tasks. 

\paragraph{Independent link prediction} 
In this link prediction task masking is irrespective of layer. That is,  we assume that for $b$-fold \acs{xval}, elements in the tensor are missing with uniform and independent probability $1/b$. Specifically, missing entry $(i,j)$ in layer $k$ does not imply that entry $(i,j)$ is missing in all layers ($M_{ijk}=0 \not \Rightarrow M_{ij\ell}=0$ for $\ell \neq k$). 
\paragraph{Tubular link prediction} 
In this link prediction task edges are always observed or missing \textit{across all layers}. That is, we assume that for $b$-fold \acs{xval}, tubes $(i,j,\cdot)$ in the tensor are missing with uniform probability $1/b$ (see \Cref{fibers} for a visualization of tensor tube fibers). Specifically, missing link $(i,j)$ in layer $k$ \textit{does} imply that link $(i,j)$ is missing in all layers ($M_{ijk}=0 \Rightarrow M_{ij\ell}=0$, $\forall \ell$). We motivate this tubular task by commenting that independent link prediction is often ``too easy,'' in the sense that if many layers are dependent then missing elements are much easier to impute when other elements from the same tube are available. Tubular link prediction captures realistic settings where one knows nothing at all about the relationship between two units $i$ and $j$ in any layer.

Given the structure of an adjacency tensor, there are (at least) two other link prediction tasks which are representative of true missingness patterns in data: one in which an entire horizontal slice of data is missing ($M_{ijk}=0 \Rightarrow M_{ip\ell}=0$ for all $p, \ell$), and one in which an entire lateral slice of data is missing ($M_{ijk}=0 \Rightarrow M_{rj\ell}=0$ for all $r, \ell$). We limit our scope to the independent and tubular link prediction tasks above, but mention these as potentially interesting missingness patterns in the context of multilayer networks.

\section{Application of the NNTuck to Synthetic and Empirical Networks}\label{empirical}
In this section we use the \acs{xval} tools discussed in \Cref{applications}, the layer dependence tests developed in \Cref{subsec:independence}, and the $\Y$ interpretability heuristics from \Cref{subsec:yinterp} to use the \acs{ntd} in application. In \Cref{subsec:syn} we generate a synthetic network example to exhibit the interpretability of the $\Y$ factor matrix when $K$ and $C$ are known. In \Cref{subsec:emp} for each empirical network presented we carry out the following steps: (i) use a \acs{xval} approach to determine model hyper-parameter choices $K$ and $C$, (ii) use \acl{lrts} with this $(K, C)$ pair to determine layer independence, redundance, or dependence, and (iii) if the network is layer dependent at level $\alpha$, examine the $\Yhat$ factor matrix.

\subsection{Synthetic network examples}
\label{subsec:syn}
In this section we define two different synthetic networks and inspect their estimated $\Yhat$ factor matrices. For both examples, we let $N = 200$, $K = 2$, $L = 4$, and define affinity matrices
\begin{equation*}
    \mat{G}_{1} = \begin{bmatrix}
    0.2 & 0.1\\
    0.1 & 0.2
    \end{bmatrix} \textrm{ and } 
    \mat{G}_{3} = \begin{bmatrix}
    0.3 & 0.01 \\
    0.01 & 0
    \end{bmatrix}.
\end{equation*}
We set the affinity matrices $\mat{G}_2$ and $\mat{G}_4$ to be linear combinations of the above affinity matrices, \[\mat{G}_2 = a \mat{G}_1 + b \mat{G}_3 \text{ and }\mat{G}_4 = c \mat{G}_1 + d \mat{G}_3,\] for different values of $a$, $b$, $c$, and $d$ between the two examples.

In both examples, we generate a multilayer network from an \acs{sbm} which assumes that an edge between nodes $i$ and $j$ in layer $\ell$ is drawn from a Poisson distribution with mean $\vec{u}_i \mat{G}_{\ell} \vec{v}_j^T$. 
We set $\vec{u}_i = \vec{v}_i$ and assign 100 nodes to the first group ($\vec{u}_i = [1, 0]$) and 100 nodes to the second group ($\vec{u}_i = [0, 1]$). Generating these synthetic networks in this way is equivalent to drawing them from $\A \sim \textrm{Poisson}(\G \times_1 \U \times_2 \V \times_3 \Y)$ for 
\begin{equation*}
    \Y = \begin{bmatrix}
        1 & 0\\
        a & b\\
        0 & 1\\
        c & d
    \end{bmatrix}
\end{equation*}
and $\G\in \Gspace^{K \by K \by 2}$ with first and second frontal slices $\mat{G}_1$ and $\mat{G}_3$, respectively. For the first network we define $a = 0.5$, $b = 0.5$, $c = 0.1,$ and $d = 0.9,$ and for the second we let $a = 1$, $b = 0$, $c = 0,$ and $d = 1$. This second synthetic network is the strata example depicted in \Cref{fig:strata} and discussed in \Cref{ex:strata}. As an aside, note that whereas in these two networks the entries of the rows of $\Y$ sum to one, this need not be the case. Actually, by allowing the rows of $\Y$ to be unnormalized we can account for heterogeneous degree distributions \textit{across layers}, just as the degree-corrected single layer \acs{sbm} in \citet{KarrerNewman} accounts for heterogeneous degree distributions across nodes.
\begin{figure}
    \centering
    \includegraphics[width = .9\textwidth]{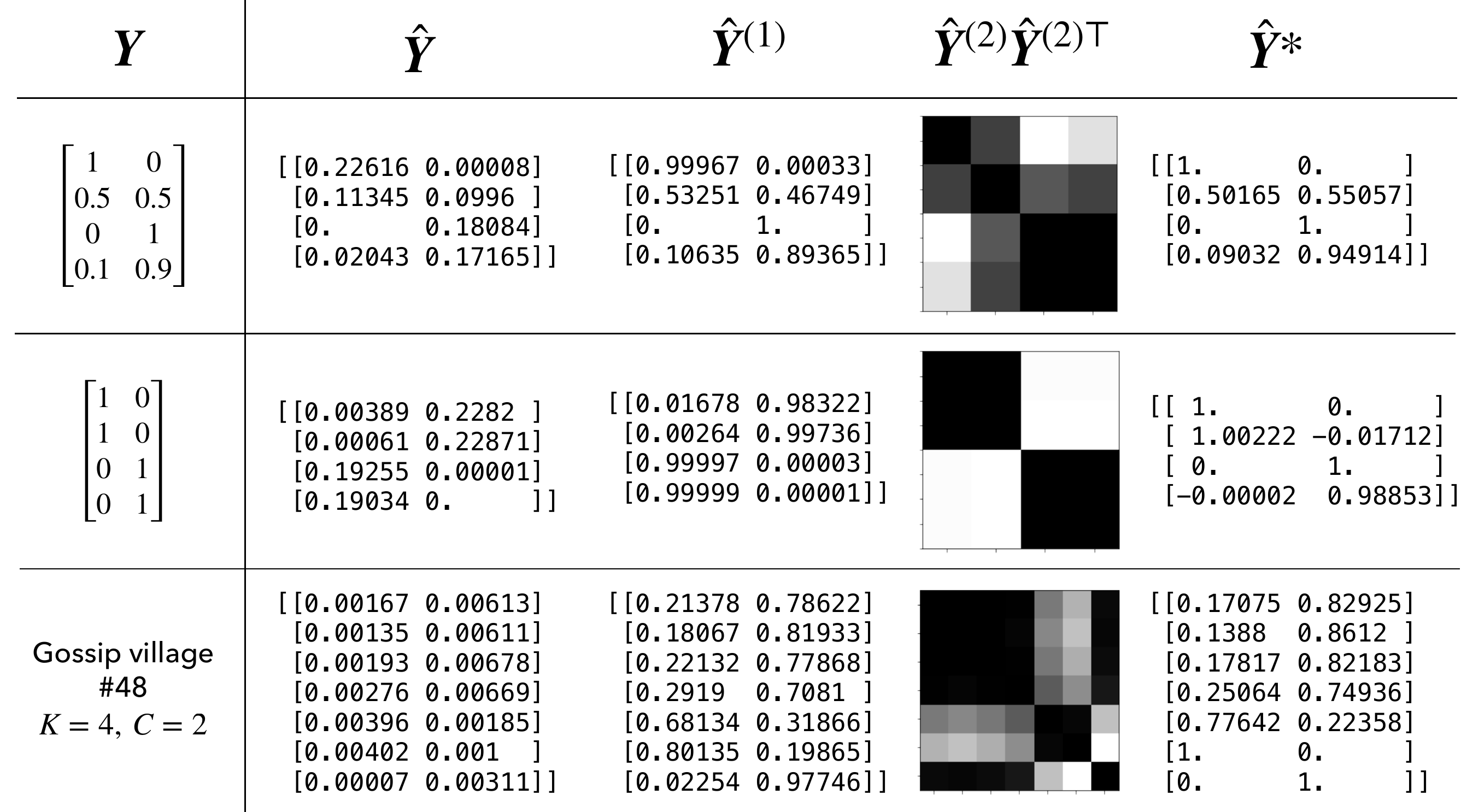}
    \caption{We reproduce the results of the methods for interpreting $\hat{\Y}$ in the \acs{ntd} of the first and second synthetic network described above as well as for the 48th village from \citet{banerjee2019} (labelled ``Gossip village 48'' in Figure \ref{fig:allauc}). For the synthetic networks, $\Y$ is the true factor matrix from which the network was generated. For all three, $\hat{\Y}$ has been estimated from the \acs{ntd} with the highest log-likelihood over 20 runs with different random initializations, $\hat{\Y}^{(1)}$ has been normalized so that the entries of each row sum to one, $\hat{\Y}^{(2)}$ has been normalized so that each row has unit 2-norm. For the synthetic networks, $\hat{\Y}^*$ is the resulting factor matrix after rewriting $\G$ in the basis of layers 1 and 3 (a process which is described in detail in \Cref{SM:masking}). Note that in the synthetic examples, all methods for interpreting $\hat{\Y}$, including simply inspecting $\hat{\Y}$, accurately represent how the layers of the network are related to one another. Specifically, note how $\hat{\Y}^*$ almost exactly recovers the ground truth of how the layers are interdependent. Focusing on $\hat{\Y}^*$ matrix for the gossip village, the two reference layers chosen are ``Who asks you for advice?'' and ``Who are your relatives?'', where the remaining layers can be understood in terms of a linear combination of these.
    }
    \label{fig:Y_syn}
\end{figure}

For both networks we estimate the \acs{ntd} with $C=2$ and $K=2$ and report the \acs{ntd} with the highest log-likelihood over 20 runs with different random initializations. We threshold the values in the resulting membership matrices $\hat{\U}$ and $\hat{\V}$ to reflect the hard membership of the generative model. Node membership is \textit{exactly} recovered for both networks and thus we focus our attention on interpreting $\hat{\Y}$. In both networks, $\hat{\Y}$ recovers the structural dependence between layers and we report the results of all three approaches for interpreting $\hat{\Y}$ in \Cref{fig:Y_syn}. 

\subsection{Empirical Multilayer Networks}\label{subsec:emp}

In this section we use the \acs{ntd} and the tools developed thus far to study several empirical datasets: the \acl{css} dataset from \citet{krackhardt1987}; a biological multilayer network from \citet{larremore2013network}; a social support multilayer network from \citet{banerjee2013};
and 112 other multilayer social support networks from \citet{banerjee2013, banerjee2019}. In Table \ref{table_dep} we include the results from the \acs{lrts} for a subset of these empirical networks and a synthetic network from \Cref{subsec:syn}. Notably, we conclude that the Malaria multilayer network has layer independence at level $\alpha = 0.05$, whereas and all of the other datasets have either layer redundance or layer dependence at the same level $\alpha$. 

Note that for the \acs{xval} tasks in the following subsections, we report the average test AUC across 50 different random initializations for each combination of $K$ and $C$. We vary $K$ from $2$ to $12$ in the Krackhardt multilayer network, and from $2$ to $20$ in the Malaria and Village multilayer networks. See \cref{SM:bigSweep} for a discussion of how we vary $K$ and to see how increasing $K$ to larger values does not impact our model selection. For the \acs{lrt} in each application we select the \acs{ntd} (for prespecified $(K,C)$) with the highest log-likelihood across 20 random initializations. See \Cref{code_repo} for computational experiments testing the variation in maximal log likelihood as a function of the number of random restarts.

\DefineShortVerb{\#}
\SaveVerb{eps}#<1e-16#
\SaveVerb{one}#1.0#
\SaveVerb{ffo}#0.451#
\begin{table}
\centering
\begin{tabular}{l|l|l|l}
\multicolumn{1}{c|}{\textbf{Dataset}}                                                  & \multicolumn{1}{c|}{\textbf{Test}}        & \multicolumn{1}{c|}{\textbf{standard LRT}}      & \multicolumn{1}{c}{\textbf{split-LRT}}                                                  \\ \hline
\multirow{3}{*}{Malaria}                                                     & \begin{tabular}[c]{@{}l@{}}$H_0$: Redundant \\ $H_1$: $C = 2$\end{tabular}    & \multicolumn{1}{l|}{\begin{tabular}[c]{@{}l@{}} $p$ \UseVerb{eps} \\ reject $H_0$\end{tabular}}    & \multicolumn{1}{l}{reject $H_0$}             \\ \cline{2-4}   
& \begin{tabular}[c]{@{}l@{}}$H_0$: Dependent $K=5, C=2$\\ $H_1$: Independent\end{tabular}  & \multicolumn{1}{l|}{\begin{tabular}[c]{@{}l@{}} $p$ \UseVerb{eps} \\ reject $H_0$\end{tabular}} & \multicolumn{1}{l}{reject $H_0$}                                                                           \\  \hline
\multirow{3}{*}{Village 0}                                                 & \begin{tabular}[c]{@{}l@{}}$H_0$: Redundant \\ $H_1$: $C = 2$\end{tabular}           & \multicolumn{1}{l|}{\begin{tabular}[c]{@{}l@{}} $p$ \UseVerb{one} \\ fail to reject $H_0$\end{tabular}} & \multicolumn{1}{l}{fail to reject $H_0$}             \\ \cline{2-4} 
& \begin{tabular}[c]{@{}l@{}}$H_0$: Dependent $K=5, C=2$\\ $H_1$: Independent\end{tabular}   & \multicolumn{1}{l|}{\begin{tabular}[c]{@{}l@{}} $p$ \UseVerb{one} \\ fail to reject $H_0$\end{tabular}} & \multicolumn{1}{l}{fail to reject $H_0$} \\  \hline
\multirow{3}{*}{\begin{tabular}[c]{@{}l@{}}Krackhardt\end{tabular}} & \begin{tabular}[c]{@{}l@{}}$H_0$: Redundant \\ $H_1$: $C = 2$\end{tabular}           & \multicolumn{1}{l|}{reject $H_0$} & \multicolumn{1}{l}{reject $H_0$}             \\ \cline{2-4} 
 & \begin{tabular}[c]{@{}l@{}}$H_0$: Dependent $K=3, C=4$\\ $H_1$: Independent\end{tabular}   & \multicolumn{1}{l|}{\begin{tabular}[c]{@{}l@{}} $p$ \UseVerb{ffo} \\ fail to reject $H_0$\end{tabular}} & \multicolumn{1}{l}{fail to reject $H_0$}         \\  \hline
\multirow{3}{*}{\begin{tabular}[c]{@{}l@{}}Synthetic\end{tabular}}  & \begin{tabular}[c]{@{}l@{}}$H_0$: Redundant \\ $H_1$: Dependent $C = 2$\end{tabular}          & \multicolumn{1}{l|}{\begin{tabular}[c]{@{}l@{}} $p$ \UseVerb{eps} \\ reject $H_0$\end{tabular}}  & \multicolumn{1}{l}{reject $H_0$}             \\ \cline{2-4} 
 & \begin{tabular}[c]{@{}l@{}}$H_0$: Dependent $K=C=2$\\ $H_1$: Independent\end{tabular}  & \multicolumn{1}{l|}{\begin{tabular}[c]{@{}l@{}} $p$ \UseVerb{one} \\ fail to reject $H_0$\end{tabular}}  & \multicolumn{1}{l}{fail to reject $H_0$} \\ \hline
\multirow{3}{*}{\begin{tabular}[c]{@{}l@{}}Gossip 48\end{tabular}}  & \begin{tabular}[c]{@{}l@{}}$H_0$: Redundant \\ $H_1$: Dependent $C = 2$\end{tabular}          & \multicolumn{1}{l|}{\begin{tabular}[c]{@{}l@{}} $p$ \UseVerb{eps} \\ reject $H_0$\end{tabular}}  & \multicolumn{1}{l}{fail to reject $H_0$}             \\ \cline{2-4} 
 & \begin{tabular}[c]{@{}l@{}}$H_0$: Dependent $K=4, C=2$\\ $H_1$: Independent\end{tabular}  & \multicolumn{1}{l|}{\begin{tabular}[c]{@{}l@{}} $p$ \UseVerb{one} \\ fail to reject $H_0$\end{tabular}}  & \multicolumn{1}{l}{fail to reject $H_0$} \\             
\end{tabular}
\caption{The standard and \acs{slrt} determinations for all datasets explored in the following sections. For the standard \acs{lrt}, the $p$-values for each test are also reported. The Village 0 support system network is determined to be layer redundant, the Malaria network is determined to have layer independence, and the Krackhardt network is determined to have layer dependence. The layer redundant test for Gossip Village 48 is the only one wherein the standard and split \acs{lrts} do not agree, and the difference is consistent with the \acs{slrt} being lower powered than the standard \acs{lrt}. Both the standard and \acs{slrt} determine that Gossip Village 48 is layer dependent at level $\alpha = 0.05$, and we explore this empirical $\hat{\Y}$ in \Cref{fig:Y_syn}.}
\label{table_dep}
\end{table}

\subsubsection{Krackhardt's Cognitive Social Structures}
\label{subsec:krack}

\begin{figure}
    \centering
    \includegraphics[width = 0.9\textwidth]{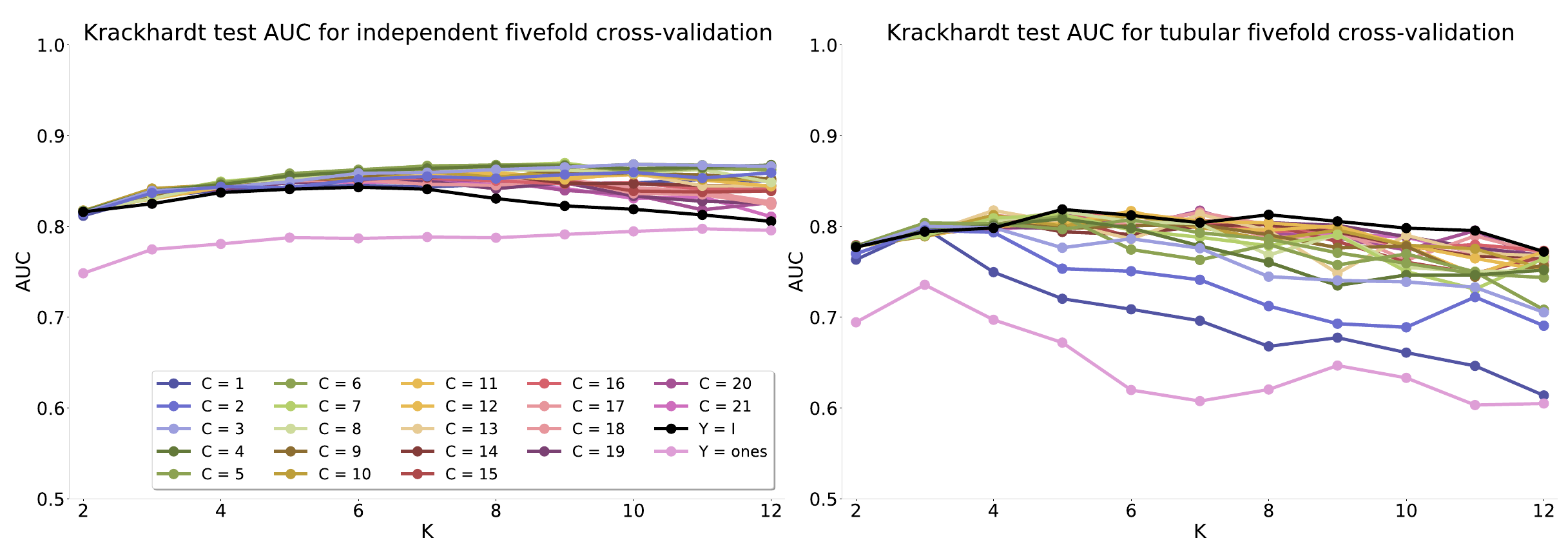}
    \caption{\acs{ntd} performance on independent (left) and tubular (right) link prediction tasks with varying latent dimensions $K$ and $C$ for Krackhardt's \acs{css} multilayer network. 
    Whereas the layer dependent \acs{ntd} with $C< L$ has a higher test-AUC in the independent task, the layer independent \acs{ntd} generally performs as well as the other models in the tubular task. Recall that in this figure, as well as in \cref{fig:mal_pred} and \cref{fig:bvil_link}, the layer independent \acs{ntd} is equivalent to \acl{mt} \citep{deBacco}.}
    \label{fig:krack_pred}
\end{figure}
The Cognitive Social Structures work from \citet{krackhardt1987} surveys 21 people in the management team at a tech firm on their perception of the advice network within the management team. Each of the 21 people were asked to answer the question ``Who would \textit{X} go to for help or advice at work?'' followed by a list of the 21 management employees (including themselves). The resulting $21 \by 21 \by 21$ multilayer network is what Krackhardt referred to as a \acf{css}, where each layer $\ell$ represents person $\ell$'s perception of who receives advice from whom in the network. The adjacency matrices for this advice \acs{css} were transcribed from the original paper for this work, and can be accessed on GitHub \citep[see][]{aguiarKrackhardt} (the \acs{css} for the friendship network is different and can be accessed in the R package \verb|cssTools|, see \citet{cssTools}). 

Interestingly, the \acs{xval} observations are different for each link prediction task. We observe a higher test-AUC associated with the layer dependent \acs{ntd} in the independent link prediction task, and  becomes more pronounced as $K$ increases. In the tubular link prediction task, however, the layer independent and layer dependent \acs{ntds} have a similar test-AUC for nearly all values of $K$. In both link prediction tasks we observe: variation in test-AUC for different values of $K$ and $C$; the layer redundant \acs{ntd} has a lower test-AUC than the layer dependent or layer independent \acs{ntds}; neither the independent nor tubular link prediction task is obviously harder than the other; and the observations from the independent link prediction task are not the same as those from the tubular link prediction task. One possible source of this difference is that the tubular link prediction task is the more difficult one, when compared to the independent task, and thus the results from this task are more representative of a model's performance. 

Based on these results, we choose $K =3$ and $C=4$ for the corresponding layer independence, redundance, and dependence tests and determine that whereas the network is not layer redundant, it is layer dependent at significance level $\alpha = 0.05$ (see \Cref{table_dep} for details). For the sake of brevity (and due to its size ($21 \by 4$)), we do not interpret the $\hat{\Y}$ matrix here. We do note that observing layer dependence in this multilayer network suggests there is latent and shared structure in the various social network perceptions in this company. Further exploration and interpretation of this finding is subject for future work.
\subsubsection{Malaria Data}\label{malaria}

This biological network was originally studied in \citet{larremore2013network}. The undirected network consists of $N = 307$ malaria parasite virulence genes connected across $L = 9$ layers. Two genes are connected if they share a genetic substring of a significant length. Each layer corresponds to a different \textit{Highly Variable Region} on the genes. For more information on the framework or motivating underlying biology, see \citet{larremore2013network}.
\begin{figure}
    \centering
    \includegraphics[width = 0.9\textwidth]{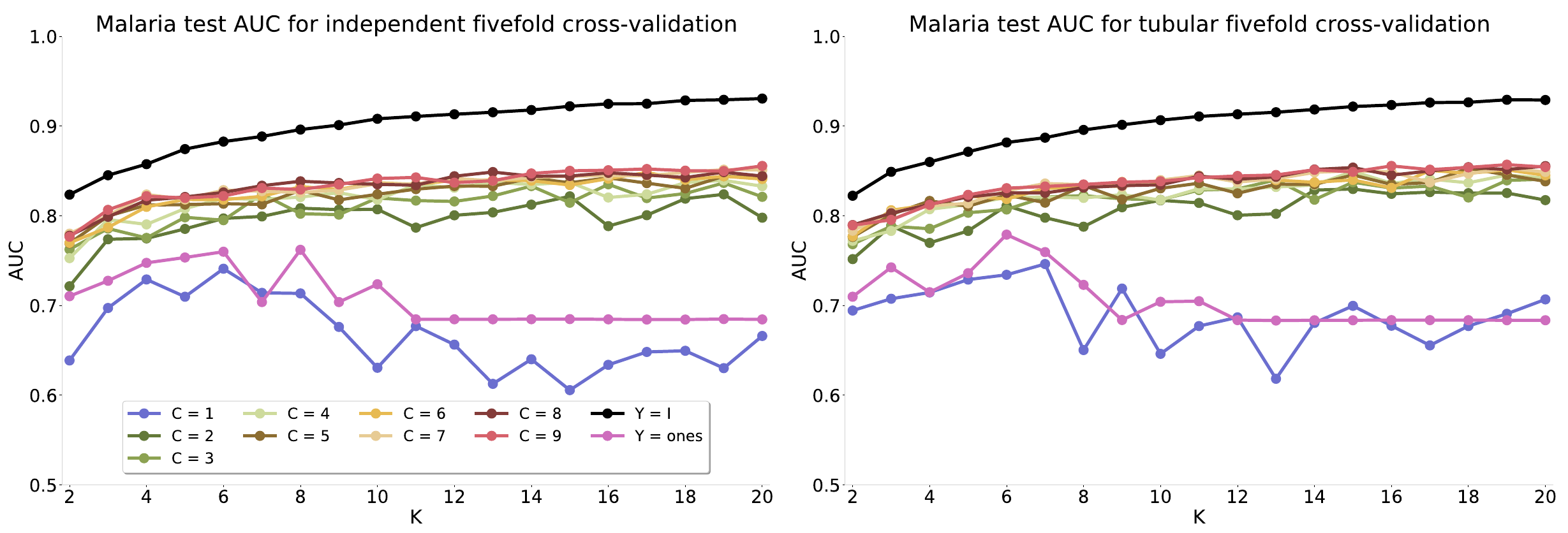}
    \caption{The test-AUC from the independent and tubular link prediction tasks in the Malaria multilayer network. The layer independent \acs{ntd} always results in a higher test-AUC than when allowing for layer dependence or layer redundance.}
    \label{fig:mal_pred}
\end{figure}

In this network the test-AUC of the layer independent \acs{ntd} is always higher than the test-AUC of either the layer dependent or layer redundant \acs{ntd}. This performance difference indicates that the core tensor cannot be deflated without losing important information about the network's layers. Interestingly, we observe that the layer dependent \acs{ntd} with $C=9$ does not perform as well as the layer independent \acs{ntd}, even though the core tensor has the same dimension in both models. While this observation may be an artifact of the underlying optimization landscape, we do not fully understand the implications or causes and it is an interesting topic for future work. Finally, we do not observe a gap in test-AUC between the independent and tubular link-prediction tasks: predicting a missing link with information about that link in other layers is just as difficult as predicting a missing link with no other information about that link in \textit{any} layer.

We therefore determine that an appropriate model choice is a layer independent \acs{ntd} with $K = 5$ and find that this network is layer independent at significance level $\alpha = 0.05$. Therefore, $\hat{\Y} = \mat{I}$ and thus does not need to be interpreted. Finding evidence of layer independence in this multilayer network is supported by a biological explanation \citep{larremore2013network} and by the findings discussed in \citet{deBacco}, namely that the diversity of Malaria genes helps them to evade the immune system.
\subsubsection{Village Social Support Network}\label{subsec:village}
\begin{figure}
    \centering
    \includegraphics[width = 0.9\textwidth]{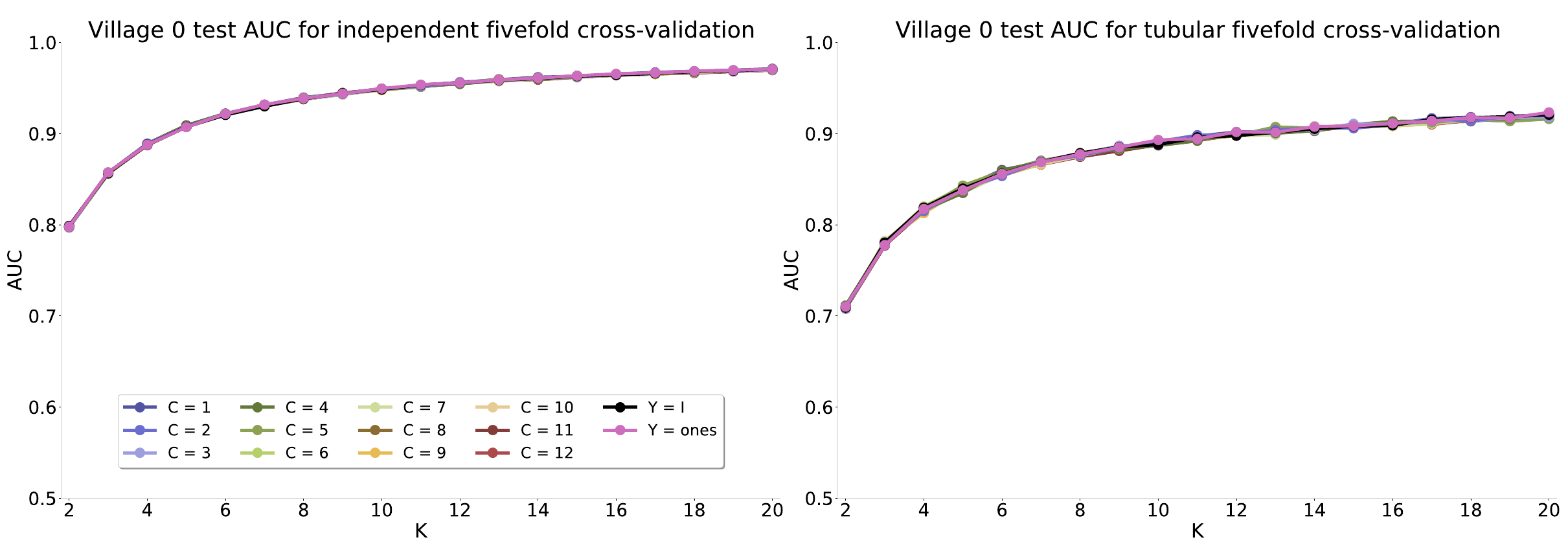}
    \caption{The test-AUC from the independent and tubular link prediction tasks for the Village 0 multilayer network. In both tasks, both the layer dependent and layer redundant \acs{ntds} perform just as well as the layer independent \acs{ntd} in terms of test-AUC.}
    \label{fig:bvil_link}
\end{figure}

The social multilayer network we consider contains different types of social interaction within a village in Karnataka, India, one of 43 \textit{microfinance villages} from \citet{banerjee2013}. We arbitrarily selected the first of the 43 villages and will henceforth refer to this village as ``Village 0''. The directed network consists of $N = 843$ individuals across $L=12$ layers. One individual is connected to another if the first indicated that they would interact in a specified way with the second. Each layer corresponds to a different type of social interaction (e.g., ``Who are the people who give you advice?'' and ``Who are your kin?''. See \Cref{SM:villageQ} for a full list of the questions.). For more information about the networks, survey instruments, or context of this data, see \citet{banerjee2013}.

Cross validation results for the Village 0 network are shown in \Cref{fig:bvil_link}. There is a slight gap in test-AUC across the two link-prediction tasks for this dataset, where the tubular link-prediction task is more difficult than the independent link-prediction task. However, in both tasks we observe that the layer redundant \acs{ntd} and the layer dependent \acs{ntds} (for all $C$) perform just as well as the layer independent \acs{ntd} in terms of test-AUC. 

The layer redundancy test confirms that this network is layer redundant at significance level $\alpha = 0.05$, consistent with the notion that the 12 layer network may indeed be such that all layer models are drawn from the same \acs{sbm}, as we saw in \Cref{ex:ones}. Considering the efforts made to collect data on these 12 different social support systems, this observation is surprising. One would expect that the distinct questions generating each layer of the network capture new information about the social network. This observation suggests otherwise, at least for the social structures that are well-modeled by stochastic block models. In this context, we echo the motivation of identifying layer interdependence in such social multilayer networks: that finding layer redundancies could justify a less extensive data-collection of future social networks in similar settings. We further explore this possibility with analysis in the next section.

\subsubsection{Collection of Village Social Support Networks} \label{multiVillage}
\begin{figure}
    \centering
    \includegraphics[width = 0.98\textwidth]{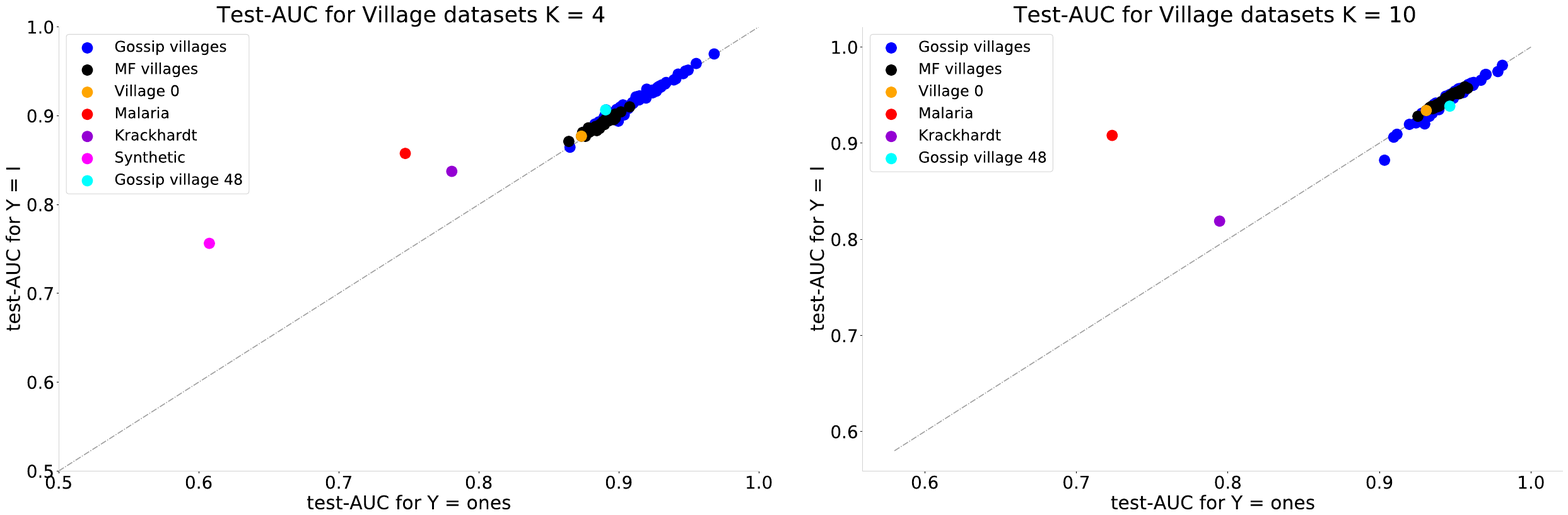}
    \caption{Comparing the test-AUC from the layer redundant \acs{ntd} to that from the layer independent \acs{ntd} with $K = 4$ (left) and $K = 10$ (right) for 113 different village multilayer networks from \citet{banerjee2013} and \citet{banerjee2019},  
    the malaria network from \citet{larremore2013network}, the \acs{css} from \citet{krackhardt1987}, and the first synthetic network from Section \ref{subsec:syn}. We discuss the data point marked ``Gossip village 48'' in further detail in Section \ref{multiVillage} and \Cref{fig:Y_syn}.}
    \label{fig:allauc}
\end{figure}
We now turn our attention to two sets of multilayer social networks representing different types of interaction in 113 villages: 43 \textit{microfinance villages} from \citet{banerjee2013} and 70 \textit{gossip villages} from \citet{banerjee2019}. The intent in studying these large collections of village networks is to see if the observation from \Cref{subsec:village}, that Village 0 is layer redundant at level $\alpha = 0.05$, is common across multiple different networks of the same type. The survey questions defining each layer for these networks are different for each data source (see \Cref{SM:villageQ}): the 12 types of social support defining 
the layers in the networks of \citet{banerjee2013} are different from the social support defining the 7 layers in the networks of \citet{banerjee2019}.

For each of these 113 villages we fix $K=4$ and $K=10$ and perform \acs{xval} under the independent link prediction task for both the layer independent and layer redundant \acs{ntd}. We plot the test-AUCs of this multi-village link prediction task in Figure \ref{fig:allauc}, where we also plot the test-AUC of the corresponding models in the Krackhardt, Village 0,
and first synthetic network. Surprisingly, we note that the test-AUC for the layer redundant \acs{ntd} is nearly equivalent to the test-AUC for the layer independent \acs{ntd} for almost all of the village networks. 
We highlight the village network with the biggest difference between the two test-AUCs, labeled ``Gossip village 48'', and estimate a layer dependent \acs{ntd} with $K=4$ and $C=2$. With this model choice, we find that the network is layer dependent at level $\alpha = 0.05$ and interpret the corresponding $\hat{\Y}$ in \Cref{fig:Y_syn}.

Finding evidence of layer redundancy in each of these 113 different village multilayer networks provides more evidence and motivation for using the \acs{ntd} as a tool for survey design. Specifically, we see evidence that each of these different types of social affiliation and support (between 7 to 12 depending on the network, see \Cref{SM:villageQ} for information on the different types of relationships in each network) are all well modeled by the same generative process. If, in a future study, the same evidence was found in initial data collection, then before scaling the study to more villages a less expensive survey could ask less questions of the participants. However, even if all layers of data are collected, knowing that they are redundant could help enhance other structural properties in the network \citep[e.g.,][]{nayar2015, taylor2016enhanced, taylor2017super}.
\section{Conclusion}
\label{conclusion}
In this work we use the \acf{ntd} with KL-divergence as an extension of the \acf{sbm} to multilayer networks. The \acs{ntd} allows for layers in the network to have latent structure, just as the \acs{sbm} allows for latent structure in the nodes of a single layer network. Using algebraic examples we show that the third factor matrix of the \acs{ntd} both captures and incorporates information about layer interdependence in multilayer networks. We show that the multiplicative updates for minimizing the KL-divergence of the \acs{ntd} are step-by-step equivalent to maximizing the log-likelihood of observing the network under the \acs{ntd} model using \acl{em}. This equivalence generalizes a previously known result about matrices and motivates the use of this algorithm in the context of the \acs{ntd}.

To use the \acs{ntd} to validate layer dependence in empirical multilayer networks, we define three \acf{lrts} to test layer independence, layer redundance, and layer dependence. Furthermore, we propose three methods for interpreting the third factor matrix of an \acs{ntd} estimated for an empirical network. We propose \acs{xval} as a means for model selection and formalize two link prediction tasks for the multilayer setting. We use \acs{xval}, the \acs{lrts}, and the approaches for interpreting $\hat{\Y}$ to study a variety of synthetic and empirical multilayer networks. In doing so, we find that the Malaria multilayer network has independent layers, 113 different social support networks are layer redundant, and Krackhardt's \acl{css} has layer dependence.

 This work also lays the groundwork for diverse future work and applications. Given the observation in \Cref{multiVillage}, that for many of the village multilayer networks we study the layers seem to be noisy observations from the same \acs{sbm}, it would be interesting to explore how other models of network formation (e.g., the choice-based dynamic models in \citet{overgoor2019}) uncover different characteristics amongst the layers that the \acs{sbm} cannot identify. As discussed in \Cref{applications}, the difference in test-AUC between the layer independent \acs{ntd} and the layer dependent \acs{ntd} with $C=L$ is not fully understood and could be addressed in future work. Furthermore, some multilayer graphs (for instance, the \verb|ogbl-wikikg2| multilayer network from \cite{hu2020open}, which a reviewer brought to our attention) have such a structure where an edge in one layer determines the presence of an edge in another layer. It is unclear if the \acs{ntd} would be well suited, or justified, to be used to model such structureed multilayer networks, and inspecting this is an interesting subject for future work. Finally, an interesting future direction is understanding how  the \acs{ntd} can be made more interpretable under a Varimax rotation, following recent connections between Varimax and factor model inference in \citet{rohe2020}.

Understanding the layer dependencies in a multilayer network can inform the development of survey design, identify redundancies, or illuminate contextual connections. Moreover, the usefulness of finding latent structure in the layers motivates the use of latent-space models as a noise-free smoothing of the observed network, as proposed by \citet{fisher2021using}. As such, there is potential to use this work to understand layer dependence in a variety of applications where domain-specific knowledge can make use of the interpretations that the  \acs{ntd} provides.

\section*{Acknowledgments and Disclosure of Funding} 
IA acknowledges support from the NSF GRFP and the Knight-Hennessy Scholars Fellowship. DT acknowledges partial support from NSF (\#DMS-2052720) and the Simons Foundation (\#578333). JU acknowledges partial support ARO (\#76582-NS-MUR) and NSF (\#2143176). We would also like to thank our reviewers for insightful suggestions and Caterina De Bacco, Eleanor Power, Dan Larremore, and Samir Khan for helpful conversations.
\newpage

\appendix
\section{Implementation Details}
\label{code_repo}
Tools for this work and an example jupyter notebook (all in python) can be found at \\ \verb|https://github.com/izabelaguiar/NNTuck|

The implementation of the multiplicative updates algorithm from \cite{kimChoi}, \Cref{nntuck:alg}, is done in python using the \verb|tensorly| backend. The  \verb|tensorly| package \citep{kossaifi2016} already has an implementation of the multiplicative updates algorithm for estimating the \acl{ntd}, although the implementation is for minimizing the least-squares loss between the tensor and its decomposition. We thus altered their implementation to include the multiplicative updates for minimizing \acs{kld}. Our implementation uses dense matrix computations, and as such, does not take advantage of the sparse structure present in multilayer networks. Although the \verb|tensorly| package does have a sparse option, their documentation describes that only \textit{memory usage} is improved by using the sparse backend, and that using the sparse backend for decompositions actually takes much longer.

\begin{figure}
    \centering
    \includegraphics[width = 0.45\textwidth]{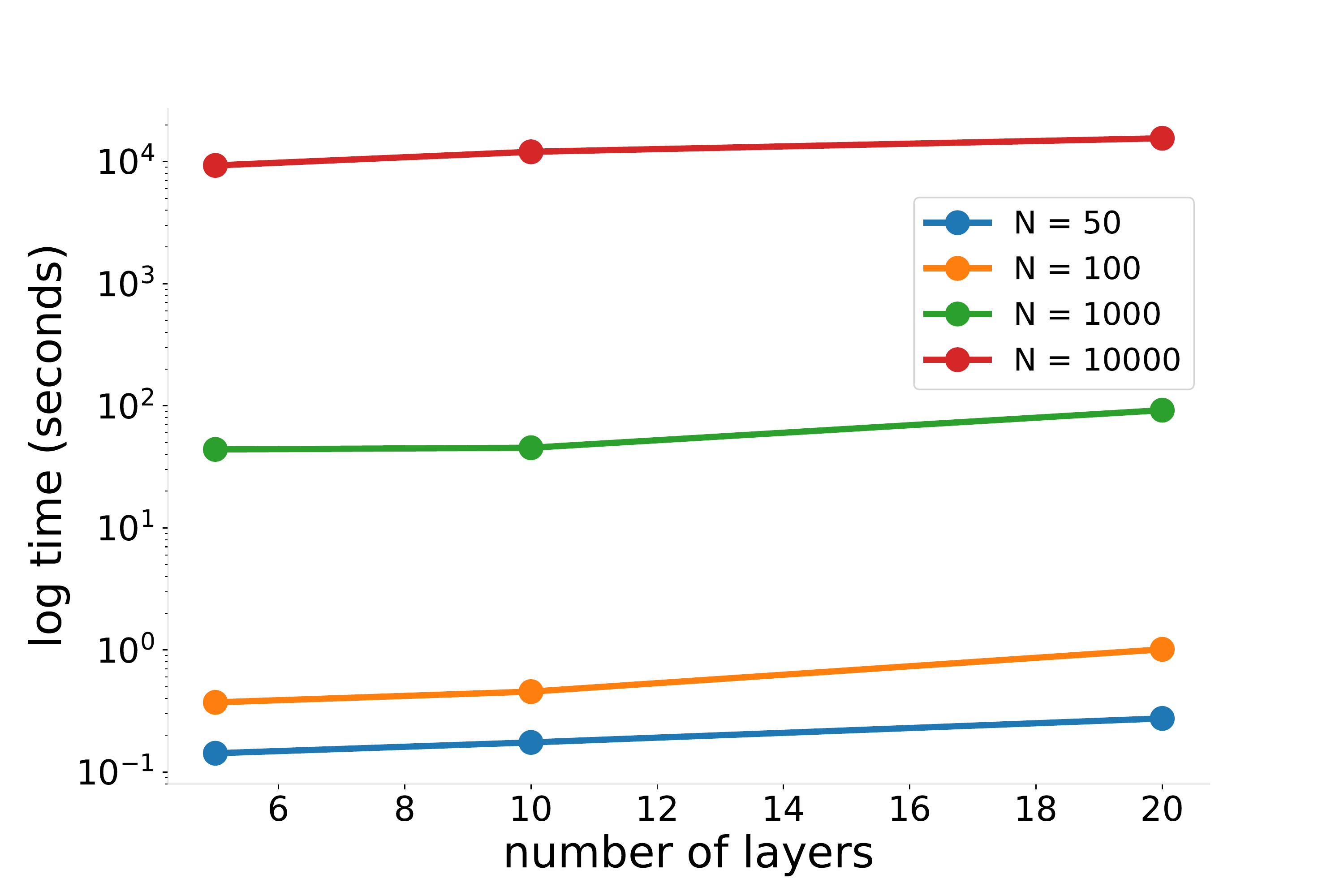}
    \includegraphics[width = 0.45\textwidth]{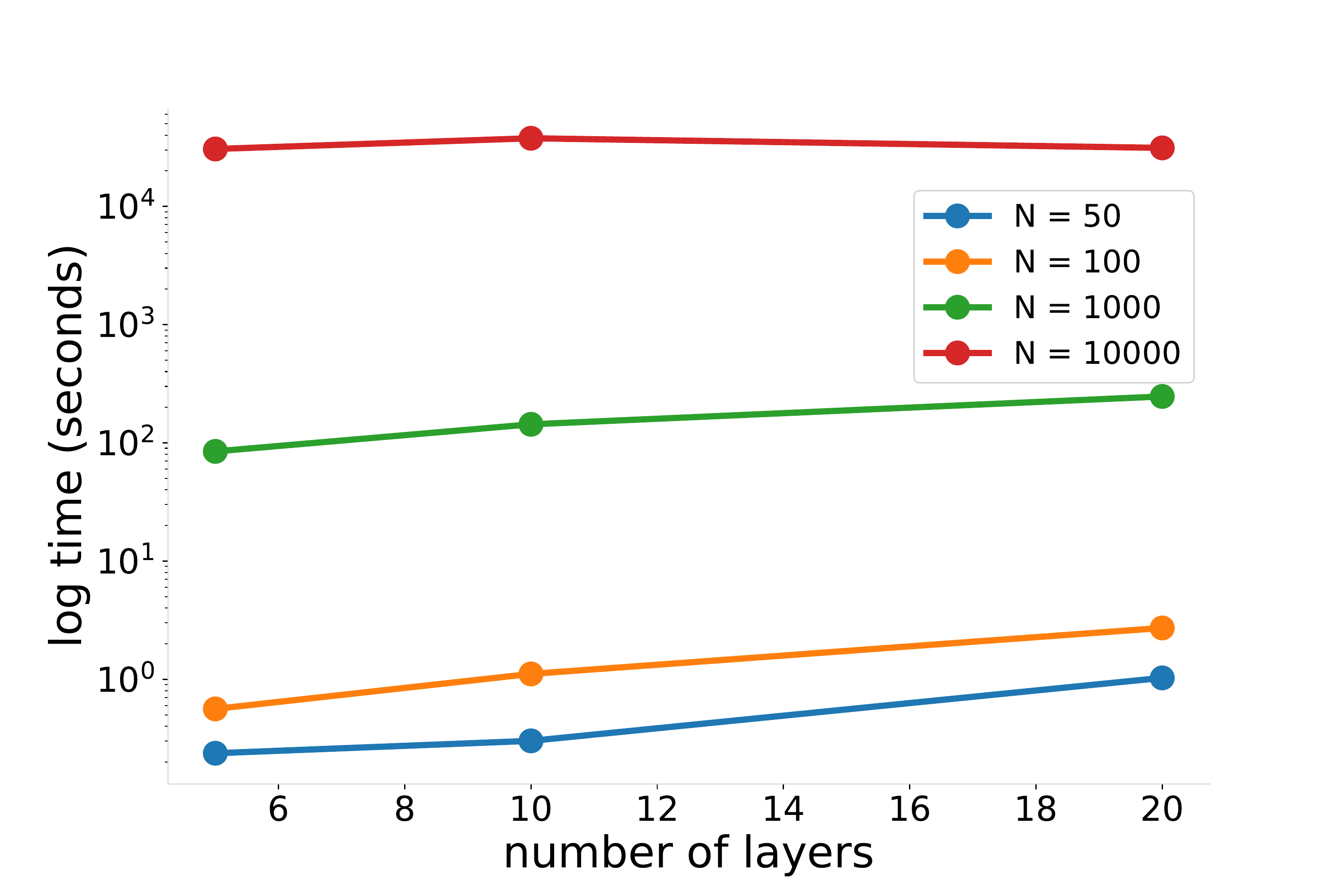}
    \caption{The time to convergence for estimating the layer redundant (left) and layer independent (right) \acs{ntd} of synthetic networks of varying sizes, averaged across 20 random initializations, plotted on a log scale.}
    \label{fig:comp}
\end{figure}

To present a scope of how computation time scales with network size, we plot the average time \Cref{nntuck:alg} takes to converge across 20 random initializations for estimating both the layer redundant and layer independent \acs{ntd} of synthetic multilayer networks with number of nodes varying from $50$ to $10,000$ and number of layers varying from $5$ to $20$. As we see in \Cref{fig:comp}, the average time to convergence for the smallest network ($N=50$, $L=5$) was 0.125 seconds, and for the largest network ($N=10,000$, $L =20$) the average convergence time was 8.66 hours. As an upper limit for the computation we were able to do before running into memory limits, we generated a synthetic multilayer network with $N=25,000$ nodes and $L=5$ layers. Over 20 random initializations of \Cref{nntuck:alg}, estimating the layer redundant \acs{ntd} and the layer independent \acs{ntd} of this network took 12.43 hours and 18.044 hours, on average, respectively.

Efficient algorithms for estimating the \acl{ntd}, such as those discussed in \cite{zhou2015efficient}, minimize least-squares loss. Future work in extending the \acs{ntd} for use in larger networks will necessitate a faster implementation of the algorithm discussed and used in this work, if not a completely different algorithm for minimizing \acs{kld}.

\begin{figure}
    \centering
    \includegraphics[width =\textwidth]{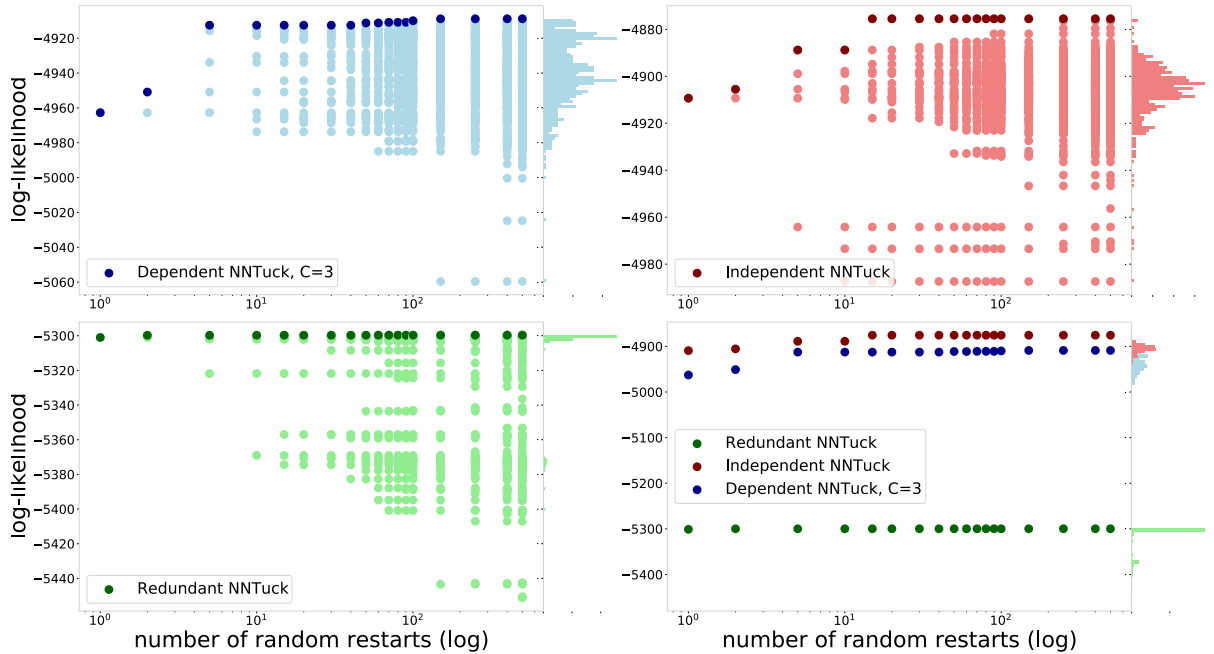}
    \caption{We show the log-likelihood of the dependent, independent, and dependent \acs{ntds} for the Krackhardt multilayer network estimated using \cref{nntuck:alg}. For each estimation we vary the number of random initializations on the horizontal axis, and for each we plot the maximal log-likelihood over that set with a bold point. In the last plot we plot the maximal log-likelihood for each \acs{ntd} together. The histogram on the right hand side of each plot shows the distribution of the log-likelihood across all 500 random initializations. We see that the maximal log-likelihood does not vary greatly when using more than 20 random initializations, and thus determine that using 20 random initializations to estimate the \acs{ntd} is appropriate.}
    \label{fig:multistart}
\end{figure}

Another implementation detail, discussed in \cref{subsec:emp}, is that we select the \acs{ntd} with the highest log-likelihood across 20 random initializations. 
In \cref{fig:multistart}, we show the variation in maximal log likelihood as a function of number of random restarts and show that 20 random initializations is sufficient for estimating the \acs{ntd}.

\section{EM and NNTuck multiplicative updates equivalence}
\label{SM:MTeqNTD}
In this section we show the equivalence between the \acrfull{em} updates and the multiplicative updates for the \acl{ntd} under a \acs{kld} loss.

Recall \Cref{nntuck_em_equiv_prop}, restated below, 
\setcounter{proposition}{0}
\begin{proposition}
Determining factor matrices $\U, \V$, and $\Y$ and the core tensor $\G$ in the \acs{ntd} by maximizing the log-likelihood using \acf{em} \cref{eq:EMupdates} is equivalent to using the multiplicative updates \cref{eq:KimChoiUpdates} to minimize KL-divergence.
\end{proposition}
Consider using \acf{em} to reach a local maximum of the log-likelihood of observing $\A$ under the model given by \eqref{Tucker},
    \begin{equation}
        \mathcal{L}(\A | \U, \V, \Y, \G) = \sum_{i, j, \alpha} \left[ a_{ij\alpha} \log \sum_{k, \ell, \rho} u_{ik}v_{j\ell}y_{\alpha \rho}g_{k \ell \rho} - \sum_{k, \ell, \rho} u_{ik}v_{j\ell}y_{\alpha \rho}g_{k \ell \rho}\right],
        \label{loglike-Tuck}
    \end{equation} 
    in which case the following update equations are used
\begin{equation}
\begin{aligned}
u_{i k} &=\frac{\sum_{j, \alpha} A_{i j \alpha} \sum_{\ell} \rho_{i j k \ell}^{(\alpha)}}{\sum_{\ell}\left(\sum_{j} v_{j \ell}\right)\left(\sum_{\alpha} g_{k \ell \alpha}\right)}, \\
v_{j \ell} &=\frac{\sum_{i, \alpha} A_{i j \alpha} \sum_{k} \rho_{i j k \ell}^{(\alpha)}}{\sum_{k}\left(\sum_{i} u_{i k}\right)\left(\sum_{\alpha} g_{k \ell \alpha}\right)}, \\
g_{k \ell \alpha} &=\frac{\sum_{i j} A_{i j\alpha} \rho_{i j k \ell}^{(\alpha)}}{\left(\sum_{i} u_{i k}\right)\left(\sum_{j} v_{j \ell}\right)},
\end{aligned}
\label{eq:EMupdates}
\end{equation}
where
\begin{equation}
    \rho_{i j k \ell}^{(\alpha)}=\frac{u_{i k} v_{j \ell} y_{\alpha c} g_{k \ell c}}{\sum_{k^{\prime} \ell^{\prime} c^{\prime}} u_{i k^{\prime}} v_{j \ell^{\prime}} y_{\alpha c^{\prime}}g_{k^{\prime} \ell^{\prime} c^{\prime}}}.
\end{equation}
Conversely, the multiplicative updates for \acf{ntd} under the KL-divergence loss as given by \citet{kimChoi} are, 
\begin{equation}
\begin{aligned}
\U &\leftarrow \U \circledast \frac{\left[\mat{A}_{(1)} / \left(\U \mat{G}_{\U}^{(1)}\right)\right] \mat{G}_{\U}^{(1) \top}}{\vec{1} \vec{z}_{\mat{U}}^{\top}},\\
\V &\leftarrow \V \circledast \frac{\left[\mat{A}_{(2)} /\left(\V \mat{G}_{\V}^{(2)}\right)\right] \mat{G}_{\V}^{(2) \top}}{\vec{1} \vec{z}_{\mat{V}}^{\top}},\\
\Y &\leftarrow \Y \circledast \frac{\left[\mat{A}_{(3)} /\left(\Y \mat{G}_{\Y}^{(3)}\right)\right] \mat{G}_{\Y}^{(3) \top}}{\vec{1} \vec{z}_{\mat{Y}}^{\top}},\\
\G &\leftarrow \G \circledast \frac{(\A / \hatA) \times_{1} \U^\top \times_{2} \V^\top \times_3 \Y^\top}{\ten{E} \times_{1} \U^\top \times_{2} \V^\top \times_3 \Y^\top}, \\
&z_{\mat{U} i} = \sum_j (\mat{G}^{(1)}_{\mat{U}})_{ij},\\
&z_{\mat{V} i} = \sum_j (\mat{G}^{(2)}_{\mat{V}})_{ij},\\
&z_{\mat{Y} i} = \sum_j (\mat{G}^{(3)}_{\mat{Y}})_{ij}.
\end{aligned}
\label{eq:KimChoiUpdates}
\end{equation}
Above, $\ten{E}$ is the all ones tensor of the same dimension as $\A$, $\circledast$ and $/$ denote elementwise multiplication and division, respectively, and the subscript $\cdot_{(\ell)}$ denotes the tensor $\ell$-unfolding. $\mat{G}_{\U}^{(1)}$, $\mat{G}_{\V}^{(2)}$, and $\mat{G}_{\Y}^{(3)}$ are defined as
\begin{equation*}
    \mat{G}^{(1)}_{\mat{U}}= \left[ \ten{G} \times_2 \mat{V} \times_3 \mat{Y} \right]_{(1)},
\mat{G}^{(2)}_{\mat{V}} = \left[ \ten{G} \times_1 \mat{U} \times_3 \mat{Y} \right]_{(2)},
\mat{G}^{(3)}_{\mat{Y}} = \left[ \ten{G} \times_1 \mat{U} \times_2 \mat{V} \right]_{(3)}.
\end{equation*}

\begin{figure}
    \centering
    \includegraphics[width = 0.55\textwidth]{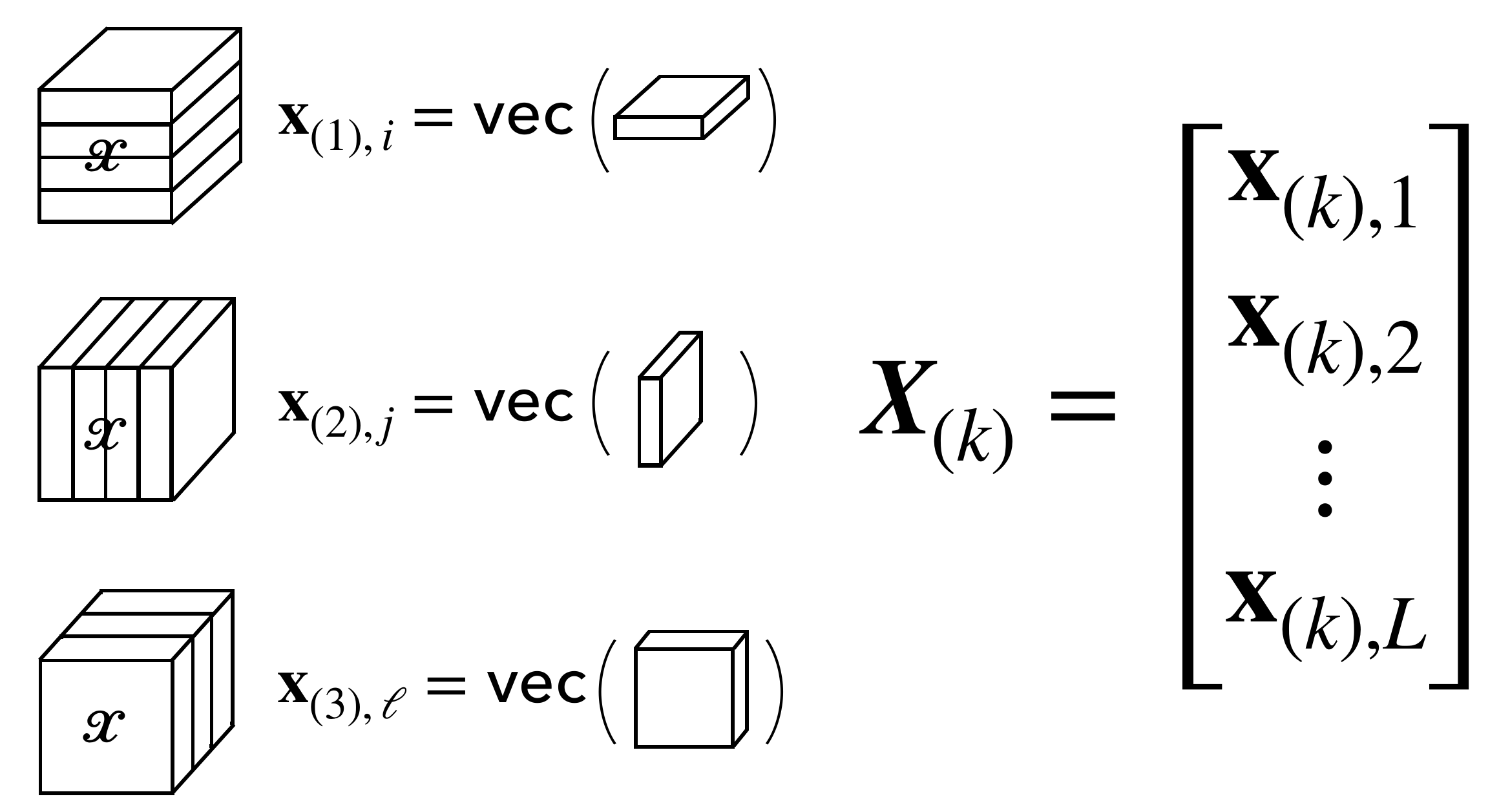}
    \caption{The $k$-unfolding of a tensor can be described by vertically stacking vectorizations of one of its slices. Above, in descending order, we see the horizontal, lateral, and frontal slices of a tensor. Vectorizing each slice and vertically stacking them gives the 1-, 2-, and 3-unfolding of tensor $\ten{X}$, respectively. For $\ten{X} \in \R^{N \by M \by L}$, then the dimensions of the three unfoldings are $\mat{X}_{\uf{1}} \in \R^{M \by NL}$, $\mat{X}_{\uf{2}} \in \R^{N \by ML}$, and $\mat{X}_{\uf{3}} \in \R^{L \by NM}$.}
    \label{fig:unfolding}
\end{figure}

We will make use of \textit{tensor unfoldings} and the \textit{tensor $n$-mode product} in the following equivalence proofs. To help guide intuition on how tensor unfoldings are used, we provide \Cref{fig:unfolding} as one visualization of the three unfoldings of a third-order tensor $\ten{X}$.
\paragraph{Equivalence of core tensor updates}
We first show that the updates to $\G$ in \cref{eq:EMupdates} are equivalent to the updates to $\G$ in \cref{eq:KimChoiUpdates}. 
\begin{proof}
The $(i, j, \alpha)$ entry of $\hatA:= \G \times_1 \U \times_2 \V \times_3 \Y$ is 
\begin{equation*}
    \hat{A}_{ij\alpha} = \sum_{k'\ell'p'}u_{ik'}v_{j\ell'}y_{\alpha p'}g_{k'\ell'p'},
\end{equation*}
and therefore the update to $g_{k\ell\alpha}$ can be rewritten as,
\begin{equation*}
\begin{aligned}
g_{k \ell p} &=\frac{\sum_{i j \alpha}\left( {A_{i j\alpha} }/{\hat{A}_{i j\alpha}}\right)u_{i k} v_{j \ell}y_{\alpha p} g_{k \ell p}}{\left(\sum_{i} u_{i k}\right)\left(\sum_{j} v_{j \ell}\right) \left(\sum_{\alpha} y_{\alpha p}\right)} \\
    & = g_{k \ell p} \cdot \frac{\sum_{i j \alpha}\left( {A_{i j\alpha} }/{\hat{A}_{i j\alpha} }\right)u_{i k} v_{j \ell} y_{\alpha p} }{\left(\sum_{i} u_{i k}\right)\left(\sum_{j} v_{j \ell}\right)\left(\sum_{\alpha} y_{\alpha p}\right)}.
\end{aligned}
\end{equation*}
Note the $(a, b, c)$ element of the tensor mode-1 product of the following, 
\begin{equation}
    \begin{aligned}
    \left[ \left(  \frac{\ten{A} }{\hatA}\right) \times_1 \U^\top \right]_{a b c} = \sum_i \left(\frac{\ten{A} }{\hatA}\right)_{ibc}u_{ia}.
    \end{aligned}
\end{equation}
Then, 
\begin{equation}
    \begin{aligned}
    \left\{ \left[ \left(  \frac{\ten{A} }{\hatA}\right) \times_1 \U^\top \right] \times_2 \V^\top \times_3 \Y^\top \right\}_{k \ell p} & = \sum_\alpha \left[ \left(  \frac{\ten{A} }{\hatA}\right) \times_1 \U^\top \times_2 \V^\top \right]_{k\ell p} y_{\alpha p}\\
    & = \sum_{j \alpha} \left[ \left(  \frac{\ten{A} }{\hatA}\right) \times_1 \U^\top \right]_{kj \alpha} v_{j \ell} y_{\alpha p}\\
    &= \sum_{j \alpha} \left( \sum_i \left(\frac{\ten{A} }{\hatA}\right)_{ij\alpha} u_{ik}\right) v_{j \ell} y_{\alpha p} \\
    & = \sum_{ij\alpha} \left(A_{ij\alpha}/\hat{A}_{ij\alpha}\right) u_{ik} v_{j \ell}y_{\alpha p}.
    \end{aligned}
\end{equation}
Now, note the $(a, b, c)$ element of the tensor mode-1 product of the following,
\begin{equation}
    \left[ \ten{E} \times_1 \U^\top \right]_{abc} = \sum_i E_{ibc} u_{ia} = \sum_i u_{ia}.
\end{equation}
Then, 
\begin{equation}
    \begin{aligned}
    \left\{ \left[ \ten{E} \times_1 \U^\top \right] \times_2 \V^\top \times_3 \Y^\top \right\}_{k \ell p} & = \sum_\alpha \left[ \ten{E} \times_1 \U^\top \times_2 \V^\top \right]_{k\ell\alpha} y_{\alpha p}\\
    & = \sum_{j\alpha} \left[ \ten{E} \times_1 \U^\top \right]_{kj\alpha} v_{j \ell}y_{\alpha p}\\
    &= \sum_{j \alpha} \left(\sum_i u_{ik}\right) v_{j\ell}y_{\alpha p}\\
    &= \left(\sum_i u_{ik}\right) \left(\sum_j v_{j \ell}\right)\left(\sum_\alpha y_{\alpha p}\right).
    \end{aligned}
\end{equation}
Therefore, focusing on the NNTuck multiplicative update of core tensor $\G$, we see that the update to the $(k, \ell, \alpha)$ element of the core tensor is, 
\begin{equation}
\begin{aligned}
g_{k\ell p} &\leftarrow g_{k\ell p} \cdot \frac{[(\A / \hatA) \times_{1} \U^\top \times_{2} \V^\top]_{k\ell\alpha}}{[\ten{E} \times_{1} \U^\top \times_{2} \V^\top \times_3 \Y^\top]_{k\ell p}} \ \ \Leftrightarrow \\
g_{k \ell p}&\leftarrow g_{k \ell p} \cdot \frac{\sum_{i j \alpha}\left( {A_{i j\alpha} }/{\hat{A}_{i j\alpha} }\right)u_{i k} v_{j \ell} y_{\alpha p} }{\left(\sum_{i} u_{i k}\right)\left(\sum_{j} v_{j \ell}\right)\left(\sum_{\alpha} y_{\alpha p}\right)},
\end{aligned}
\end{equation}
which is equivalent to the update to the $(k, \ell, \alpha)$ element of $\G$ in \cref{eq:EMupdates}.
\end{proof}
\paragraph{Equivalence of factor matrix $\U$ updates}
For showing the equivalence in updates to factor matrix $\U$, the following identity is used multiple times. For tensor $\mathcal{X}$,
\begin{equation}
    \sum_{j \alpha}X_{ij\alpha} = \sum_j \mat{X}_{(1) ij},
    \label{eq:unfoldID}
\end{equation}
where $\mat{X}_{(1)}$ denotes the $1$-unfolding of $\ten{X}$.
\begin{proof}
Consider the EM update to $u_{ik}$ from \ref{eq:EMupdates}. Again,  
\begin{equation*}
     \hat{A}_{ij\alpha} = \sum_{k'\ell'p'}u_{ik'}v_{j\ell'}y_{\alpha p'}g_{k'\ell'p'},
\end{equation*}
and so the EM update to $u_{ik}$ can be rewritten as,
\begin{equation}
\begin{aligned}
u_{ik} &=\frac{\sum_{j\alpha}\left( {A_{i j\alpha} }/{\hat{A}_{i j\alpha} }\right)\sum_{\ell p} u_{i k} v_{j \ell} y_{\alpha p} g_{k \ell p}}{\sum_{\ell p} \left(\sum_{j} v_{j \ell}\right)\left(\sum_{\alpha} y_{\alpha p} g_{k \ell p}\right)} \\
    & = u_{i k} \cdot \frac{\sum_{j\alpha}\left( {A_{i j\alpha} }/{\hat{A}_{i j\alpha} }\right)\sum_{\ell p} v_{j \ell} y_{\alpha p} g_{k \ell p}}{\sum_{\ell p} \left(\sum_{j} v_{j \ell}\right) \left(\sum_{ \alpha } y_{\alpha p} g_{k \ell p}\right)}.
\end{aligned}
\label{eq:emUup1}
\end{equation}
Now note that from the above equation we can identify out 
\begin{equation}
    \sum_{\ell p} v_{j \ell} y_{\alpha p} g_{k \ell p} = [\G \times_2 \V \times_3 \Y]_{kj\alpha}.
    \label{vw}
\end{equation}
For vector $\vec{z}$ such that $z_{i} = \sum_j [\G \times_2 \V \times_3 \Y]_{(1) i j}$, note that 
\begin{equation}
\begin{aligned}
    \left[ \vec{1}\vec{z}^{\top} \right]_{ik} = (1)(z_k) = z_k &=[\G \times_2 \V \times)3 ]_{(1) k j}\\
    &= \sum_{j \alpha}[\G \times_2 \V \times_3 \Y]_{kj\alpha}\\
    &= \sum_{j\alpha} \sum_{\ell p} g_{k\ell p} v_{j \ell} y_{\alpha p}\\
    & = \sum_{\ell p} \sum_{j\alpha} g_{k\ell p} y_{\alpha p} v_{j \ell} \\
    & = \sum_{\ell p} \left(\sum_{j} v_{j \ell}\right)\left(\sum_{\alpha} y_{\alpha p} g_{k \ell\alpha}\right).
\end{aligned}
\label{denom}
\end{equation}
Substituting \eqref{vw} and \eqref{denom} into \cref{eq:emUup1}, we have that
\begin{equation}
    \begin{aligned}
    u_{ik} & = u_{i k} \cdot \frac{\sum_{j\alpha}\left( {A_{i j\alpha} }/{\hat{A}_{i j\alpha} }\right)[\G \times_2 \V \times_3 \Y]_{kj\alpha} }{[\vec{1}\vec{z}^\top]_{ik}} \\
     & = u_{i k} \cdot \frac{\sum_{j\alpha}\left( {A_{i j\alpha} }/{\hat{A}_{i j\alpha} }\right)[\G \times_2 \V \times_3 \Y]_{ (1)kj} }{[\vec{1}\vec{z}^\top]_{ik}} \\
     & = u_{i k} \cdot \frac{\sum_{j\alpha}\left( {A_{i j\alpha} }/{\hat{A}_{i j\alpha} }\right)[\G \times_2 \V \times_3 \Y]^\top_{ (1)jk} }{[\vec{1}\vec{z}^\top]_{ik}}.
    \end{aligned}
\end{equation}
Then using \cref{eq:unfoldID} we have that,
\begin{equation}
    \begin{aligned}
    &\sum_{j\alpha}A_{ij\alpha} = \sum_j \mat{A}_{(1)ij},\\
    &\sum_{j\alpha}\hat{A}_{ij\alpha} = \sum_j \hat{\mat{A}}_{(1)ij} = \sum_j (\U[\G \times_2 \V \times_3 \Y]_{(1)})_{ij},\\
    &\sum_{j \alpha} [\G \times_2 \V \times_3 \Y]_{kj \alpha} = \sum_j [\G \times_2 \V \times_3 \Y]_{(1)kj} = \sum_j [\G \times_2 \V \times_3 \Y]_{(1)jk}^\top.
    \end{aligned}
\label{steps}
\end{equation}

In the second line of \cref{steps} above, we use that 
\begin{align*}
    \hatA &= \G \times_1 \U \times_2 \V  \times_3 \Y \\ 
    &= \G \times_2 \V \times_3 \Y \times_1 \U \\
    &= (\G \times_2 \V \times_3 \Y) \times_1 \U,
\end{align*}
and thus using the identity $\ten{Y} = \ten{X}\times_{n}\mat{B} \Rightarrow \mat{Y}_{(n)} = \mat{B}\mat{X}_{(n)},$ we get $\mat{\hat{A}}_{(1)}=\U[\G \times_2 \V \times_3 \Y]_{(1)}$.

Therefore, 
\begin{equation}
    \begin{aligned}
    u_{ik} &= u_{i k} \cdot \frac{\sum_j ( \mat{A}_{(1)}/(\U[\G \times_2 \V \times_3 \Y]_{(1)}))_{ij} [\G \times_2 \V \times_3 \Y]_{(1)jk}^\top }{[\vec{1}\vec{z}^\top]_{ik}}\\
    & = u_{i k} \cdot \frac{ \left[( \mat{A}_{(1)}/(\U[\G \times_2 \V \times_3 \Y]_{(1)})) [\G \times_2 \V \times_3 \Y]_{(1)}^\top \right]_{ik}}{[\vec{1}\vec{z}^\top]_{ik}}.
    \end{aligned}
\end{equation}
This is equivalent to the $ik$ update of $\U$ given by \cref{eq:KimChoiUpdates}.
\end{proof}
\paragraph{Equivalence of factor matrix $\V$ updates} In connecting the updates to factor matrix $\V$, the following identity is used multiple times. For tensor $\mathcal{X}$,
\begin{equation}
    \sum_{i \alpha}X_{ij\alpha} = \sum_i \mat{X}_{(2) ji},
\end{equation}
where $\mat{X}_{(1)}$ denotes the $(1)$-unfolding of $\ten{X}$.
\begin{proof}
Consider the EM update to $v_{j\ell}$ given by \cref{eq:EMupdates}. Because 
\begin{equation*}
    \hat{A}_{ij\alpha} = \sum_{k'\ell'p'}u_{ik'}v_{j\ell'}y_{\alpha p'}g_{k'\ell'p'},
\end{equation*}
then the EM update to $v_{j\ell}$ can be rewritten as,
\begin{equation}
\begin{aligned}
v_{j\ell} &=\frac{\sum_{i\alpha}\left( {A_{i j\alpha} }/{\hat{A}_{i j\alpha} }\right)\sum_{kp} u_{i k} v_{j \ell} y_{\alpha p}g_{k \ell p}}{\sum_{kp} \left(\sum_{i} u_{ik}\right)\left(\sum_{\alpha} y_{\alpha p}g_{k \ell p} \right)} \\
    & = v_{j\ell} \cdot \frac{\sum_{i\alpha}\left( {A_{i j\alpha} }/{\hat{A}_{i j\alpha} }\right)\sum_{kp} u_{i k} y_{\alpha p}g_{k \ell p}}{\sum_{kp} \left(\sum_{j} u_{ik}\right)\left(\sum_{\alpha} y_{\alpha p}g_{k \ell p} \right)}.
\end{aligned}
\label{v_em_updates}
\end{equation}
Now note that we can again identify out
\begin{equation}
    \sum_{kp} u_{ik} y_{\alpha p} g_{k \ell p} = [\G \times_1 \U \times_3 \Y]_{i \ell\alpha}.
    \label{uw}
\end{equation}

And again, for vector $\vec{z}$ such that $z_j = \sum_i [\G \times_1 \U \times_2 \Y]_{(2) j i}$, note that 
\begin{equation}
\begin{aligned}
    \left[\vec{1}\vec{z}^\top\right]_{j \ell} = (1)(z_\ell) = z_\ell &=\sum_i[\G \times_1 \U \times_2 \Y]_{(2) \ell i}\\
    &= \sum_{i \alpha}[\G \times_1 \U \times_3 \Y]_{i \ell\alpha}\\ 
    &= \sum_{i\alpha} \sum_{kp} u_{ik}y_{\alpha p}g_{k\ell p} \\
    & = \sum_{kp} \sum_{i\alpha} u_{ik} y_{\alpha p} g_{k\ell p} \\
    & = \sum_{kp} \left(\sum_{i} u_{ik}\right)\left(\sum_{\alpha} y_{\alpha p}g_{k \ell p}\right).
\end{aligned}
\label{denomv}
\end{equation}
Substituting together \eqref{uw} and \eqref{denomv} into the EM updates \cref{v_em_updates}, we have that
\begin{equation}
    v_{j\ell} = v_{j\ell} \cdot \frac{\sum_{i\alpha}\left( {A_{i j\alpha}}/{\hat{A}_{i j\alpha} }\right)[\G \times_1 \U \times_3 \Y]_{i \ell\alpha} }{[\vec{1}\vec{z}^\top]_{j \ell}}.
\end{equation}
Then,
\begin{equation}
    \begin{aligned}
    \sum_{i\alpha}A_{ij\alpha} = \sum_i \mat{A}_{(2)ji}&,\\
    \sum_{i\alpha}\hat{A}_{ij\alpha} = \sum_i \hat{\mat{A}}_{(2)ji} = \sum_i (\V[\G \times_1 \U \times_3 \Y]_{(2)})_{ji}&,\\
    \sum_{i \alpha} [\G \times_1 \U \times_3 \Y]_{i \ell\alpha} = \sum_i [\G \times_1 \U \times_3 \Y]_{(2)\ell i} = \sum_i [\G \times_1 \U \times_3 \Y]_{(2) i\ell}^\top&.
    \end{aligned}
    \label{stepsv}
\end{equation}
In the second line of \ref{stepsv} above, we use that 
\begin{equation*}
     \hatA = \G \times_1 \U \times_2 \V  \times_3 \Y = \G \times_1 \U \times_3 \Y \times_2 \V = (\G \times_1 \U \times_3 \Y) \times_2 \V,
\end{equation*}
and thus using the identity $\ten{Y} = \ten{X}\times_{n}\mat{B} \Rightarrow \mat{Y}_{(n)} = \mat{B}\mat{X}_{(n)},$ we get $\mat{\hat{A}}_{(2)}=\V[\G \times_1 \U \times_3 \Y]_{(2)}$.

Therefore, 
\begin{equation}
    \begin{aligned}
    v_{j \ell} &= v_{j \ell} \cdot \frac{\sum_i ( \mat{A}_{(2)}/(\V[\G \times_1 \U \times_3 \Y]_{(2)}))_{ji} [\G \times_1 \U \times_3 \Y]_{(2)i \ell}^\top }{[\vec{1}\vec{z}^\top]_{j \ell}}\\
    & = v_{j \ell} \cdot \frac{ \left[( \mat{A}_{(2)}/(\V[\G \times_1 \U \times_3 \Y]_{(2)})) [\G \times_1 \U \times_3 \Y]_{(2)}^\top \right]_{j \ell}}{[\vec{1}\vec{z}^\top]_{j \ell}}
    \end{aligned}
\end{equation}
This is equivalent to the $j\ell$ update of $\V$ given by \cref{eq:KimChoiUpdates}. 
\end{proof}
Showing the equivalence for the updates to $\Y$ follow exactly as those above, and thus we do not reproduce it here. Therefore, we have shown that the EM updates given by \cref{eq:EMupdates} are step-by-step equivalent to the multiplicative updates given by \cref{eq:KimChoiUpdates}.
\section{Masked updates}
\label{SM:masking}
In this section we discuss the algorithmic changes to the multiplicative updates from  \citet{kimChoi} for the \acl{ntd} under a \acs{kld} loss to allow for masking, as used during cross-validated model evaluation. This section describes the tensor completion problem wherein we wish to build the \acs{ntd} using only observed entries.

Consider that some of the entries of adjacency tensor $\A$ are unobserved. We wish to build the \acs{ntd} of $\A$ using \textit{only} the observed entries, but we want the reconstruction $\hatA$ to predict the unobserved entries. This \textit{tensor completion} problem is how we train on 80\% of the entries of $\hatA$ for the five-fold \acs{xval} in \Cref{applications}.
\begin{figure}
    \centering
    \includegraphics[width = 0.6\textwidth]{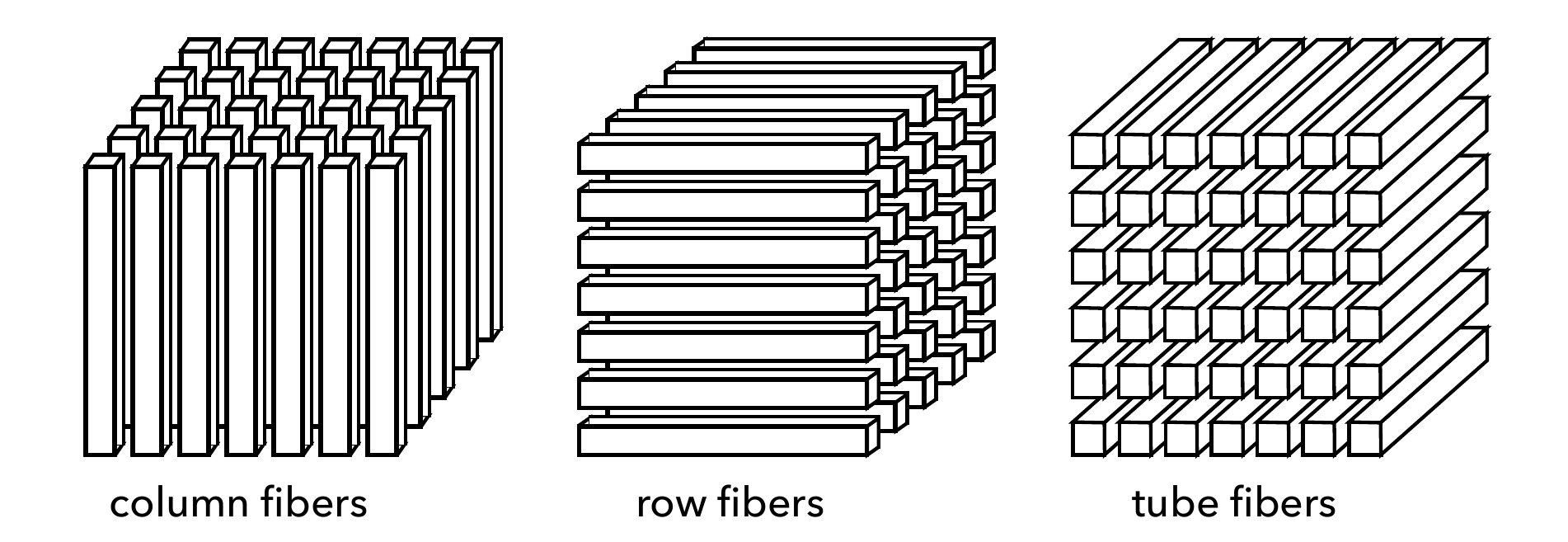}
    \caption{The row, column, and tube \textit{fibers} of a third-order tensor. This figure has been adapted from Figure 2 in the tensor review by \citet{Kolda2009}. In the \textit{tubular} link prediction task in \Cref{applications}, entire tubes of the adjacency tensor are missing i.i.d. with uniform probability.}
    \label{fibers}
\end{figure}
Assume that there is a set $\mathcal{I}$ such that
\begin{equation}
    \mathcal{I} := \{ (i, j, \alpha) \mid A_{ij\alpha} \text{ is unobserved}\}.
\end{equation}
We want to rederive the update rules from \citet{kimChoi} for nonnegative Tucker decomposition to only account for the observed entries of $\A$. To do so, we introduce a \textit{masking tensor}, $\ten{M}$ such that,
\begin{equation*}
    M_{ij\alpha} = \begin{cases}
    0 & \text{ if } (i, j, \alpha) \in \mathcal{I}\\
    1 & \text{ if } (i, j, \alpha) \notin \mathcal{I}.\\
    \end{cases}
\end{equation*}
Then re-deriving the update rules from the log-likelihood and EM approach, and after re-tensorizing them, the update rules for factor matrices $\mat{U}, \mat{V}, \mat{Y}$ and for core tensor $\ten{G}$ are,
\begin{equation}
    \U = \U \circ \frac{[\mat{M}_{(1)} \circ \mat{A}\uf{1}/ \mat{\hat{A}}_{(1)}] [\G\times_2 \V\times_3\Y]\uf{1}^\top}{\mat{M}\uf{1}[\G\times_2 \V\times_3 \Y]\uf{1}^\top},
    \label{U_update}
\end{equation}
\begin{equation}
    \V = \V \circ \frac{[\mat{M}\uf{2} \circ \mat{A}\uf{2}/ \mat{\hat{A}}_{(2)}] [\G\times_1 \U\times_3\Y]\uf{2}^\top}{\mat{M}\uf{2}[\G\times_1 \U\times_3 \Y]\uf{2}^\top},
    \label{V_update}
\end{equation}
\begin{equation}
    \Y = \Y \circ \frac{[\mat{M}\uf{3} \circ \mat{A}\uf{3}/ \mat{\hat{A}}_{(3)}] [\G\times_1 \U\times_2\V]\uf{3}^\top}{\mat{M}\uf{3}[\G\times_1 \U\times_2 \V]\uf{3}^\top},
    \label{Y_update}
\end{equation}
\begin{equation}
    \G = \G \circ \frac{[\ten{M} \circ \A/ \hatA] \times_1  \U^\top\times_2\V^\top \times_3 \Y^\top}{\ten{M}\times_1 \U\top \times_2 \V^\top \times_3 \Y^\top}.
    \label{G_update}
\end{equation}
In all of the above, $\circ$ denotes elementwise multiplication and $/$ denotes elementwise division.

\section{Variation in latent structure parameter \textit{K}}
\label{SM:bigSweep}
Here, we discuss in further detail the decisions in varying $K$ as we do in the cross-validation tasks from main text. The parameter $K$ defines the dimension of the latent structure in the node set of the network. As such, $K$ can vary from $1$ to $N$, however in the cross-validation tasks in the main text we only vary it from $2$ to $20$ (or from $2$ to $12$ for the Krackhardt \citeyear{krackhardt1987} dataset). In interpreting the ends of the possible range of $K$ values, we note that $K=1$ assumes that each node in the multilayer network belongs to the same latent space. Conversely, $K=N$ allows each node to belong to its own latent space (and, therefore, assumes that there is no latent structure in the node set of the network). To consider how test-AUC varies with $K$ values beyond those which we considered in the main text, we here extend the range of $K$ in two of our multilayer network datasets.

\begin{figure}
    \centering
    \includegraphics[width = 0.45\textwidth]{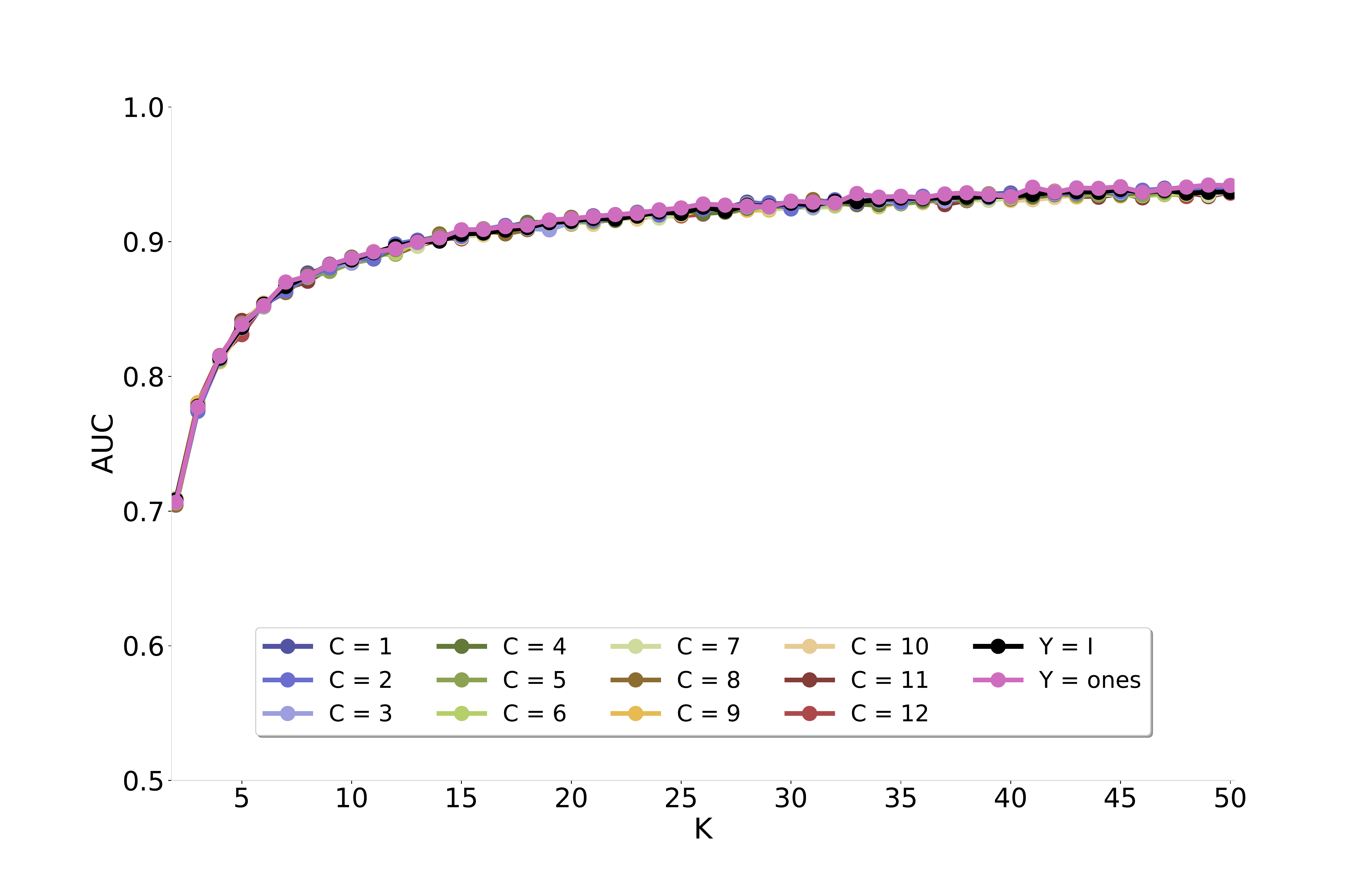}
    \includegraphics[width = 0.45\textwidth]{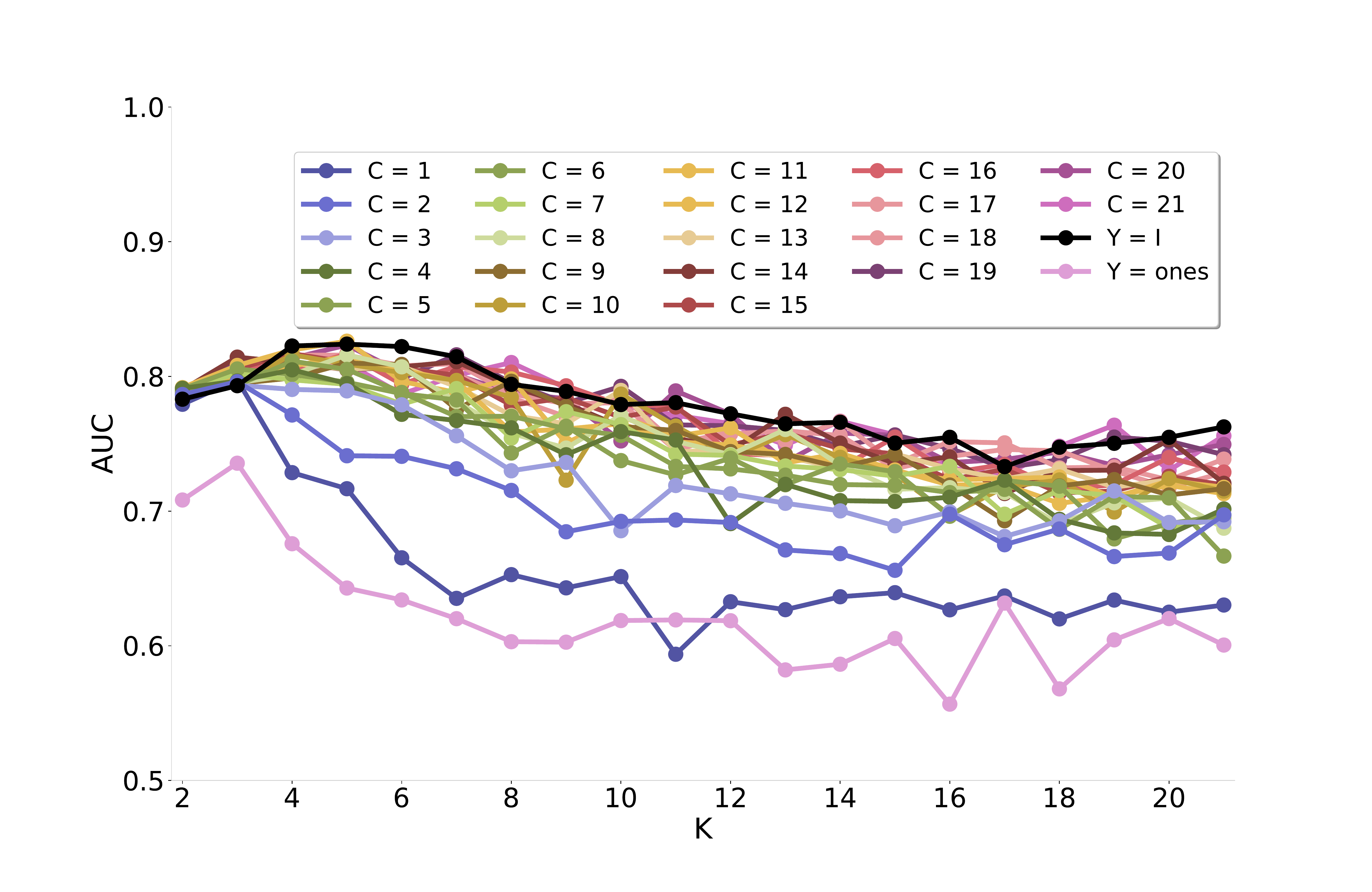}
    \caption{The test-AUC for the fivefold tubular cross-validation task for expanded range of $K$ for (left) the Village 0 multilayer network from \cite{banerjee2013} and (right) the cognitive social structure multilayer network from \cite{krackhardt1987}.}
    \label{fig:bigK}
\end{figure}

First, we consider how test-AUC varies for larger values of $K$ in the Village 0 multilayer social network. In this dataset, there are $843$ nodes, and therefore the largest possible value of $K$ is $843$. However, because it is computationally unreasonable to consider all parameter combinations of $K$ and $C$ for such a large $K$, and because the model where $K$ is this large lacks interpretability (in village networks like the one examined here, the latent node space tends to identify caste membership, see, e.g., \cite{deBacco}), we increase our range of $K$ to be from $K=2$ to $K=50$. We plot the test-AUC for the fivefold tubular cross-validation task for this expanded range of $K$ in \Cref{fig:bigK} (left). We see that test-AUC improves with increasing $K$, but that this increase follows the pattern of increase we saw with the smaller range of $K$. More importantly, the larger $K$ values do not differentiate the layer redundant \acs{ntd} from the layer dependent or layer independent \acs{ntd} model specifications: considering larger values of $K$ does not change the conclusions we draw in the main text. 
 
Next, we consider extending our range of $K$ to its maximum of $K=N$ for the cognitive social structure multilayer network from \cite{krackhardt1987}. 
Although letting $K=N$ assumes that there is no latent structure in the nodes of the network, we take the opportunity in this dataset with a relatively small node set to vary $K$ to its maximum. 
We plot the test-AUC for the fivefold tubular cross-validation task in \Cref{fig:bigK} (right). 
Again, we see that test-AUC doesn't improve when considering larger values of $K$ and, notably, our conclusions from the main text do not change with this increased variation.

\section{Interpretability of the \texorpdfstring{$\Y$}{Y} factor matrix}
\label{SM:Yinterp}
In this section we show the steps necessary to rewrite the core tensor $\G \in \Gspace^{\Gdims}$ of the \acs{ntd} in the basis of $C$ unique reference layers, and how to find the corresponding $\Yhat^*$ matrix, as discussed in \Cref{subsec:yinterp}.
Take as given a set of $C$ unique reference layers, denoted $r^{\text{*}} = \{ r_1, \dots , r_C\}$ where $r_i \in \{1, \dots, L\}$. Define $\G^{\text{*}}$ whose frontal slices are $\G^{\text{*}}_\ell = \sum_{i = 1}^ C y_{r_\ell, i} \mat{G}_\ell$ for $\ell = 1, \dots, C$. Define matrix $\Y^{\text{*}}$ such that rows $r^{\text{*}}$ of $\Y^{\text{*}}$ are identity rows. Specifically, for row $\ell$ of the matrix,
$$ \vec{y}^*_\ell = 
\begin{cases}
e_\ell & \textrm{ if } \ell \in r^*, \\
\vec{y}^*_\ell  & \textrm{ otherwise.}
\end{cases}$$
Define $\bar{r} := \{\ell \mid \ell \notin r^*\}.$ Then row $\vec{y}^*_\ell$ for $\ell \in \bar{r}$ satisfies,
\begin{equation*}
\begin{aligned}
y_{\ell, 1} \mat{G}_1 + y_{\ell, 2} \mat{G}_2 + \cdots + y_{\ell, C} \mat{G}_C = y^*_{\ell, 1} \left(\sum_{i=1}^C y_{r_1, i} \mat{G}_i\right) + y^*_{\ell, 2} \left(\sum_{i=1}^C y_{r_2, i} \mat{G}_i\right) \\ + \cdots + y^*_{\ell, C} \left(\sum_{i=1}^C y_{r_C, i} \mat{G}_i\right). 
\end{aligned}
\end{equation*}
This gives us the system of equations,
\begin{equation*}
\begin{aligned}
y_{\ell, 1} &= y^*_{\ell 1} y_{r_1, 1} + y^*_{\ell, 2} y_{r_2, 1} + \cdots y^*_{\ell, C} y_{r_C, 1} \\
y_{\ell, 2} &= y^*_{\ell 2} y_{r_1, 2} + y^*_{\ell, 2} y_{r_2, 2} + \cdots y^*_{\ell, C} y_{r_C, 2} \\
\vdots \\
y_{\ell, C} &= y^*_{\ell C} y_{r_1, C} + y^*_{\ell, 2} y_{r_2, C} + \cdots y^*_{\ell, C} y_{r_C, C}. \\
\end{aligned}
\end{equation*}
Note that the first equation is the inner product between the first column of $\Y$ subsetted to the rows in $r^*$ with the unknown vector $\vec{y^*}$. That is,
\begin{equation*}
y_{\ell, 1} = \vec{y}_{r^*, 1}^\top \vec{y}_\ell^{*\top}.
\end{equation*}
Then let $\Y_{r^*}$ be the matrix $\Y$ subsetted to the rows in $r^*$. Then the $\ell$th row of matrix $\Y^*$, denoted $\vec{y}^*_\ell$, satisfies the linear system
\begin{equation}
\Y_{r^*}^\top \vec{y}^{*\top}_\ell = \vec{y}_\ell^\top.
\end{equation}
Because there are $(L-C)$ unknown rows of matrix $\Y^*$, there will be $(L-C)$ such linear systems for each $\ell \in \bar{r}.$

Let $\Y^*_{\bar{r}}$ be the matrix $\Y^*$ subsetted to the rows \textit{not }in $r^*$. Then,
\begin{equation}
\Y_{r^*}^\top \Y^{*\top}_{\bar{r}} = \Y_{\bar{r}}^T \,\,\, \Longleftrightarrow \,\,\, 
\Y_{\bar{r}} = \Y^{*}_{\bar{r}}\Y_{r^*}.
\end{equation}
Note that $\Ysub{r^*} \in \R^{C \by C}$ and is invertible if and only if the rows of $\Y$ defined by $r^*$ are not linearly dependent. This highlights the importance in choosing the ``correct'' reference rows. Even though in practice it's unlikely that any two rows, even if poorly chosen, will be exactly linearly dependent, if they are close then $\Ysub{r^*}$ will not be invertible and the transformed $\Y^*$ matrix will not be interpretable.

The best reference layers are often determined by domain knowledge. Absent a principled approach but seeking to make the matrix interpretable, we propose the following heuristic for choosing the best reference layers: choose the $C$ layers such that $\text{det}(\Y_{r^*})$ is furthest from zero. Although searching over the entire space amounts to finding the determinant of $\binom{L}{C}$ different submatrices and isn't practical, one can instead compare the determinant of a handful of different submatrices, where reference layers can be chosen with a combination of (perhaps weak) domain expertise and by inspection of the $\Y$ matrix.

\section{Likelihood ratio test with varying network size}\label{SM:LRT_test}
In this section we explore the relationship between the power of the standard \acl{lrt} and the number of samples\textemdash in our context, the number of nodes and layers in the network. As discussed in \Cref{subsec:independence}, the analysis of Wilks' theorem, which provides the foundation for the \acs{lrt}, is asymptotic. Thus, it is important to explore how the size of the multilayer networks we study impact the power of the statistical tests we use to evaluate layer interdependence.

\begin{figure}
    \centering
    \includegraphics[width = .7\textwidth]{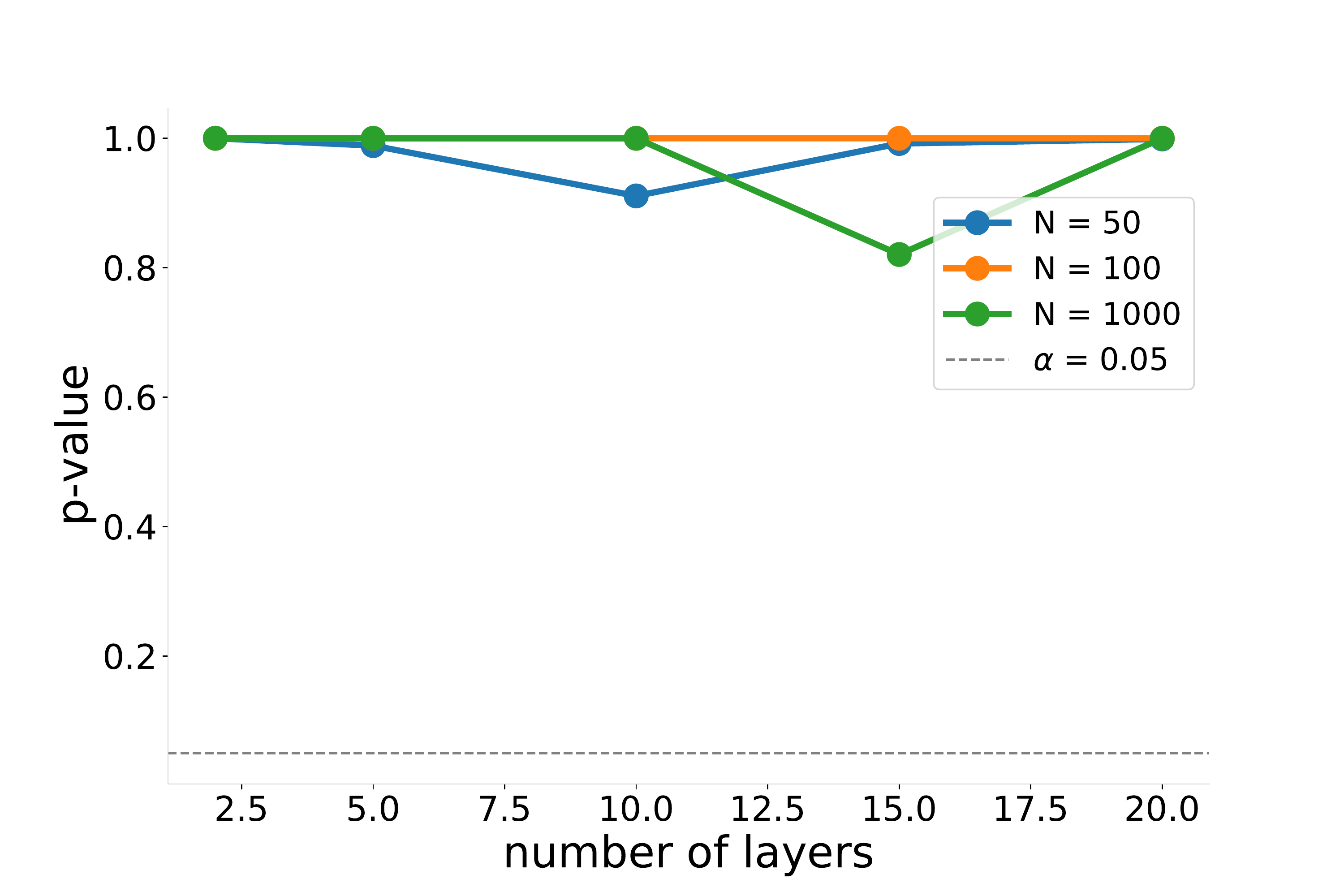}
    \caption{The $p$-values corresponding to the layer redundance \acs{lrt} on synthetic, layer redundant multilayer networks of varying size, both in number of nodes and number of layers. The dashed grey line corresponds to a $p$-value of $0.05$, at which we would reject the null (layer redundant) hypothesis with significance at $0.05$. For every $p$-value above the dashed grey line, we would fail to reject the null hypothesis. Recall that the layer redundance \acs{lrt} compares the layer redundant \acs{ntd} to the layer dependent \acs{ntd} with $C = 2$. In this plot we see that the \acs{lrt} (correctly) fails to reject the null hypothesis for all synthetic networks of varying $L$ and $N$.}
    \label{fig:lrt_tests}
\end{figure}

To explore this, we conduct the following numerical experiment. We generate multiple layer redundant multilayer networks where we vary the number of nodes from $50$ to $1000$ and number of layers from $2$ to $20$. Each layer of the multilayer network is drawn from the same $K=2$ \acs{sbm}, where the first half of the nodes belong to the first latent node space and the second half of the nodes belong to the second latent node space. 
We then estimate a layer redundant \acs{ntd} and a $C=2$ layer dependent \acs{ntd}, each by using the estimation with the highest log-likelihood over 20 random initializations. We use the log-likelihoods from each estimation to conduct a layer redundance \acs{lrt}. We plot the $p$-value from each test in \Cref{fig:lrt_tests}.

We see that we (correctly) fail to reject $H_0$ for all of the synthetic networks of varying $L$ and $N$. The $p$-value corresponding to the \acs{lrt} is consistently far above $\alpha = 0.05$, with no discernible pattern corresponding to either the number of nodes $N$ or the number of layers $L$. Although the concept of a $p$-value is not the same for the \acs{slrt}, conducting the same experiments using the \acs{slrt} we also fail to reject $H_0$ for all of the synthetic networks.

\section{Likelihood ratio test without regularity conditions} \label{SM:UI}
In this section we briefly discuss the split-likelihood ratio test from \citet{wasserman2020} as it relates to the \acs{lrt} for the \acs{ntd}. The \acs{slrt} requires no regularity conditions, and is thus appealing in the setting at hand, wherein Wilks' theorem is not satisfied: the \acs{ntd} is both non-identifiable and has a non-convex log-likelihood. 

To describe the \acs{slrt} we first define the following notation. The sets $D_0$ and $D_1$ split the data into two sets, each roughly equal in size, represented by masking tensors $\ten{M}_0\in \{0,1\}^{\Adims}$ and $\ten{M}_1\in \{0,1\}^{\Adims}$ where
\begin{equation*}
\ten{M}_0(i,j,k) = 
    \begin{cases}
    1 & \textrm{if } (i,j,k) \in D_0,\\
    0 & \textrm{if } (i,j,k) \in D_1
    \end{cases}
    \textrm{ and }
    \ten{M}_1(i,j,k) = 
    \begin{cases}
    1 & \textrm{if } (i,j,k) \in D_1,\\
    0 & \textrm{if } (i,j,k) \in D_0
    \end{cases}
    .
\end{equation*}
The hypotheses for the \acs{slrt} are
\begin{align*}
&H_0\textrm{: The network comes from the nested (layer redundant or dependent) \acs{ntd},}\\
&H_1\textrm{: The network comes from the full (layer independent) \acs{ntd}}. 
\end{align*}
Parameter $\hat{\theta}_1 = [\hat{\G}_1, \hat{\U}_1, \hat{\V}_1, \mat{I}]$ is \textit{any} estimator under the layer independent \acs{ntd} estimated only on $D_1$ using masking tensor $\ten{M}_1$. Parameter $\hat{\theta}_0= [\hat{\G}_0, \hat{\U}_0, \hat{\V}_0, \hat{\Y}_0]$ is the \textit{maximum likelihood estimator} under the nested \acs{ntd} estimated only on $D_0$ using masking tensor $\ten{M}_0$. 
The null hypothesis is rejected at significance level $\alpha$ if
\begin{equation}
\label{eq:slrt}
    \log \frac{\mathcal{L}_0(\hat{\theta}_1)}{\mathcal{L}_0(\hat{\theta}_0)} > \log{1/\alpha},
\end{equation}
where 
\begin{equation}
    \mathcal{L}(\theta) = \prod_{(i,j,k) \in D_0} p_\theta(\A_{ijk})
\end{equation}
is the likelihood associated with $\theta$ only measured over the set $D_0$. Then, under no regularity conditions, the probability that we falsely reject the null hypothesis is,
\begin{equation}
    \textrm{Pr}_{H_0}(reject) \leq \alpha.
\end{equation}
In applying the \acs{slrt} to testing for layer redundance, dependence, and independence in multilayer networks, we use the same tests as discussed in \Cref{deflation} but using the \acs{slrt} defined in \cref{eq:slrt} to reject or fail to reject the null hypothesis. Our results from both the standard and \acs{slrt} are in \Cref{table_dep}.

We estimate $\hat{\theta}_1$ with the layer independent \acs{ntd} with the highest log-likelihood over 20 random initializations of \Cref{nntuck:alg} (see \Cref{code_repo}). We estimate $\hat{\theta}_0$ with the redundant or dependent \acs{ntd} with the highest log-likelihood over 50 random initializations of \Cref{nntuck:alg}. We increase the number of random initializations for estimating the nested \acs{ntd} because the analysis in \citet{wasserman2020} requires $\hat{\theta}_0$ to be the maximum likelihood estimator (MLE), something we cannot guarantee due to non-convexity. A more suitable approach to adapting the \acs{slrt} to this setting would be to instead find $\hat{\theta}_0$ corresponding to the maximum of a (convex) proper relaxation to the likelihood of the \acs{ntd}. An interesting topic of future work is finding a multilayer extension of the semidefinite relaxation of the MLE for the (single layer) \acl{sbm} developed by \citet{amini2018}.

\section{Questions generating the social multilayer networks}
\label{SM:villageQ}
We reproduce the questions generating the multilayer social support networks from the \citet{banerjee2013} and \citet{banerjee2019}. For specific information about the context, original findings, or survey instruments of the research, please refer to the original papers.

\paragraph{Microfinance Villages \citep{banerjee2013}}
The 43 multilayer networks from this research, representing different villages in Karnataka, India, were the result of asking individuals about 12 different types of support. Each layer in the resulting network corresponds to the following relationships.
\begin{lstlisting}
    1: Those from whom the respondent would borrow money
    2: Those to whom the respondent gives advice
    3: Those from whom the respondent gets advice
    4: Those from whom the respondent would borrow material goods
    5: Those to whom the respondent would lend material goods
    6: Those to whom the respondent would lend money
    7: Those from whom the respondent receives medical advice
    8: Non-relatives with whom the respondent socializes
    9: Kin in the village
    10: Those whom the respondent goes to pray with
    11: Those who visit the respondent's home
    12: Those whose homes the respondent visits
\end{lstlisting}

\paragraph{Gossip Villages \citep{banerjee2019}} The 70 multilayer networks from villages in Karnataka, India, were generated from asking individuals the following seven questions, each of which corresponds to a layer in the resulting network.
\begin{lstlisting}
    1. Whose house do you go to in your free time? 
    2. Who comes to your house in their free time?
    3. If you urgently needed kerosene, rice other groceries or money, who do you borrow them from? 
    4. Who comes to your house if he or she needed to borrow kerosene, rice, other groceries or money? 
    5. Who do you ask for advice on matters pertaining to health/finance/farming? 
    6. Who asks you for advice on matters pertaining to health/finance/farming? 
    7. Besides people living in your household, state names of your relatives who are living in this village. 
\end{lstlisting}

\vskip 0.2in
\bibliography{references}

\begin{thebibliography}{85}
\providecommand{\natexlab}[1]{#1}
\providecommand{\url}[1]{\texttt{#1}}
\expandafter\ifx\csname urlstyle\endcsname\relax
  \providecommand{\doi}[1]{doi: #1}\else
  \providecommand{\doi}{doi: \begingroup \urlstyle{rm}\Url}\fi

\bibitem[Aguiar(2021)]{aguiarKrackhardt}
Izabel~P Aguiar.
\newblock Transcription of the 21 adjacency matrices in the appendix of {K}rackhardt's 1987 "cognitive social structures", Jun 2021.
\newblock URL \url{https://github.com/izabelaguiar/krackhardt/}.

\bibitem[Airoldi et~al.(2008)Airoldi, Blei, Fienberg, and Xing]{Airoldi2008}
Edoardo~M Airoldi, David~M Blei, Stephen~E Fienberg, and Eric~P Xing.
\newblock Mixed membership stochastic blockmodels.
\newblock \emph{Journal of Machine Learning research}, 9\penalty0 (Sep):\penalty0 1981--2014, 2008.

\bibitem[Al-Sharoa et~al.(2018)Al-Sharoa, Al-Khassaweneh, and Aviyente]{al2018tensor}
Esraa Al-Sharoa, Mahmood Al-Khassaweneh, and Selin Aviyente.
\newblock Tensor based temporal and multilayer community detection for studying brain dynamics during resting state f{MRI}.
\newblock \emph{IEEE Transactions on Biomedical Engineering}, 66\penalty0 (3):\penalty0 695--709, 2018.

\bibitem[Altenburger and Ugander(2021)]{altenburger2021which}
Kristen~M Altenburger and Johan Ugander.
\newblock Which node attribute prediction task are we solving? within-network, across-network, or across-layer tasks.
\newblock In \emph{Proceedings of the International AAAI Conference on Web and Social Media}, volume~15, pages 38--48, 2021.

\bibitem[Amini and Levina(2018)]{amini2018}
Arash~A Amini and Elizaveta Levina.
\newblock On semidefinite relaxations for the block model.
\newblock \emph{The Annals of Statistics}, 46\penalty0 (1):\penalty0 149--179, 2018.

\bibitem[Bader et~al.(2007)Bader, Harshman, and Kolda]{bader2007temporal}
Brett~W Bader, Richard~A Harshman, and Tamara~G Kolda.
\newblock Temporal analysis of semantic graphs using {ASALSAN}.
\newblock In \emph{Seventh IEEE International Conference on Data Mining}, pages 33--42. IEEE, 2007.

\bibitem[Ball et~al.(2011)Ball, Karrer, and Newman]{BallKarrerNewman2011}
Brian Ball, Brian Karrer, and Mark~EJ Newman.
\newblock Efficient and principled method for detecting communities in networks.
\newblock \emph{Physical Review E}, 2011.

\bibitem[Banerjee et~al.(2013)Banerjee, Chandrasekhar, Duflo, and Jackson]{banerjee2013}
Abhijit Banerjee, Arun~G Chandrasekhar, Esther Duflo, and Matthew~O Jackson.
\newblock The diffusion of microfinance.
\newblock \emph{Science}, 341\penalty0 (6144):\penalty0 1236498, 2013.

\bibitem[Banerjee et~al.(2019)Banerjee, Chandrasekhar, Duflo, and Jackson]{banerjee2019}
Abhijit Banerjee, Arun~G Chandrasekhar, Esther Duflo, and Matthew~O Jackson.
\newblock Using gossips to spread information: Theory and evidence from two randomized controlled trials.
\newblock \emph{The Review of Economic Studies}, 86\penalty0 (6):\penalty0 2453--2490, 2019.

\bibitem[Battiston et~al.(2014)Battiston, Nicosia, and Latora]{battiston2014structural}
Federico Battiston, Vincenzo Nicosia, and Vito Latora.
\newblock Structural measures for multiplex networks.
\newblock \emph{Physical Review E}, 89\penalty0 (3):\penalty0 032804, 2014.

\bibitem[Boccaletti et~al.(2014)Boccaletti, Bianconi, Criado, Del~Genio, G{\'o}mez-Gardenes, Romance, Sendina-Nadal, Wang, and Zanin]{boccaletti2014structure}
Stefano Boccaletti, Ginestra Bianconi, Regino Criado, Charo~I Del~Genio, Jes{\'u}s G{\'o}mez-Gardenes, Miguel Romance, Irene Sendina-Nadal, Zhen Wang, and Massimiliano Zanin.
\newblock The structure and dynamics of multilayer networks.
\newblock \emph{Physics Reports}, 544\penalty0 (1):\penalty0 1--122, 2014.

\bibitem[Breiger et~al.(1975)Breiger, Boorman, and Arabic]{Breiger1975}
R.~Breiger, S.~Boorman, and P.~Arabic.
\newblock An algorithm for clustering relational data with applications to social network analysis and comparison with multidimensional scaling.
\newblock \emph{Journal of Mathematical Psycholoy}, 12:\penalty0 328--383, 1975.

\bibitem[Carlen et~al.(2022)Carlen, de~Dios~Pont, Mentus, Chang, Wang, and Porter]{Carlen2019}
Jane Carlen, Jaume de~Dios~Pont, Cassidy Mentus, Shyr-Shea Chang, Stephanie Wang, and Mason~A Porter.
\newblock Role detection in bicycle-sharing networks using multilayer stochastic block models.
\newblock \emph{Network Science}, 10\penalty0 (1):\penalty0 46--81, 2022.

\bibitem[Carroll and Chang(1970)]{carrollchang}
J~Douglas Carroll and Jih-Jie Chang.
\newblock Analysis of individual differences in multidimensional scaling via an n-way generalization of “{E}ckart-{Y}oung” decomposition.
\newblock \emph{Psychometrika}, 35\penalty0 (3):\penalty0 283--319, 1970.

\bibitem[Chen et~al.(2020)Chen, Moustaki, and Zhang]{chen2020note}
Yunxiao Chen, Irini Moustaki, and Haoran Zhang.
\newblock A note on likelihood ratio tests for models with latent variables.
\newblock \emph{Psychometrika}, 85\penalty0 (4):\penalty0 996--1012, 2020.

\bibitem[Chi and Kolda(2012)]{chi2012}
Eric~C Chi and Tamara~G Kolda.
\newblock On tensors, sparsity, and nonnegative factorizations.
\newblock \emph{SIAM Journal on Matrix Analysis and Applications}, 33\penalty0 (4):\penalty0 1272--1299, 2012.

\bibitem[Cichocki and Zdunek(2007)]{cichocki2007}
Andrzej Cichocki and Rafal Zdunek.
\newblock Multilayer nonnegative matrix factorization using projected gradient approaches.
\newblock \emph{International Journal of Neural Systems}, 17\penalty0 (06):\penalty0 431--446, 2007.

\bibitem[Clauset et~al.(2008)Clauset, Moore, and Newman]{clauset2008}
Aaron Clauset, Cristopher Moore, and Mark~EJ Newman.
\newblock Hierarchical structure and the prediction of missing links in networks.
\newblock \emph{Nature}, 453\penalty0 (7191):\penalty0 98--101, 2008.

\bibitem[De~Bacco et~al.(2017)De~Bacco, Power, Larremore, and Moore]{deBacco}
Caterina De~Bacco, Eleanor~A Power, Daniel~B Larremore, and Cristopher Moore.
\newblock Community detection, link prediction, and layer interdependence in multilayer networks.
\newblock \emph{Physical Review E}, 2017.

\bibitem[De~Domenico(2022)]{dedomenicoWeb}
Manlio De~Domenico.
\newblock Datasets released for reproducibility, 2022.
\newblock URL \url{https://manliodedomenico.com/data.php}.

\bibitem[De~Domenico and Biamonte(2016)]{de2016spectral}
Manlio De~Domenico and Jacob Biamonte.
\newblock Spectral entropies as information-theoretic tools for complex network comparison.
\newblock \emph{Physical Review X}, 6\penalty0 (4):\penalty0 041062, 2016.

\bibitem[De~Domenico et~al.(2013)De~Domenico, Sol{\'e}-Ribalta, Cozzo, Kivel{\"a}, Moreno, Porter, G{\'o}mez, and Arenas]{de2013mathematical}
Manlio De~Domenico, Albert Sol{\'e}-Ribalta, Emanuele Cozzo, Mikko Kivel{\"a}, Yamir Moreno, Mason~A Porter, Sergio G{\'o}mez, and Alex Arenas.
\newblock Mathematical formulation of multilayer networks.
\newblock \emph{Physical Review X}, 3\penalty0 (4):\penalty0 041022, 2013.

\bibitem[De~Domenico et~al.(2014)De~Domenico, Sol{\'e}-Ribalta, G{\'o}mez, and Arenas]{de2014navigability}
Manlio De~Domenico, Albert Sol{\'e}-Ribalta, Sergio G{\'o}mez, and Alex Arenas.
\newblock Navigability of interconnected networks under random failures.
\newblock \emph{Proceedings of the National Academy of Sciences}, 111\penalty0 (23):\penalty0 8351--8356, 2014.

\bibitem[De~Domenico et~al.(2015)De~Domenico, Nicosia, Arenas, and Latora]{de2015structural}
Manlio De~Domenico, Vincenzo Nicosia, Alexandre Arenas, and Vito Latora.
\newblock Structural reducibility of multilayer networks.
\newblock \emph{Nature Communications}, 6\penalty0 (1):\penalty0 1--9, 2015.

\bibitem[Dong et~al.(2020)Dong, Hu, Wang, Sun, and Tang]{dong2020heterogeneous}
Yuxiao Dong, Ziniu Hu, Kuansan Wang, Yizhou Sun, and Jie Tang.
\newblock Heterogeneous network representation learning.
\newblock In \emph{IJCAI}, volume~20, pages 4861--4867, 2020.

\bibitem[F{\'e}votte and Cemgil(2009)]{fevotte}
C{\'e}dric F{\'e}votte and A~Taylan Cemgil.
\newblock Nonnegative matrix factorizations as probabilistic inference in composite models.
\newblock In \emph{2009 17th European Signal Processing Conference}, pages 1913--1917. IEEE, 2009.

\bibitem[Finn et~al.(2019)Finn, Silk, Porter, and Pinter-Wollman]{finn2019use}
Kelly~R Finn, Matthew~J Silk, Mason~A Porter, and Noa Pinter-Wollman.
\newblock The use of multilayer network analysis in animal behaviour.
\newblock \emph{Animal Behaviour}, 149:\penalty0 7--22, 2019.

\bibitem[Fisher and Pinter-Wollman(2021)]{fisher2021using}
David~N Fisher and Noa Pinter-Wollman.
\newblock Using multilayer network analysis to explore the temporal dynamics of collective behavior.
\newblock \emph{Current Zoology}, 67\penalty0 (1):\penalty0 71--80, 2021.

\bibitem[Gallotti and Barthelemy(2015)]{gallotti2015multilayer}
Riccardo Gallotti and Marc Barthelemy.
\newblock The multilayer temporal network of public transport in {G}reat {B}ritain.
\newblock \emph{Scientific Data}, 2\penalty0 (1):\penalty0 1--8, 2015.

\bibitem[Gauvin et~al.(2014)Gauvin, Panisson, and Cattuto]{gauvin2014detecting}
Laetitia Gauvin, Andr{\'e} Panisson, and Ciro Cattuto.
\newblock Detecting the community structure and activity patterns of temporal networks: a non-negative tensor factorization approach.
\newblock \emph{PloS one}, 9\penalty0 (1):\penalty0 e86028, 2014.

\bibitem[Ghasemian et~al.(2020)Ghasemian, Hosseinmardi, Galstyan, Airoldi, and Clauset]{ghasemian2020}
Amir Ghasemian, Homa Hosseinmardi, Aram Galstyan, Edoardo~M Airoldi, and Aaron Clauset.
\newblock Stacking models for nearly optimal link prediction in complex networks.
\newblock \emph{Proceedings of the National Academy of Sciences}, 117\penalty0 (38):\penalty0 23393--23400, 2020.

\bibitem[Gopalan et~al.(2013)Gopalan, Hofman, and Blei]{gopalan2013scalable}
Prem Gopalan, Jake~M Hofman, and David~M Blei.
\newblock Scalable recommendation with poisson factorization.
\newblock \emph{arXiv preprint arXiv:1311.1704}, 2013.

\bibitem[Harshman(1970)]{harshman1970}
Richard~A Harshman.
\newblock Foundations of the {PARAFAC} procedure: Models and conditions for an explanatory multi-mode factor analysis.
\newblock \emph{UCLA Working Papers in Phonetics}, 16:\penalty0 1--84, 1970.

\bibitem[Harshman(1978)]{harshman1978}
Richard~A Harshman.
\newblock Models for analysis of asymmetrical relationships among {N} objects or stimuli.
\newblock In \emph{First Joint Meeting of the Psychometric Society and the Society of Mathematical Psychology, Hamilton, Ontario, 1978}, 1978.

\bibitem[Harshman and Lundy(1996)]{harshman1996}
Richard~A Harshman and Margaret~E Lundy.
\newblock Uniqueness proof for a family of models sharing features of {T}ucker's three-mode factor analysis and {PARAFAC/CANDECOMP}.
\newblock \emph{Psychometrika}, 61\penalty0 (1):\penalty0 133--154, 1996.

\bibitem[Hien and Gillis(2021)]{hien2020}
Le~Thi~Khanh Hien and Nicolas Gillis.
\newblock Algorithms for nonnegative matrix factorization with the {K}ullback--{L}eibler divergence.
\newblock \emph{Journal of Scientific Computing}, 87\penalty0 (3):\penalty0 1--32, 2021.

\bibitem[Holland et~al.(1983)Holland, Laskey, and Leinhardt]{Holland1983}
Paul~W Holland, Kathryn~Blackmond Laskey, and Samuel Leinhardt.
\newblock Stochastic blockmodels: First steps.
\newblock \emph{Social Networks}, 5\penalty0 (2):\penalty0 109--137, 1983.

\bibitem[Hu et~al.(2020)Hu, Fey, Zitnik, Dong, Ren, Liu, Catasta, and Leskovec]{hu2020open}
Weihua Hu, Matthias Fey, Marinka Zitnik, Yuxiao Dong, Hongyu Ren, Bowen Liu, Michele Catasta, and Jure Leskovec.
\newblock Open graph benchmark: Datasets for machine learning on graphs.
\newblock \emph{Advances in neural information processing systems}, 33:\penalty0 22118--22133, 2020.

\bibitem[Kao and Porter(2018)]{kao2018layer}
Ta-Chu Kao and Mason~A Porter.
\newblock Layer communities in multiplex networks.
\newblock \emph{Journal of Statistical Physics}, 173\penalty0 (3):\penalty0 1286--1302, 2018.

\bibitem[Karrer and Newman(2011)]{KarrerNewman}
Brian Karrer and Mark~EJ Newman.
\newblock Stochastic blockmodels and community structure in networks.
\newblock \emph{Physical Review E}, 83\penalty0 (1):\penalty0 016107, 2011.

\bibitem[Kasai(2018)]{kasai}
Hiroyuki Kasai.
\newblock Stochastic variance reduced multiplicative update for nonnegative matrix factorization.
\newblock In \emph{2018 IEEE International Conference on Acoustics, Speech and Signal Processing}, pages 6338--6342, 2018.

\bibitem[Kim and Choi(2007)]{kimChoi}
Yong-Deok Kim and Seungjin Choi.
\newblock Nonnegative {T}ucker decomposition.
\newblock In \emph{IEEE Conference on Computer Vision and Pattern Recognition}, 2007.

\bibitem[Kivel{\"a} et~al.(2014)Kivel{\"a}, Arenas, Barthelemy, Gleeson, Moreno, and Porter]{kivela2014multilayer}
Mikko Kivel{\"a}, Alex Arenas, Marc Barthelemy, James~P Gleeson, Yamir Moreno, and Mason~A Porter.
\newblock Multilayer networks.
\newblock \emph{Journal of Complex Networks}, 2\penalty0 (3):\penalty0 203--271, 2014.

\bibitem[Kolda and Bader(2009)]{Kolda2009}
Tamara~G Kolda and Brett~W Bader.
\newblock Tensor decompositions and applications.
\newblock \emph{SIAM Review}, 51\penalty0 (3):\penalty0 455--500, 2009.

\bibitem[Kossaifi et~al.(2016)Kossaifi, Panagakis, Anandkumar, and Pantic]{kossaifi2016}
Jean Kossaifi, Yannis Panagakis, Anima Anandkumar, and Maja Pantic.
\newblock Tensorly: Tensor learning in python.
\newblock \emph{arXiv preprint arXiv:1610.09555}, 2016.

\bibitem[Krackhardt(1987)]{krackhardt1987}
David Krackhardt.
\newblock Cognitive social structures.
\newblock \emph{Social Networks}, 9\penalty0 (2):\penalty0 109--134, 1987.

\bibitem[Larremore et~al.(2013)Larremore, Clauset, and Buckee]{larremore2013network}
Daniel~B Larremore, Aaron Clauset, and Caroline~O Buckee.
\newblock A network approach to analyzing highly recombinant malaria parasite genes.
\newblock \emph{PLoS Computational Biology}, 9\penalty0 (10):\penalty0 e1003268, 2013.

\bibitem[Lee and Seung(2000)]{LeeSeung}
Daniel Lee and H~Sebastian Seung.
\newblock Algorithms for non-negative matrix factorization.
\newblock \emph{Advances in Neural Information Processing Systems}, 13, 2000.

\bibitem[Lee and Seung(1999)]{nmf1999}
Daniel~D Lee and H~Sebastian Seung.
\newblock Learning the parts of objects by non-negative matrix factorization.
\newblock \emph{Nature}, 401\penalty0 (6755):\penalty0 788--791, 1999.

\bibitem[Liben-Nowell and Kleinberg(2007)]{liben2007}
David Liben-Nowell and Jon Kleinberg.
\newblock The link-prediction problem for social networks.
\newblock \emph{Journal of the American Society for Information Science and Technology}, 58\penalty0 (7):\penalty0 1019--1031, 2007.

\bibitem[Mucha et~al.(2010)Mucha, Richardson, Macon, Porter, and Onnela]{mucha2010community}
Peter~J Mucha, Thomas Richardson, Kevin Macon, Mason~A Porter, and Jukka-Pekka Onnela.
\newblock Community structure in time-dependent, multiscale, and multiplex networks.
\newblock \emph{Science}, 328\penalty0 (5980):\penalty0 876--878, 2010.

\bibitem[Nayar et~al.(2015)Nayar, Miller, Geyer, Caceres, Smith, and Nadakuditi]{nayar2015}
Himanshu Nayar, Benjamin~A Miller, Kelly Geyer, Rajmonda~S Caceres, Steven~T Smith, and Raj~Rao Nadakuditi.
\newblock Improved hidden clique detection by optimal linear fusion of multiple adjacency matrices.
\newblock In \emph{2015 49th Asilomar Conference on Signals, Systems and Computers}, pages 1520--1524. IEEE, 2015.

\bibitem[Nickel et~al.(2011)Nickel, Tresp, and Kriegel]{Nickel2011}
Maximilian Nickel, Volker Tresp, and Hans-Peter Kriegel.
\newblock A three-way model for collective learning on multi-relational data.
\newblock In \emph{International Conference on Machine Learning}, volume~11, pages 809--816, 2011.

\bibitem[Overgoor et~al.(2019)Overgoor, Benson, and Ugander]{overgoor2019}
Jan Overgoor, Austin Benson, and Johan Ugander.
\newblock Choosing to grow a graph: modeling network formation as discrete choice.
\newblock In \emph{The World Wide Web Conference}, pages 1409--1420, 2019.

\bibitem[Paatero and Tapper(1994)]{nmf1994}
Pentti Paatero and Unto Tapper.
\newblock Positive matrix factorization: A non-negative factor model with optimal utilization of error estimates of data values.
\newblock \emph{Environmetrics}, 5\penalty0 (2):\penalty0 111--126, 1994.

\bibitem[Paul and Chen(2016)]{paul2016consistent}
Subhadeep Paul and Yuguo Chen.
\newblock Consistent community detection in multi-relational data through restricted multi-layer stochastic blockmodel.
\newblock \emph{Electronic Journal of Statistics}, 10\penalty0 (2):\penalty0 3807--3870, 2016.

\bibitem[Power(2017)]{power2017social}
Eleanor~A Power.
\newblock Social support networks and religiosity in rural {S}outh {I}ndia.
\newblock \emph{Nature Human Behaviour}, 1\penalty0 (3):\penalty0 1--6, 2017.

\bibitem[Racz and Sridhar(2021)]{racz2021correlated}
Miklos Racz and Anirudh Sridhar.
\newblock Correlated stochastic block models: Exact graph matching with applications to recovering communities.
\newblock \emph{Advances in Neural Information Processing Systems}, 34, 2021.

\bibitem[Roethlisberger and Dickson(1939)]{roethlisberger1939and}
Fritz~Jules Roethlisberger and William~J Dickson.
\newblock \emph{Management and the Worker}, volume~5.
\newblock Psychology Press, 1939.

\bibitem[Rohe and Zeng(2020)]{rohe2020}
Karl Rohe and Muzhe Zeng.
\newblock Vintage factor analysis with varimax performs statistical inference.
\newblock \emph{arXiv preprint arXiv:2004.05387}, 2020.

\bibitem[Sampson(1969)]{sampson_crisis_1969}
Samuel~F Sampson.
\newblock \emph{Crisis in a cloister}.
\newblock PhD thesis, Ph. D. Thesis. Cornell University, Ithaca, 1969.

\bibitem[Schein et~al.(2015)Schein, Paisley, Blei, and Wallach]{Schein2015}
Aaron Schein, John Paisley, David~M Blei, and Hanna Wallach.
\newblock Bayesian poisson tensor factorization for inferring multilateral relations from sparse dyadic event counts.
\newblock In \emph{Proceedings of the 21th ACM SIGKDD International Conference on Knowledge Discovery and Data Mining}, pages 1045--1054, 2015.

\bibitem[Schein et~al.(2016)Schein, Zhou, Blei, and Wallach]{schein2016bayesian}
Aaron Schein, Mingyuan Zhou, David Blei, and Hanna Wallach.
\newblock Bayesian poisson {T}ucker decomposition for learning the structure of international relations.
\newblock In \emph{International Conference on Machine Learning}, pages 2810--2819. PMLR, 2016.

\bibitem[Spector et~al.(2023)Spector, Candes, and Lei]{spectordiscussion}
Asher Spector, Emmanuel Candes, and Lihua Lei.
\newblock A discussion of {T}se and {D}avidson (2022) ``{A} note on universal inference''.
\newblock \emph{Stat}, 2023.

\bibitem[Stanley et~al.(2016)Stanley, Shai, Taylor, and Mucha]{stanley2016}
Natalie Stanley, Saray Shai, Dane Taylor, and Peter~J Mucha.
\newblock Clustering network layers with the strata multilayer stochastic block model.
\newblock \emph{IEEE transactions on network science and engineering}, 3\penalty0 (2):\penalty0 95--105, 2016.

\bibitem[Strieder and Drton(2022)]{strieder2022choice}
David Strieder and Mathias Drton.
\newblock On the choice of the splitting ratio for the split likelihood ratio test.
\newblock \emph{Electronic Journal of Statistics}, 16\penalty0 (2):\penalty0 6631--6650, 2022.

\bibitem[Sun et~al.(2009)Sun, Papadimitriou, Lin, Cao, Liu, and Qian]{sun2009multivis}
Jimeng Sun, Spiros Papadimitriou, Ching-Yung Lin, Nan Cao, Shixia Liu, and Weihong Qian.
\newblock Multivis: Content-based social network exploration through multi-way visual analysis.
\newblock In \emph{Proceedings of the 2009 SIAM International Conference on Data Mining}, pages 1064--1075. SIAM, 2009.

\bibitem[Tarr{\'e}s-Deulofeu et~al.(2019)Tarr{\'e}s-Deulofeu, Godoy-Lorite, Guimera, and Sales-Pardo]{tarres2019tensorial}
Marc Tarr{\'e}s-Deulofeu, Antonia Godoy-Lorite, Roger Guimera, and Marta Sales-Pardo.
\newblock Tensorial and bipartite block models for link prediction in layered networks and temporal networks.
\newblock \emph{Physical Review E}, 99\penalty0 (3):\penalty0 032307, 2019.

\bibitem[Taylor et~al.(2016)Taylor, Shai, Stanley, and Mucha]{taylor2016enhanced}
Dane Taylor, Saray Shai, Natalie Stanley, and Peter~J Mucha.
\newblock Enhanced detectability of community structure in multilayer networks through layer aggregation.
\newblock \emph{Physical review letters}, 116\penalty0 (22):\penalty0 228301, 2016.

\bibitem[Taylor et~al.(2017)Taylor, Caceres, and Mucha]{taylor2017super}
Dane Taylor, Rajmonda~S Caceres, and Peter~J Mucha.
\newblock Super-resolution community detection for layer-aggregated multilayer networks.
\newblock \emph{Physical Review X}, 7\penalty0 (3):\penalty0 031056, 2017.

\bibitem[Taylor et~al.(2021)Taylor, Porter, and Mucha]{taylor2021tunable}
Dane Taylor, Mason~A Porter, and Peter~J Mucha.
\newblock Tunable eigenvector-based centralities for multiplex and temporal networks.
\newblock \emph{Multiscale Modeling \& Simulation}, 19\penalty0 (1):\penalty0 113--147, 2021.

\bibitem[Torrey and Shavlik(2010)]{torrey2010transfer}
Lisa Torrey and Jude Shavlik.
\newblock Transfer learning.
\newblock In \emph{Handbook of Research on Machine Learning Applications and Trends: Algorithms, Methods, and Techniques}, pages 242--264. IGI global, 2010.

\bibitem[Tse and Davison(2022)]{tse2022note}
Timmy Tse and Anthony~C Davison.
\newblock A note on universal inference.
\newblock \emph{Stat}, 2022.

\bibitem[Tucker(1966)]{tucker1966}
Ledyard~R Tucker.
\newblock Some mathematical notes on three-mode factor analysis.
\newblock \emph{Psychometrika}, 31\penalty0 (3):\penalty0 279--311, 1966.

\bibitem[Valles-Catala et~al.(2016)Valles-Catala, Massucci, Guimera, and Sales-Pardo]{valles2016}
Toni Valles-Catala, Francesco~A Massucci, Roger Guimera, and Marta Sales-Pardo.
\newblock Multilayer stochastic block models reveal the multilayer structure of complex networks.
\newblock \emph{Physical Review X}, 6\penalty0 (1):\penalty0 011036, 2016.

\bibitem[Wang and Zou(2018)]{wang2018new}
Dingjie Wang and Xiufen Zou.
\newblock A new centrality measure of nodes in multilayer networks under the framework of tensor computation.
\newblock \emph{Applied Mathematical Modelling}, 54:\penalty0 46--63, 2018.

\bibitem[Wang and Zeng(2019)]{wang2019multiway}
Miaoyan Wang and Yuchen Zeng.
\newblock Multiway clustering via tensor block models.
\newblock \emph{Advances in neural information processing systems}, 32, 2019.

\bibitem[Wasserman et~al.(2020)Wasserman, Ramdas, and Balakrishnan]{wasserman2020}
Larry Wasserman, Aaditya Ramdas, and Sivaraman Balakrishnan.
\newblock Universal inference.
\newblock \emph{Proceedings of the National Academy of Sciences}, 117\penalty0 (29):\penalty0 16880--16890, 2020.

\bibitem[White et~al.(1976)White, Boorman, and Breiger]{White1976}
Harrison~C White, Scott~A Boorman, and Ronald~L Breiger.
\newblock Social structure from multiple networks. {I.} blockmodels of roles and positions.
\newblock \emph{American Journal of Sociology}, 81\penalty0 (4):\penalty0 730--780, 1976.

\bibitem[Wilks(1938)]{wilks1938}
Samuel~S Wilks.
\newblock The large-sample distribution of the likelihood ratio for testing composite hypotheses.
\newblock \emph{The Annals of Mathematical Statistics}, 9\penalty0 (1):\penalty0 60--62, 1938.

\bibitem[Wills and Meyer(2020)]{wills2020metrics}
Peter Wills and Fran{\c{c}}ois~G Meyer.
\newblock Metrics for graph comparison: a practitioner’s guide.
\newblock \emph{PloS one}, 15\penalty0 (2):\penalty0 e0228728, 2020.

\bibitem[Wu et~al.(2019)Wu, He, Zhang, Chen, Sun, Liu, Zhang, and Poor]{wu2019tensor}
Mincheng Wu, Shibo He, Yongtao Zhang, Jiming Chen, Youxian Sun, Yang-Yu Liu, Junshan Zhang, and H~Vincent Poor.
\newblock A tensor-based framework for studying eigenvector multicentrality in multilayer networks.
\newblock \emph{Proceedings of the National Academy of Sciences}, 116\penalty0 (31):\penalty0 15407--15413, 2019.

\bibitem[Yenigun et~al.(2016)Yenigun, Ertan, and Siciliano]{cssTools}
Deniz Yenigun, Gunes Ertan, and Michael Siciliano.
\newblock \emph{cssTools: Cognitive Social Structure Tools}, 2016.
\newblock URL \url{https://CRAN.R-project.org/package=cssTools}.
\newblock R package version 1.0.

\bibitem[Zhao et~al.(2012)Zhao, Levina, and Zhu]{zhao2012consistency}
Yunpeng Zhao, Elizaveta Levina, and Ji~Zhu.
\newblock Consistency of community detection in networks under degree-corrected stochastic block models.
\newblock \emph{The Annals of Statistics}, 40\penalty0 (4):\penalty0 2266--2292, 2012.

\bibitem[Zhou et~al.(2015)Zhou, Cichocki, Zhao, and Xie]{zhou2015efficient}
Guoxu Zhou, Andrzej Cichocki, Qibin Zhao, and Shengli Xie.
\newblock Efficient nonnegative tucker decompositions: Algorithms and uniqueness.
\newblock \emph{IEEE Transactions on Image Processing}, 24\penalty0 (12):\penalty0 4990--5003, 2015.

\end{thebibliography}
\bibliographystyle{plainnat}
\end{document}